\documentclass[aps,journal=prb,twocolumn,noeprint]{revtex4-1}
\usepackage{graphicx}
\usepackage{epstopdf}
\usepackage{bmpsize}
\usepackage{amsmath,amsfonts,amsthm,bm}
\usepackage{tikz}
\usetikzlibrary{decorations.pathreplacing}
\usepackage{subcaption}
\usepackage[bookmarks=false]{hyperref}
\hypersetup{colorlinks=true, citecolor=blue, urlcolor=blue, linkcolor=blue}

\makeatletter
\renewcommand{\maketag@@@}[1]{\hbox{\m@th\normalsize\normalfont#1}}%
\makeatother

\graphicspath{{figs/}}

\begin{document}

\title{Coherent time-dependent oscillations and temporal correlations in triangular triple quantum dots}
\author{Samuel L. Rudge}
\affiliation{Institute of Physics, Albert-Ludwigs University Freiburg, Freiburg, 79100, Germany}
\author{Daniel S. Kosov}
\affiliation{College of Science and Engineering, James Cook University, Townsville, QLD, 4814, Australia }

\begin{abstract}
\noindent The fluctuation behavior of triple quantum dots (TQDs) has, so far, largely focused on current cumulants in the long-time limit via full counting statistics. Given that (TQDs) are non-trivial open quantum systems with many interesting features, such as Aharonov-Bohm interference and coherent population blocking, new fluctuating-time statistics, such as the waiting time distribution (WTD), may provide more information than just the current cumulants alone. In this paper, we use a Born-Markov master equation to calculate the standard and higher-order WTDs for coherently-coupled TQDs  arrayed in triangular ring geometries for several transport regimes. In all cases we find that the WTD displays coherent oscillations that correspond directly to individual time-dependent dot occupation probabilities, a result also reported recently in Ref.\cite{Engelhardt2019}. Our analysis, however, goes beyond the single-occupancy and single waiting time regimes, investigating waiting time behavior for TQDs occupied by multiple electrons and with finite electron-electron interactions. We demonstrate that, in these regimes of higher occupancy, quantum coherent effects introduce correlations between successive waiting times, which we can tune via an applied magnetic field. We also show that correlations can be used to distinguish between TQD configurations that have identical FCS and that dark states can be tuned with Aharonov-Bohm interference for more complicated regimes than single-occupancy.
\end{abstract}

\maketitle

\section{Introduction}

Due to their small size, great chemical variability, fast transistor signals, and advanced assembly, molecular-sized devices offer the potential to supersede contemporary silicon electronic components \cite{Guo2007,Scheer2010,Xiang2016}. These advantages are inherent to nanoscale electronic junction design, in which a nanostructure, such as a molecule or atomic bridge, is chemically bonded to macroscopic metal electrodes. Beyond just technological advantages, furthermore, nanoscale electronic junctions are excellent systems for investigating nonequilibrium quantum physics and are host to many novel phenomena, such as cotunneling \cite{Koch2006}, vibrationally coupled transport \cite{Schinabeck2016,Schinabeck2018}, and interference effects \cite{Ihn2007,Strambini2009,Kobayashi2002}. 

Triple quantum dots (TQDs) are open quantum systems with particularly interesting and complex features, which we will explore in this paper. Experimentally, TQDs are usually formed by placing a 2-dimensional electron gas (2DEG) below the surface of a heterostructure, such as GaAs/AlGaAs, depositing metal gates arrayed in an appropriate geometry, and then depleting electrons below the surface via a negative voltage \cite{Gaudreau2006,Korkusinski2007,Gaudreau2007,Gaudreau2009}. This process is not an easy task, however, and, while singular quantum dot synthesis has been viable for decades, it is only in the last 15 years that experimental techniques have become sophisticated enough to create coherently-coupled doubly \cite{vanderWiel2002} and triply \cite{Gaudreau2006,Korkusinski2007,Gaudreau2007,Gaudreau2009,Schroer2007,Rogge2008,Amaha2008,Amaha2013} coupled systems. Further experiments have shown, among other things, that TQDs are strong candidates for qubit creation in quantum computing \cite{PioroLadriere2008,Takakura2010,Russ2017,Laird2010,Sanchez2014,Luczak2014,Luczak2016}.

\begin{figure}
\begin{centering}
\begin{subfigure}[b]{0.49\columnwidth}
\includegraphics[width = \textwidth]{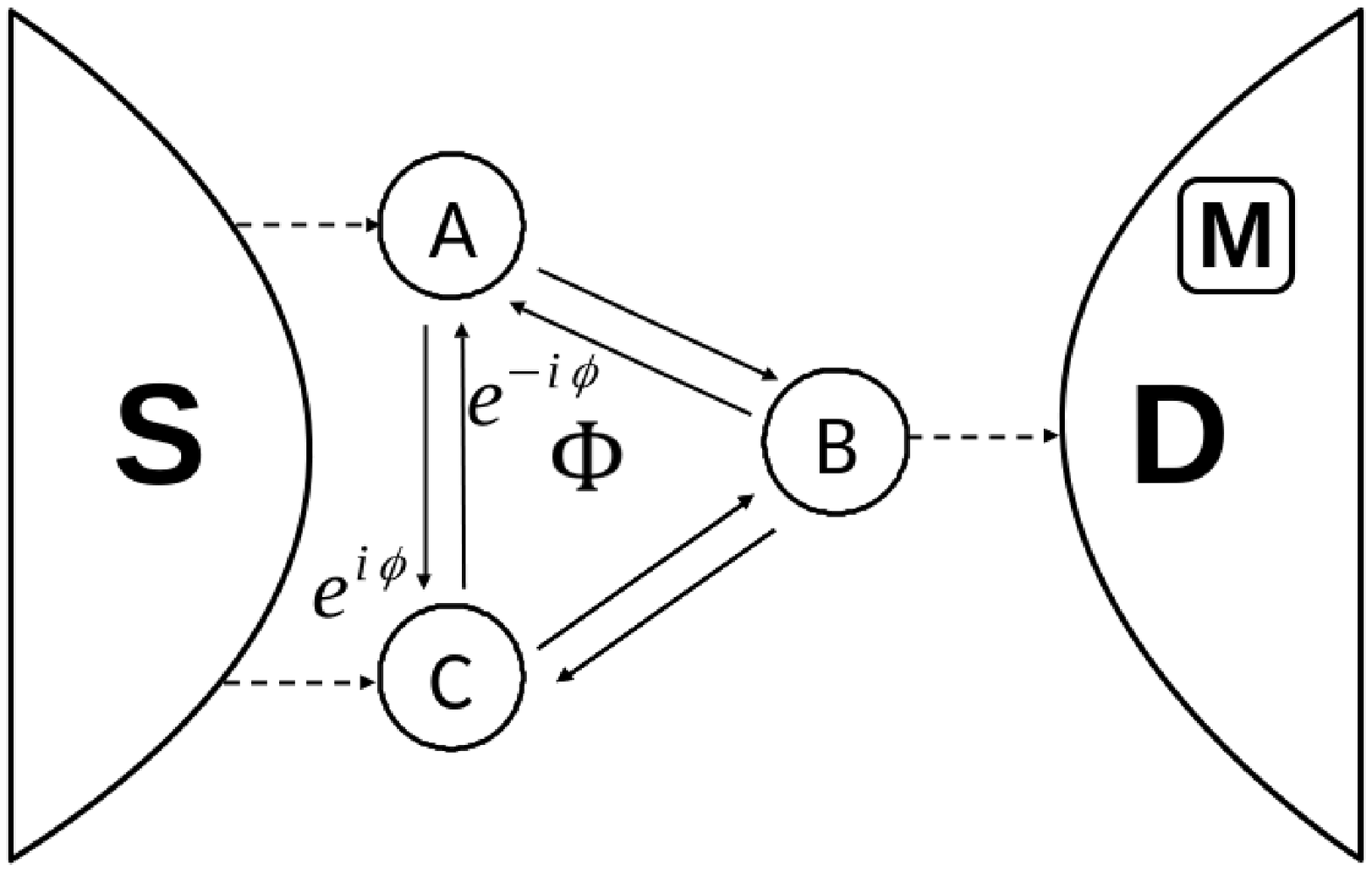}
\caption{}
\label{fig: 1a}
\end{subfigure}
\begin{subfigure}[b]{0.49\columnwidth}
\includegraphics[width = \textwidth]{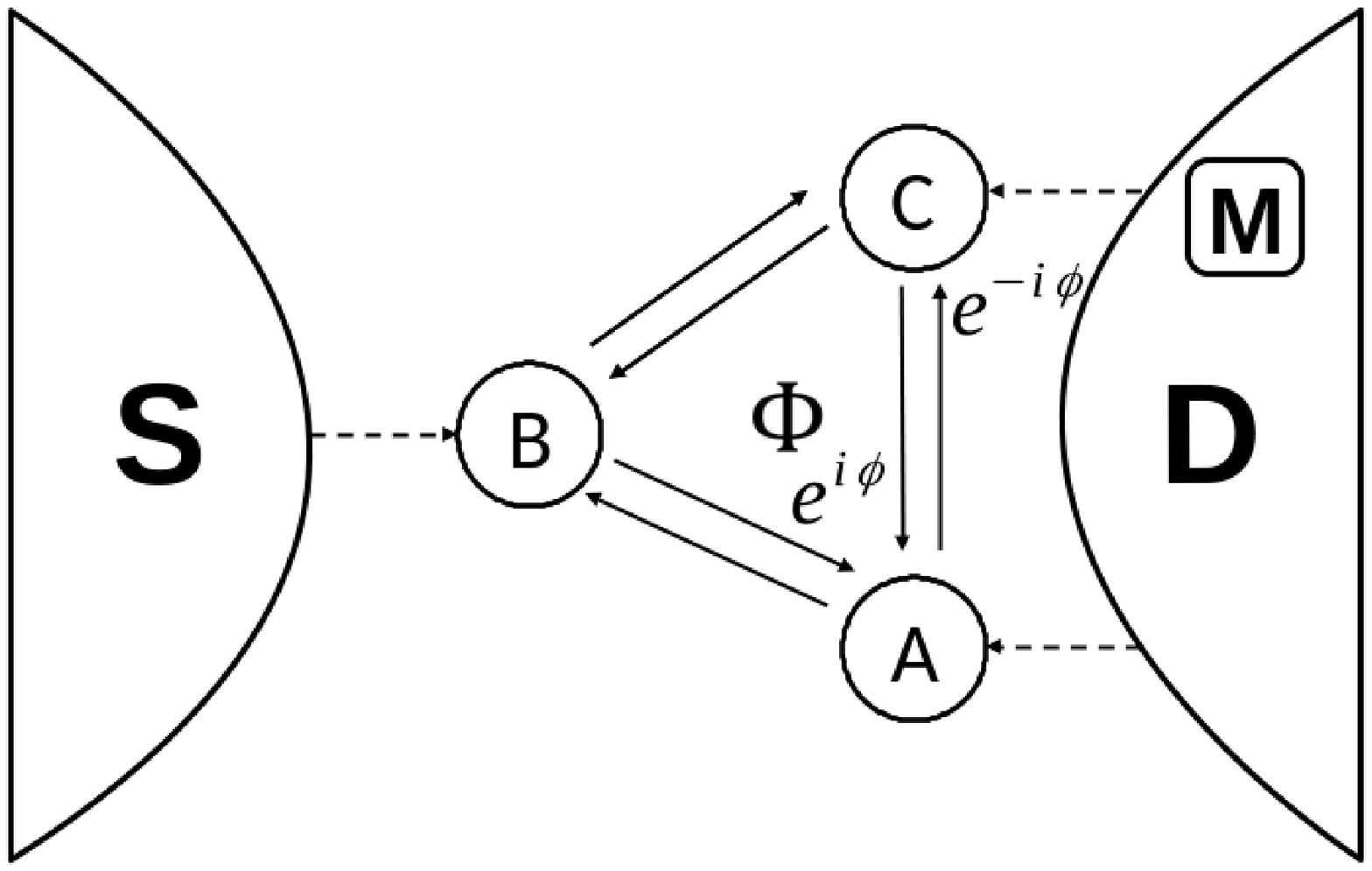}
\caption{}
\label{fig: 1b}
\end{subfigure}
\caption{Schematics of the two triangular TQD geometries. In (a) dots $A$ and $C$ are coupled to the source, while dot $B$ alone is coupled to the drain, and (b) contains the opposite configuration, obtained by coupling dots $A$ and $C$ to the drain and dot $B$ to the source. A magnetic flux, $\Phi$, is applied perpendicular to the interior of the TQD, which induces a phase difference, $\phi$, for the two different paths an electron can take through the configuration: a manifestation of the Aharonov-Bohm effect. We always choose a gauge in which this phase difference is entirely factored onto the hopping strength between dots $A$ and $C$.}
\end{centering}
\end{figure}

Since coherent multi-dot synthesis is a relatively new development, the accompanying theory has in many cases far outstripped experiment. Due to the multiple geometries, large number of tunable parameters, and complex Fock space, TQDs admit a wide variety of novel phenomena, such as quantum phase transitions \cite{Mitchell2009}, charge frustration \cite{Seo2013}, Kondo physics \cite{Jiang2005,Zitko2006,Numata2009,Cheng2017,Lopez2013,Vernek2009,Oguri2011,Kuzmenko2006a}, and spin-entangled current \cite{Saraga2003}. Of these possible configurations, in this paper we are concerned with TQDs are arranged in a triangular geometry, as in Fig.(\ref{fig: 1a}) and Fig.(\ref{fig: 1b}). In these geometries, the three dots form a closed loop such that if one applies a magnetic flux, $\Phi$, perpendicular to the loop interior, one can induce Aharonov-Bohm interference as an electron transports through the system.

For triangular TQDs, the Aharonov-Bohm effect is often analyzed in conjunction with coherent population trapping. This occurs when the coupling parameters of the TQD are tuned so as to form a ``dark" state \cite{Groth2006,Michaelis2006}: a coherent superposition of dot states that block the current. In the configuration depicted in Fig.(\ref{fig: 1a}), and when the occupancy of the three dots is limited to one electron, Emary \cite{Emary2007b} demonstrated that there exists certain parameters for which $|\psi_{\text{dark}}\rangle = a|A\rangle + c|C\rangle$ is an eigenstate of the Hamiltonian, decoupling the $B$ dot, and thus the drain. He also showed that a Aharonov-Bohm interference lifts this coherent blockade and produces oscillations in the stationary current. Most importantly for us, he found that, at the peak of destructive interference, the Fano factor is super-Poissonian, possibly due to avalanche tunneling. 

The dark state is not just a feature of the single electron regime; P\"{o}ltl et al. \cite{Poltl2009,Poltl2013} and Busl et al. \cite{Busl2010} have shown that for double-occupancy of the TQD, with each dot now modeled as an Anderson impurity, spin effects could produce coherent electron trapping if the inter-dot coupling $U_{\nu\nu '}$ equaled the intra-dot coupling $U_{\nu\nu}$. Dark states are evidently an interesting quantum phenomenon, but they can also produce current rectification and negative differential resistance \cite{Kostyrko2009}: two promising technological applications. If not penetrated by a magnetic flux, a strong molecule-electrode coupling can lift the coherent blockade, as Weymann et al. \cite{Weymann2011} showed by including cotunneling processes in a triangular TQD, although Noiri et al. \cite{Noiri2017} has found that cotunneling can introduce an additional spin blockade in serially coupled TQDs. Coherent population trapping also occurs for higher occupancies \cite{Niklas2017}; for example, under triple-occupancy a coherent spin blockade forms for certain parameters under an applied electric field \cite{Bulka2011}.

There are clearly interesting current fluctuations in a triangular TQD, which have so far been investigated mainly via the current cumulants in the long-time limit \cite{Lai2018}. We propose to use fluctuating-time statistics, like the waiting time distribution (WTD) \cite{Brandes2008} and its cumulants, to analyze coherent charge transport through triangular TQDs in a variety of regimes, and in particular those systems considered by Emary \cite{Emary2007b} and P\"{o}ltl et al. \cite{Poltl2009,Poltl2013} for correlations arising from quantum interference in the dark state and moderated by the Aharonov-Bohm effect. WTDs are a relatively recent addition to the fluctuation statistics toolbox and have consistently proven to contain complementary information not available from the long-time FCS alone \cite{Srinivas2010,Thomas2013,Kosov2017b,Dasenbrook2015,Albert2011,Albert2012,Albert2016,Albert2014,Chevallier2016,Haack2014,Ptaszynski2017a,Ptaszynski2017,Ptaszynski2018,Stegmann2020}; their ability to describe short-time behavior is particularly relevant when correlations exist between successive waiting times. Previous work on WTDs for TQDs has not investigated correlations \cite{Welack2009}; however, within only the last two years Engelhardt and Cao \cite{Engelhardt2019} have shown that the WTD for a similar TQD configuration displays oscillatory behavior, which corresponds directly to the relevant dot occupancies. This is similar in some respect to the WTDs of other coherent systems, such as quantum dot spin-valves \cite{Davis2021,Stegmann2018}. We find similar results for the geometries in Fig.(\ref{fig: 1a}) and Fig.(\ref{fig: 1b}), albeit for more complicated transport regimes.

In Section \ref{sec: Theory}, we first briefly outline the general Born-Markov master equation (BMME) and the superoperator form necessary to calculate all fluctuation statistics. We then discuss in depth all TQD models and transport regimes used in our analysis in Section \ref{sec: TQD models and transport regimes}. Section \ref{sec: Fluctuation statistics}, meanwhile, introduces the FCS and WTD, as well as a discussion on renewal and non-renewal behavior. We present results for all three transport scenarios in Section \ref{sec: TQD Results} well as an explanation for each, with the conclusions contained in Section \ref{sec: Conclusion}. Relevant derivations are displayed in Appendix \ref{app: Master equation derivation} and \ref{app: TQD triple-occupancy}.

Throughout the paper we use natural units: $\hbar = e = k_{B} = 1$.

\section{Theory}\label{sec: Theory}

Although we exclusively investigate the electrode-TQD-electrode configurations displayed in Fig.(\ref{fig: 1a}) and Fig.(\ref{fig: 1b}), these two schematics are specific examples of the general design paradigm in molecular electronics: a nanostructure coupled to source and drain metal electrodes. In such electrode-nanostructure-electrode devices, the fluctuation statistics naturally originate from the difference between the number of electrons transferred forward to the drain from the TQD in time $t$, $n_{F}(t)$, and the number of electrons back transferred from the drain to the TQD in time $t$, $n_{B}(t)$:
\begin{align}
n(t) & = n_{F}(t) - n_{B}(t),
\end{align}
where $n(t)$ is commonly referred to as the jump number \cite{Ptaszynski2018}. To incorporate the jump number into the dynamics of the open quantum system, we follow standard approaches and assume that there is a measuring device in the drain, where back-action is negligible, which records individual electron tunneling events. Since $n(t)$ is a time-dependent stochastic variable, we actually seek the distribution of the total number of transferred electrons, $P(n,t)$.

We will connect $P(n,t)$ to the system dynamics via $\boldsymbol{\rho}(t)$, the reduced density matrix of the nanostructure. First, we resolve $\boldsymbol{\rho}(t)$ upon the jump number to define $\boldsymbol{\rho}^{(n)}(t)$, the reduced density matrix of the nanostructure conditioned upon $n$ electrons being recorded by the detector by time $t$. Considering that the general nanostructure states, $\{|q\rangle\}$, form a basis of the nanostructure subspace, then $P(n,t)$ is 
\begin{align}
P(n,t) & = \text{Tr}_{q}\left[\boldsymbol{\rho}^{(n)}(t)\right]. \label{eq: probability density from density matrix}
\end{align}

What remains now is to choose a method by which we can calculate $\boldsymbol{\rho}^{(n)}(t)$; we will use a BMME in the style of Li et al. \cite{Li2005a,Li2005b}.

\subsection{General master equation}\label{subsec: Master equation}

First, we decompose the total Hamiltonian into component parts,
\begin{align}
H = H_{Q} + H_{S} + H_{D} + H_{M} + V, \label{eq: total hamiltonian}
\end{align}
where $H_{Q}$ is the nanostructure Hamiltonian, $H_{S}$ and $H_{D}$ are the source and drain electrode Hamiltonians, respectively, $V$ describes the nanostructure-electrode interaction, and $H_{M}$ is the Hamiltonian of the measuring device, an electron detector in the drain. Section \ref{sec: TQD models and transport regimes} introduces the specific TQD Hamiltonian, but for now let us consider the general nanostructure applied in Eq.\eqref{eq: probability density from density matrix} with the accompanying annihilation, $a_{q}^{}$, and creation, $a^{\dag}_{q}$, operators. Similarly, the detector Hamiltonian's exact form is not essential, but it can be generally written as
\begin{align}
H_{M} = \sum_{n} \varepsilon_{n} | n\rangle \langle n |, \label{eq: detector Hamiltonian}
\end{align}
where $|n\rangle$ is the state of the measurement device when the jump number is $n$ at time $t$. The electrodes, modeled as non-interacting electrons, are denoted by $\alpha \in \{S,D\}$,
\begin{align}
H_{\alpha} & = \sum_{\mathbf{k}_{\alpha}} \varepsilon_{\mathbf{k}_{\alpha}} a^{\dag}_{\mathbf{k}_{\alpha}}a^{}_{\mathbf{k}_{\alpha}},
\end{align}
where the $a^{}_{\mathbf{k}_{\alpha}}$ and $a^{\dag}_{\mathbf{k}_{\alpha}}$ operators annihilate and create electrons in electrode $\alpha$ with wavevector $\mathbf{k}_{\alpha}$ and energy $\varepsilon_{\mathbf{k}_{\alpha}}$, respectively. The Hamiltonians of these individual parts together form $H_{0} = H_{S} + H_{D} + H_{M} + H_{Q}$, the Hamiltonian of the uncoupled electrode-nanostructure-electrode system.

In contrast, the interaction Hamiltonian, which is composed of a source and drain contribution, $V = V_{S} + V_{D}$, contains the linear dynamics of electron tunneling across the configuration:
\begin{widetext}
\begin{align}
V & = \sum_{\mathbf{k}_{S},q} \: t_{\mathbf{k}_{S},q}\left(a^{\dag}_{\mathbf{k}_{S}}a^{}_{q} + a^{\dag}_{q}a^{}_{\mathbf{k}_{S}}\right) + \sum_{\mathbf{k}_{D},q} \: t_{\mathbf{k}_{D},q}\left(\mathcal{M}^{\dag}a^{\dag}_{\mathbf{k}_{D}}a^{}_{q} + \mathcal{M}a^{\dag}_{q}a^{}_{\mathbf{k}_{D}}\right)  \label{eq: general interaction hamiltonian} 
\end{align}
\end{widetext}
Included in $V_{D}$ are measurement operators acting in the detector Hilbert space that increase,  
\begin{align}
\mathcal{M}^{\dag} & = \sum_{n = -\infty}^{\infty} \: |n\rangle \langle n-1 |,
\end{align}
and decrease, 
\begin{align}
\mathcal{M} & = \sum_{n = -\infty}^{\infty} \: | n-1 \rangle \langle n |,
\end{align}
the number of electrons detected by the measuring device, according to the corresponding drain tunneling operators.

Neglecting the nanostructure for the moment, let us consider the source, drain, and measuring device collectively as the bath, denoted by subscript ``$B$" and defined by the Hilbert space
\begin{align}
B & = S \otimes D \otimes M.
\end{align}
Alternatively, as Li et al. \cite{Li2005a,Li2005b} do, we can write the bath Hilbert subspace as a tensor product of the electrodes and many states of the detector,
\begin{align}
B & = \lim_{N \rightarrow \infty} B^{(-N)} \otimes \hdots \otimes B^{(-1)} \otimes B^{(0)} \otimes B^{(1)} \otimes \hdots \otimes B^{(N)},
\end{align}
where $B^{(n)} = S \otimes D \otimes | n \rangle \langle n |$ is the Hilbert space of the electrodes conditioned upon $n$ electrons being detected by the measuring device. 

Following from this bath-system demarcation, the reduced density matrix of the nanostructure is defined by tracing out the bath degrees of freedom from the total density matrix, $\boldsymbol{\rho}_{T}(t)$: $\boldsymbol{\rho}(t) = \text{Tr}_{B}\left[\boldsymbol{\rho}_{T}(t)\right]$. Similarly, the $n$-resolved reduced density matrix of the nanostructure is 
\begin{align}
\boldsymbol{\rho}^{(n)}(t) & = \text{Tr}_{B^{(n)}}\left[\boldsymbol{\rho}_{T}(t)\right] = \text{Tr}_{B}\left[|n\rangle\langle n|\boldsymbol{\rho}_{T}(t)\right]. \label{n-resolved trace}
\end{align}

Since the total density matrix of the system follows the Liouville-von Neumann equation of motion,
\begin{align}
i \frac{d\boldsymbol{\rho}_{T,I}}{dt} & = [V_{I}(t),\boldsymbol{\rho}_{T}(t)], \label{Liouville equation for total DM}
\end{align}
written here in the interaction picture, where operators are $A_{I}(t) = e^{i H_{0}t} A(t) e^{-iH_{0}t}$, we can obtain a time-evolution equation for the $n$-resolved reduced density matrix of the nanostructure as 
\begin{align}
\dot{\boldsymbol{\rho}}^{(n)}_{I}(t) & = - \int^{t}_{0} d\tau \: \text{Tr}_{B^{(n)}}\left[V_{I}(t),\left[V_{I}(t-\tau)\boldsymbol{\rho}_{T,I}(t-\tau)\right]\right].
\label{First n master equation}
\end{align}
In Eq.\eqref{First n master equation}, we have expanded the Liouville-von Neuman equation to second-order in $V_{I}(t)$ and performed the trace from Eq.\eqref{n-resolved trace}. 

Similarly to a standard BMME approach \cite{Breuer2002}, we now expand the commutators in Eq.\eqref{First n master equation} and apply weak-coupling between the electrodes and the nanostructure. Rather than the regular Born approximation, however, this is enforced by assuming that the total density matrix follows the ansatz \cite{Li2005a,Li2005b}
\begin{align}
\boldsymbol{\rho}_{T}(t) & \simeq \sum_{n = -\infty}^{\infty} \: \boldsymbol{\rho}^{(n)}(t)\boldsymbol{\rho}_{S}\boldsymbol{\rho}_{D}| n \rangle \langle n |. \label{eq: ansatz Born}
\end{align}
This ansatz choice is equivalent to the original Born approximation because all the bath subspaces $B^{(n)}$ are orthogonal to each other, which in turn is true because the electrode Hamiltonians conserve charge.
After leaving the interaction picture and then applying Markovianity, steps which are relegated to Appendix \ref{app: Master equation derivation} for brevity, the master equation, written here in the basis of eigenstates of $H_{Q}$, is
\begin{widetext}
\begin{align}
i\dot{\rho}^{(n)}_{mn} & = \omega_{mn}\rho^{(n)}_{mn} + \sum_{qq'}\sum_{kl}  \: \Big(\sum_{\alpha}\left[\Sigma_{\alpha}^{>}(\omega_{lk}) \langle m|a^{\dag}_{q}|k\rangle\langle k| a_{q'}|l\rangle \rho^{(n)}_{ln} + \Sigma_{\alpha}^{<}(\omega_{kl})^{*} \langle m|a_{q}|k\rangle\langle k|a^{\dag}_{q'}|l\rangle\rho^{(n)}_{ln}\right] \nonumber \\
& \:\:\:\: + \Sigma_{S}^{<}(\omega_{nl}) \langle m |a^{\dag}_{q}|k\rangle \rho^{(n)}_{kl}\langle l |a_{q'}|n\rangle + \Sigma_{S}^{>}(\omega_{ln})^{*} \langle m|a_{q}|k\rangle\rho^{(n)}_{kl}\langle l |a^{\dag}_{q'}|n\rangle + \Sigma_{D}^{<}(\omega_{nl}) \langle m |a^{\dag}_{q}|k\rangle \rho^{(n+1)}_{kl}\langle l |a_{q'}|n\rangle \nonumber \\
& \:\:\:\: + \Sigma_{D}^{>}(\omega_{ln})^{*} \langle m|a_{q}|k\rangle\rho^{(n-1)}_{kl}\langle l |a^{\dag}_{q'}|n\rangle - (m\leftrightarrow n,k \leftrightarrow l)^{*}\Big). \label{n resolved ME final}
\end{align}
\end{widetext}
In Eq.\eqref{n resolved ME final}, we have introduced the lesser and greater self-energies of electrode $\alpha$ as 
\begin{align}
\Sigma^{<}_{\alpha}(\omega) & =  \Delta^{<}_{\alpha}(\omega) + \frac{i}{2}\gamma^{\alpha}n_{F}(\omega - \mu_{\alpha}) \\
\Sigma^{>}_{\alpha}(\omega) & =  -\Delta^{>}_{\alpha}(\omega) - \frac{i}{2}\gamma^{\alpha}[1-n_{F}(\omega - \mu_{\alpha})],
\end{align}
respectively, with real components given by the Lamb shift,
\begin{align}
\Delta^{<}_{\alpha}(\omega) & = \frac{\gamma^{\alpha}}{2\pi} \lim_{\eta \rightarrow 0} \int^{\infty}_{-\infty} d\varepsilon \: \frac{(\varepsilon - \omega)n_{F}(\varepsilon - \mu_{\alpha})}{(\omega - \varepsilon)^{2} + \eta^{2}} \label{eq: lesser lamb shift} \\
\Delta^{>}_{\alpha}(\omega) & = \frac{\gamma^{\alpha}}{2\pi} \lim_{\eta \rightarrow 0} \int^{\infty}_{-\infty} d\varepsilon \: \frac{(\varepsilon - \omega)[1-n_{F}(\varepsilon - \mu_{\alpha})]}{(\omega - \varepsilon)^{2} + \eta^{2}}, \label{eq: greater lamb shift} 
\end{align}
and imaginary components given by the electrode-system coupling strength,
\begin{align}
\gamma^{\alpha} & = 2\pi |t_{\omega}|^{2} \rho_{\alpha}, \label{eq: frequency dependent coupling}
\end{align}
where in Eq.\eqref{eq: lesser lamb shift}-Eq.\eqref{eq: frequency dependent coupling} we have implicitly applied the wide-band limit in the electrodes. The self-energies also depend on the energy separation between eigenstates of $H_{Q}$: $\omega_{mn} = E_{m} - E_{n}$. Although in principle one can calculate the self-energies outside of the wide-band limit and for an arbitrary density of states, we choose to use this regime because it makes the calculations particularly tractable; we can now calculate the Lamb shifts analytically, as shown in Appendix \ref{app: Analytic Lamb shifts}.

These considerations are not necessary for the WTD theory introduced later on, however. This remains general for essentially any choice of the electrode density of states. As we will see, WTD theory does require unidirectional tunneling, so we are also forced to take a large bias voltage limit for all calculations.

In general, we must also be careful to remain in the parameter regime for which Born-Markov theory is applicable. Mostly, this means restricting the system-electrode coupling, $\gamma^{\alpha}$, to small values in comparison to the temperature. For example, in all plots in Sec.(\ref{sec: TQD Results}), we choose $\gamma^{\alpha} = 0.01\text{meV}$, while the temperature is $k_{B}T = 0.075\text{meV}$. This has the dual effect of ensuring that the perturbative expansion around the interaction Hamiltonian, $V$, is appropriate, while also enforcing bath Markovianity. Furthermore, since we terminate the perturbative expansion at second-order in $V$, we are actually also restricted to the sequential tunneling regime. If we wanted to investigate higher-order transport processes, such as cotunneling, then we would need to continue the expansion to include higher-order terms as well. Based on our previous experience with cotunneling in WTDs calculated from BMMEs, we would expect that this would add a small quantitative correction to the results \cite{Rudge2018}. 

If one wanted to move out of the weak coupling regime entirely, however, then evidently the Born-Markov approach is insufficient. In this case, one would need a method that can treat the system-bath interaction exactly, like a scattering matrix approach or non-equilibrium Green's functions, both of which have well-developed WTD theory \cite{Tang2014,Albert2012}. They are limited, however, to either a non-interacting or weakly interacting system; we will see in Sec.(\ref{sec: TQD models and transport regimes}) that a regime of particular interest to us contains strong inter- and intra-dot Coulomb interactions. To include both strong system interactions and a strong system-electrode coupling is, of course, non-trivial, and one must turn to numerically exact methods, such as the hierarchical equation of motion (HEOM) \cite{Schinabeck2016}.

\subsection{Liouville space}

In all practical calculations, we will actually work in Liouville space, rather than the Hilbert space of the TQD subsystem. To do so, we unfold $\rho^{(n)}(t)$ from an $N \times N$ density matrix to a vector, $\mathbf{P}(n,t)$, of length $N^{2}$ and with the general form
\begin{widetext}
\begin{align}
\mathbf{P}(n,t) & = [\underbrace{P_{1}(n,t),\hdots,P_{n}(n,t),\hdots,P_{N}(n,t)}_{\text{Populations}}\bigg|\underbrace{\rho_{12}(n,t),\hdots,\rho_{mn}(n,t),\hdots,\rho_{N,N-1}(n,t)}_{\text{Coherences}}]^{T}.  \label{eq: unfolded DM}
\end{align}
\end{widetext}
The population probability densities are $P_{k}(n,t) = \rho_{kk}^{(n)}(t)$. The trace, now, is defined as 
\begin{align}
\text{Tr}\left[ \boldsymbol{\rho}^{(n)}(t) \right] & = \left(\mathbf{I},\mathbf{P}(n,t)\right),
\end{align}
where the round brackets denote an inner product and $\mathbf{I} = \left[1,1,\hdots,1,0,0,\hdots,0\right]$ is a row vector in which the first $N$ elements are unity and the last $N(N-1)$ elements are zero.

The probability vector in Eq.\eqref{eq: unfolded DM} obeys the time-evolution equation 
\begin{align}
\dot{\mathbf{P}}(n,t) & = \sum_{n'}\mathbf{L}(n-n')\mathbf{P}(n,t) \label{n-resolved ME} \\ 
& = \mathbf{L}_{0}\mathbf{P}(n,t) + \mathbf{J}_{F}\mathbf{P}(n-1,t) + \mathbf{J}_{B}\mathbf{P}(n+1,t), \label{n-resolved ME expanded}
\end{align}
which collects the dynamics described by the Markovian master equation from Eq.\eqref{n resolved ME final} into one ``superoperator" $\mathbf{L}(n-n')$: the Liouvillian. It is composed of $\mathbf{J}_{F}$ and $\mathbf{J}_{B}$, quantum jump operators that describe tunneling to and from the drain, respectively, and $\mathbf{L}_{0} = \mathbf{L} - \mathbf{J}_{F} - \mathbf{J}_{B}$, which contains the remaining dynamics. In order to calculate the FCS and WTD, it is expedient to perform a Fourier transform, introducing the counting field $\chi$ \cite{Bagrets2003,Nazarov1999}:
\begin{align}
\mathbf{P}(\chi,t) & = \sum_{n=-\infty}^{\infty}e^{in\chi}\mathbf{P}(n,t), \label{Fourier transformed probability vector} \\
\mathbf{P}(n,t) & = \frac{1}{2\pi}\int_{0}^{2\pi} d\chi e^{-in\chi} \mathbf{P}(\chi,t). \label{Inverse Fourier transformed probability vector}
\end{align}

Under this transformation, the time-evolution equation is
\begin{align}
\dot{\mathbf{P}}(\chi,t) & = \mathbf{L}(\chi)\mathbf{P}(\chi,t), \text{ where }  \label{Chi dependent ME} \\
\mathbf{L}(\chi) & = \mathbf{L}_{0} + \mathbf{J}_{F}e^{i\chi} + \mathbf{J}_{B}e^{-i\chi}, \label{Full Liouvillian Chi}
\end{align}
with solution
\begin{align}
\mathbf{P}(\chi,t) & = e^{\mathbf{L}(\chi)t}\bar{\mathbf{P}}, \label{Chi dependent ME solution}
\end{align}
assuming that all measurement starts in the stationary state, denoted by probability vector $\bar{\mathbf{P}}$.

Finally in this section, we note that if we remove the coherent elements from the unfolded density matrix, $\mathbf{P}(\chi,t)$, and the Liouvillian, $\mathbf{L}(\chi)$, then we are left with a rate equation for the populations only. Such an approach would be equivalent to a T-matrix method, where the transition rates between the populations are calculated directly using Fermi's Golden Rule \cite{Bruus2002,Timm2008}. We would expect the rate equation to reproduce the full master equation only when transport does not include coherent effects.

\section{TQD models and transport regimes}\label{sec: TQD models and transport regimes}

We perform all theory on a TQD arranged in either of the triangular geometries in Fig.(\ref{fig: 1a}) and Fig.(\ref{fig: 1b}). The three dots, to which we give the index $\nu \in \{A,B,C\}$, are each modeled with a single available orbital that is at most accessible to two electrons with opposite spin, $\sigma \in \left\{\uparrow,\downarrow\right\}$. The system Hamiltonian, for either geometry, is 
\begin{widetext}
\begin{align}
H_{Q} & = \sum_{\sigma}\sum_{\nu} \: \varepsilon^{}_{\nu}a^{\dag}_{\nu,\sigma}a^{}_{\nu,\sigma} - \sum_{\nu \neq \nu'} t_{\nu\nu',\sigma} a_{\nu,\sigma}^{\dag}a^{}_{\nu ' ,\sigma} + \sum_{\nu} \: U_{\nu\nu}n_{\nu,\uparrow}n_{\nu,\downarrow} + \sum_{\nu < \nu'}\sum_{\sigma\sigma'} \: U_{\nu\nu '}n_{\nu,\sigma}n_{\nu',\sigma'}. \label{TQD Hamiltonian}
\end{align}
\end{widetext}
where $\varepsilon_{\nu}$ is the energy of each dot, $U_{\nu\nu}$ and $U_{\nu\nu '}$ are the intra- and inter-dot Coulomb repulsions, respectively, and $t_{\nu\nu '} = t^{*}_{\nu '\nu}$ is the hopping parameter for tunneling from dot $\nu '$ to dot $\nu$. As usual, the $a^{\dag}_{\nu,\sigma}$ and $a^{}_{\nu,\sigma}$ operators create and annihilate an electron on dot $\nu$ with energy $\varepsilon_{\nu}$ and spin $\sigma$, respectively, while $n_{\nu,\sigma} = a^{\dag}_{\nu,\sigma}a^{}_{\nu,\sigma}$ is the corresponding particle number operator. Aharonov-Bohm interference interacts with the dynamics by inducing a phase difference, $\phi$, between different paths around the TQD: $\phi = \oint \mathbf{A}\cdot d\ell = 2\pi\Phi/\Phi_{0}$, where $\mathbf{A}$ is the magnetic vector potential and $\Phi_{0} = \frac{h}{e}$ is the magnetic flux quantum \cite{Aharonov1959,Ihn2007,Strambini2009,Kobayashi2002}. In principle, this adds a phase shift to each inter-dot coupling, $t_{\nu\nu'} \rightarrow t_{\nu\nu'}e^{i\phi_{\nu\nu'}}$. However, one can perform a gauge transformation such that $a^{\dag}_{\nu} \rightarrow a^{\dag}_{\nu}e^{i\tilde{\phi}_{\nu}}$ and then choose $\phi_{AB} = -(\tilde{\phi}_{A} - \tilde{\phi}_{B})$ and $\phi_{BC} = -(\tilde{\phi}_{B} - \tilde{\phi}_{C})$. Since the total phase, $\phi = \phi_{AC} + \phi_{BA} + \phi_{CB}$, is conserved under such a gauge transformation, the phase difference is now factored entirely onto the coupling between dot $A$ and dot $C$: $t_{AC} = |t_{AC}|e^{i\phi}$, with $t_{BA}\text{ and }t_{CB}$ always real.

For the configuration in Fig.(\ref{fig: 1a}), where both the $A$ and $C$ dots are coupled to the source and the $B$ dot alone is coupled to the drain, the interaction Hamiltonian is 
\begin{widetext}
\begin{align}
V & = \sum_{\mathbf{k}_{S}}\sum_{\nu = \{A,C\}} \: t^{}_{\mathbf{k}_{S},\nu,\sigma} \left(a^{\dag}_{\mathbf{k}_{S}}a^{}_{\nu,\sigma} + a^{\dag}_{\nu,\sigma}a^{}_{\mathbf{k}_{S}}\right) + \sum_{\mathbf{k}_{D}} \: t^{}_{\mathbf{k}_{D},B,\sigma}\left(\mathcal{M}^{\dag} a^{\dag}_{\mathbf{k}_{D}}a^{}_{B,\sigma} + \mathcal{M}^{}a^{\dag}_{B,\sigma}a^{}_{\mathbf{k}_{D}}\right).
\end{align}
\end{widetext}
Conversely, the interaction Hamiltonian for the configuration in Fig.(\ref{fig: 1b}), where the $B$ dot is coupled to the source and the $A$ and $C$ dots are coupled to the drain, is
\begin{widetext}
\begin{align}
V & = \sum_{\mathbf{k}_{D}}\sum_{\nu = \{A,C\}} \: t_{\mathbf{k}_{D},\nu,\sigma}\left(\mathcal{M}^{\dag}a^{\dag}_{\mathbf{k}_{D}}a^{}_{\nu,\sigma} + \mathcal{M}^{}a^{\dag}_{\nu,\sigma}a^{}_{\mathbf{k}_{D}}\right) + \sum_{\mathbf{k}_{S}} \: t_{\mathbf{k}_{S},B,\sigma}\left(a^{\dag}_{\mathbf{k}_{S}}a^{}_{B,\sigma} + a^{\dag}_{B,\sigma}a^{}_{\mathbf{k}_{S}}\right).
\end{align}
\end{widetext}

At this point in the theory, we cannot apply the TQD Hamiltonian defined in Eq.\eqref{TQD Hamiltonian} to the $n$-resolved master equation, because it is written in the basis of dot states and is consequently not diagonal. Although for restrictive transport regimes there exist analytic diagonalizations of the TQD Hamiltonian \cite{Emary2007b}, we can always numerically diagonalize $H_{Q}$ via its eigenstates:
\begin{align}
\tilde{H}_{Q} & = \sum_{k k'} \: |m_{k}\rangle \langle m_{k}| H_{Q} |m_{k'}\rangle \langle m_{k'}|\nonumber \\
& = \sum_{k=1}^{N} \: E_{k}|m_{k}\rangle \langle m_{k}|,
\end{align}
where $E_{k}$ is the eigenenergy of eigenstate $|m_{k}\rangle$. We can also write the dot states, $|d_{i}\rangle$, in the new basis,
\begin{align}
|d_{i}\rangle & = \sum_{k = 1}^{N} \: |m_{k}\rangle \langle m_{k}|d_{i}\rangle,
\end{align}
and, since the dot states also span the system space, compute the inverse transformation:
\begin{align}
|m_{k}\rangle & = \sum_{i = 1}^{N} \: |d_{i}\rangle \langle d_{i}|m_{k}\rangle. \label{Molecular eigenstates reverse}
\end{align}

With no parameter restrictions, however, the full Fock space is quite large and numerical diagonalization is a formidable task. There can be a maximum of six electrons occupying the configuration, so $N = \sum\limits_{k=0}^{6} \frac{6!}{k!(6-k)!} = 64$ and the resulting density matrix has $4096$ elements. To reduce the complexity and computational requirements, many theoretical investigations instead focus on limiting regimes where the dimensionality is much smaller \cite{Emary2007b,Poltl2009,Poltl2013}; several of which we will consider here.

\subsection{Spin-independent triple and single-occupancy}

In the first transport scenario, we assume that the intra-dot Coulomb repulsion is large: $U_{\nu\nu} \rightarrow \infty$. Under this limit, each dot in the configuration can be occupied by only one excess electron, which we label the \textit{triple-occupancy} regime. Spanning the system are ten dot states: the configuration can be empty, $|0\rangle$; a single electron may occupying any of the three dots, $|A\rangle$, $|B\rangle$, or $|C\rangle$; two electrons may be occupying any two of the dots, $|AB\rangle$, $|AC\rangle$, or $|BC\rangle$; or all three dots are occupied, $|ABC\rangle=|3\rangle$.

Although $|0\rangle$ and $|3\rangle$ are invariant under the diagonalization, the transformed basis has three new single-occupancy,
\begin{align}
|1i\rangle & = c_{1i,A} |A\rangle + c_{1i,B} |B\rangle + c_{1i,C} |C\rangle
\end{align}
and double-occupancy,
\begin{align}
|2i\rangle & = c_{2i,AB} |AB\rangle + c_{2i,AC} |AC\rangle + c_{2i,BC} |BC\rangle,
\end{align}
states, where $i = 1,2,3$. From Eq.\eqref{Molecular eigenstates reverse}, the coefficients are evidently $c_{1i,\nu} = \langle \nu|1i\rangle$ and $c_{2i,\nu \nu'} = \langle \nu\nu '|2i\rangle$. In addition to this large reduction in dimensionality, all eigenstates with different electron occupancy are orthogonal and thus decouple in the master equation; the remaining $20$ elements unfold into the probability vector
\begin{widetext}
\begin{align}
\mathbf{P}(\chi,t) & = \left[P_{0}(\chi,t),P_{11}(\chi,t),P_{12}(\chi,t),P_{13}(\chi,t),P_{21}(\chi,t),P_{22}(\chi,t),P_{23}(\chi,t),P_{3}(\chi,t), \right. \nonumber \\
& \:\:\:\:\: \rho_{11,12}(\chi,t),\rho_{11,13}(\chi,t),\rho_{12,13}(\chi,t),\rho_{12,11}(\chi,t),\rho_{13,11}(\chi,t),\rho_{13,12}(\chi,t), \nonumber \\
& \:\:\:\:\: \left.\rho_{21,22}(\chi,t),\rho_{21,23}(\chi,t),\rho_{22,23}(\chi,t),\rho_{22,21}(\chi,t),\rho_{23,21}(\chi,t),\rho_{23,22}(\chi,t) \right]^{T}. \label{triple-occupancy probability vector}
\end{align}
\end{widetext}

After applying a Fourier transform to Eq.\eqref{n resolved ME final}, as well as the rotating wave approximation so that $q \neq q'$ terms are neglected, the master equation is
\begin{widetext}
\begin{align}
\dot{\rho}_{mn}(\chi,t) & = -i \omega_{mn} \rho_{mn}(\chi,t) - i\sum_{k\ell} \Big[\sum_{\nu = \{A,C\}}\Big( \Sigma_{S}^{<}(\omega_{k\ell})^{*} \langle m |  a^{}_{\nu} | k \rangle \langle k | a^{\dag}_{\nu} | \ell \rangle \rho_{\ell n} + \Sigma_{S}^{>}(\omega_{\ell k}) \langle m |  a^{\dag}_{\nu} | k \rangle \langle k | a^{}_{\nu} | \ell \rangle \rho_{\ell n}  \nonumber \\ 
& \:\:\:\:\:\: + \Sigma_{S}^{<}(\omega_{n \ell}) \langle m |  a^{\dag}_{\nu} | k \rangle \rho_{k\ell} \langle \ell | a^{}_{\nu} | n \rangle + \Sigma_{S}^{>}(\omega_{\ell n})^{*} \langle m |  a^{}_{\nu} | k \rangle \rho_{k\ell} \langle \ell | a^{\dag}_{\nu} | n \rangle  - \Sigma_{S}^{<}(\omega_{\ell k}) \rho_{m k } \langle k |  a^{}_{\nu} | \ell \rangle \langle \ell | a^{\dag}_{\nu} | n \rangle \nonumber \\
& \:\:\:\:\:\: - \left. \Sigma_{S}^{>}(\omega_{k \ell})^{*} \rho_{m k } \langle k |  a^{\dag}_{\nu} | \ell \rangle \langle \ell | a^{}_{\nu} | n \rangle - \Sigma_{S}^{<}(\omega_{m k})^{*} \langle m |  a^{\dag}_{\nu} | k \rangle \rho_{k \ell} \langle \ell | a^{}_{\nu} | n \rangle  - \Sigma_{S}^{>}(\omega_{k m}) \langle m |  a^{}_{\nu} | k \rangle \rho_{k \ell} \langle \ell | a^{\dag}_{\nu} | n \rangle \right) \nonumber \\
& \:\:\:\:\:\: + \Sigma_{D}^{<}(\omega_{k\ell})^{*} \langle m |  a^{}_{B} | k \rangle \langle k | a^{\dag}_{B} | \ell \rangle \rho_{\ell n} +  \Sigma_{D}^{>}(\omega_{\ell k}) \langle m |  a^{\dag}_{B} | k \rangle \langle k | a^{}_{B} | \ell \rangle \rho_{\ell n} + \Sigma_{D}^{<}(\omega_{n \ell}) e^{-i\chi}\langle m |  a^{\dag}_{B} | k \rangle \rho_{k\ell} \langle \ell | a^{}_{B} | n \rangle  \nonumber \\ 
& \:\:\:\:\:\: + \Sigma_{D}^{>}(\omega_{\ell n})^{*} e^{i\chi}\langle m |  a^{}_{B} | k \rangle \rho_{k\ell} \langle \ell | a^{\dag}_{B} | n \rangle  - \Sigma_{D}^{<}(\omega_{\ell k}) \rho_{m k } \langle k |  a^{}_{B} | \ell \rangle \langle \ell | a^{\dag}_{B} | n \rangle - \Sigma_{D}^{>}(\omega_{k\ell})^{*} \rho_{m k } \langle k |  a^{\dag}_{B} | \ell \rangle \langle \ell | a^{}_{B} | n \rangle \nonumber \\
& \:\:\:\:\:\: - \Sigma_{D}^{<}(\omega_{m k})^{*}e^{-i\chi} \langle m |  a^{\dag}_{B} | k \rangle \rho_{k \ell} \langle \ell | a^{}_{B} | n \rangle - \Sigma_{D}^{>}(\omega_{k m})e^{i\chi} \langle m |  a^{}_{B} | k \rangle \rho_{k \ell} \langle \ell | a^{\dag}_{B} | n \rangle \Big)\Big]. \label{TQD ME}
\end{align}
\end{widetext}
The corresponding master equation for the configuration in Fig.(\ref{fig: 1b}) is similar, except that the drain self-energies now lie under the summation of $A,C$, and the source self-energies do not. For triple-occupancy, there are four types of coupled density matrix elements, $\rho_{0,0}$, $\rho_{1i,1j}$, $\rho_{2i,2j}$, and $\rho_{3,3}$, which we display in Appendix \ref{app: TQD triple-occupancy}.

If we also take the inter-dot repulsion as large, $U_{\nu\nu '} \rightarrow \infty$, then only one dot may be occupied at a time: the \textit{single-occupancy} regime. In this case, the probability vector is much smaller,
\begin{align}
\mathbf{P}(\chi,t) & = \left[P_{0}(\chi,t),P_{11}(\chi,t),P_{12}(\chi,t),P_{13}(\chi,t),\right.\nonumber\\
& \:\:\:\:\:\rho_{11,12}(\chi,t),\rho_{11,13}(\chi,t),\rho_{12,13}(\chi,t), \nonumber \\
& \:\:\:\:\:\left.\rho_{12,11}(\chi,t),\rho_{13,11}(\chi,t),\rho_{13,12}(\chi,t)\right]^{T},
\end{align}
and analytic solutions are available \cite{Emary2007b}.

\subsection{Spin-dependent double-occupancy}

In the \textit{double-occupancy} regime, the intra-dot Coulomb repulsion is finite, but the inter-dot repulsion is large; $U_{\nu \nu'} \rightarrow \infty$ and each dot can be occupied by two electrons of opposite spin, but only two electrons are allowed in the entire TQD configuration. There are now $6$ single electron states and $15$ double electron states, for a total of $22$ states spanning the system and $262$ coupled density matrix elements. To simplify, we also assume a large bias voltage, $V_{SD} \rightarrow \infty$, so that tunneling can only be from the source to the TQD or from the TQD to the drain, and high temperature limit, so that we can use a Lindblad master equation. For the configuration in Fig.(\ref{fig: 1a}) this is
\begin{align}
\frac{d\boldsymbol{\rho}}{dt} & = -i \left[H_{Q},\boldsymbol{\rho}\right] + \nonumber \\ 
&  \sum_{\sigma,\nu = \{A,C\}} \gamma^{S}\left(a^{\dag}_{\nu,\sigma} \boldsymbol{\rho} a^{}_{\nu,\sigma} - \frac{1}{2} a^{}_{\nu,\sigma} a^{\dag}_{\nu,\sigma}\boldsymbol{\rho} - \frac{1}{2}\boldsymbol{\rho} a^{}_{\nu,\sigma}a^{\dag}_{\nu,\sigma} \right) + \nonumber \\
& \sum_{\sigma} \gamma^{D}\left(a^{}_{B,\sigma} \boldsymbol{\rho} a^{\dag}_{B,\sigma}e^{i\chi} - \frac{1}{2} a^{\dag}_{B,\sigma} a^{}_{B,\sigma}\boldsymbol{\rho} - \frac{1}{2}\boldsymbol{\rho} a^{\dag}_{B,\sigma}a^{}_{B,\sigma} \right),
\end{align}
and for the configuration in Fig.(\ref{fig: 1b}) we exchange $\nu \in \{A,C\} \leftrightarrow B$ and $\gamma^{S} \leftrightarrow \gamma^{D}$.

We are justified in making this final assumption because all results are calculated from the WTD, which requires unidirectional tunneling and, therefore, a large bias voltage.

\section{Fluctuation statistics} \label{sec: Fluctuation statistics}

\subsection{Full counting statistics}

The FCS are cumulants of the distribution of total current, $\langle\langle I^{(k)} \rangle\rangle$, and can easily be derived from the distribution of total transferred charge as long as $P(n,t)$ satisfies a large deviation principle \cite{Budini2011}. In this case, cumulants of $P(n,t)$ scale linearly in time,
\begin{align}
\langle\langle n(t) ^{k} \rangle\rangle & \approx \langle\langle I^{k} \rangle\rangle t, \label{cumulant scaling time}
\end{align}
and the current cumulants are the asymptotic growth rates:
\begin{align}
\lim_{t\rightarrow\infty}\langle\langle I(t) ^{k} \rangle\rangle & = e^{k}\frac{d}{dt}\langle\langle n(t)^{k} \rangle\rangle \\
& \approx \langle\langle I^{k} \rangle\rangle. \label{current cumulants final}
\end{align}

In this large deviation, or long-time limit, Bagrets and Nazarov \cite{Bagrets2003} have shown that the cumulant generating function of $P(n,t)$ is
\begin{align}
\lim_{t\rightarrow \infty} K(\chi,t) & = t\Lambda_{\text{max}}(\chi), \label{eq: CGF}
\end{align}
where $\Lambda_{\text{max}}(\chi)$ is the eigenvalue of $\mathbf{L}(\chi)$ with the largest real part. Combining Eq.\eqref{cumulant scaling time} and Eq.\eqref{eq: CGF} yields the zero-frequency current cumulants:
\begin{align}
\langle\langle I^{(k)} \rangle\rangle & = (-i)^{k} \frac{\partial^{k}}{\partial \chi^{k}}\Lambda_{\text{max}}\Bigg|_{\chi=0}.
\end{align}

The fluctuation statistics of triangular TQDs have largely focused on the current cumulants as defined above, and in particular on the Fano factor, $F$, which scales the noise to that of a Poissonian process. It is defined from the first cumulant, the average stationary current $\langle I \rangle$, and the second cumulant, which is related to the zero-frequency noise as $\langle\langle I^{(2)}\rangle\rangle = \frac{1}{2}\mathcal{S}(0)$:
\begin{align}
F & = \frac{\mathcal{S}(0)}{2\langle I \rangle} = \frac{\langle\langle I ^{2}\rangle\rangle}{\langle I \rangle}. \label{FF n definition}
\end{align}

\subsection{Waiting time distribution}

We will also examine the fluctuation statistics of the TQD models in Fig.(\ref{fig: 1a}) and Fig.(\ref{fig: 1a}) with a fluctuating-time approach, the WTD, which has received only limited theoretical attention \cite{Engelhardt2019,Welack2009}. In the context of electron transport through the general electrode-nanostructure-electrode paradigm we are using, the WTD is defined as the probability density that two quantum jumps are separated by a time delay $\tau$, conditioned on the probability density of the initial quantum jump.

In the stationary state and in the limit of unidirectional tunneling, the distribution of waiting times between successive jumps to the drain is   
\begin{align}
w(\tau) & = \frac{\left(\mathbf{I},\mathbf{J}^{F}_{D}e^{\left(\mathbf{L} - \mathbf{J}^{F}_{D}\right)\tau}\mathbf{J}^{F}_{D}\bar{\mathbf{P}}\right)}{\left(\mathbf{I},\mathbf{J}^{F}_{D}\bar{\mathbf{P}}\right)}, \label{brandes definition}
\end{align}
where $\mathbf{J}^{F}_{D}$ is the quantum jump operator \cite{Brandes2008} containing all transitions from the system to the drain. It naturally contains all terms with an $e^{i\chi}$ attached to them in Eq.\eqref{n resolved ME final}.

To compare to the FCS, we will need to use the waiting time cumulants, defined from the WTD Laplace transform,
\begin{align}
\tilde{w}(z) & = \int^{\infty}_{0} d\tau \: e^{z\tau}w(\tau) \\
& = \frac{\left(\mathbf{I},\mathbf{J}_{l}\left[z - \left(\mathbf{L} - \mathbf{J}^{F}_{D}\right)\right]^{-1}\mathbf{J}^{F}_{D}\bar{\mathbf{P}}\right)}{\left(\mathbf{I},\mathbf{J}^{F}_{D}\bar{\mathbf{P}}\right)},
\end{align}
which conveniently defines a cumulant generating function:
\begin{align}
\langle\langle \tau^{k} \rangle\rangle = \left. (-1)^{k} \frac{d^{k}}{dz^{k}}\ln \tilde{w}(z)\right|_{z=0}. \label{WTD cumulants}
\end{align}

Apart from the standard WTD in Eq.\eqref{brandes definition}, to calculate correlations we will also need the two-time joint WTD $w(\tau_{1},\tau_{2})$:
\begin{align}
w(\tau_{1},\tau_{2}) & = \frac{\left(\mathbf{I},\mathbf{J}^{F}_{D}e^{\left(\mathbf{L} - \mathbf{J}^{F}_{D}\right)\tau_{2}}\mathbf{J}^{F}_{D}e^{\left(\mathbf{L} - \mathbf{J}^{F}_{D}\right)\tau_{1}}\mathbf{J}^{F}_{D}\bar{\mathbf{P}}\right)}{\left(\mathbf{I},\mathbf{J}^{F}_{D}\bar{\mathbf{P}}\right)},
\end{align}
alongside the second-order average waiting time 
\begin{align}
\langle \tau_{1}\tau_{2} \rangle & = \left. \frac{\partial}{\partial z_{1}}\frac{\partial}{\partial z_{2}}\ln \tilde{w}_{2}(z_{1},z_{2})\right|_{z_{1}=z_{2}=0}. \label{eq: WTD cumulants second-order}
\end{align}

\subsection{Renewal and non-renewal theory}

The WTD is a useful theoretical tool to compare alongside the FCS because it provides information on short-time physics that the large deviation limit of the FCS does not. As an example, correlations between successive waiting times $\tau_{1}$ and $\tau_{2}$ imply that the system does not entirely `renew' itself after each tunneling, and are indicative of non-trivial underlying transport dynamics \cite{Rudge2016a,Rudge2016b,Rudge2018,Rudge2019a,Rudge2019c,Ptaszynski2017,Ptaszynski2017a}. Waiting time correlations are largely studied via the linear Pearson correlation coefficient:
\begin{align}
p  & = \frac{\langle\tau_{1}\tau_{2}\rangle - \langle\tau\rangle^{2}}{\langle\langle\tau^{2}\rangle\rangle}, \label{Correlation coefficient}
\end{align} 
The Pearson correlation coefficient, roughly speaking, calculates an average of waiting time correlations over the entire landscape of $\tau_{1}$ and $\tau_{2}$. While useful, this approach can wash out non-zero correlations occurring only for specific regimes of $\tau_{1}$ and $\tau_{2}$, potentially obscuring important information. Consequently, we will also calculate the quantity
\begin{align}
C(\tau_{1},\tau_{2}) & = \frac{w(\tau_{1},\tau_{2}) - w(\tau_{1})w(\tau_{2})}{w(\tau_{1})w(\tau_{2})},
\end{align}
which describes correlations for each combination of $\tau_{1}$ and $\tau_{2}$. $C(\tau_{1},\tau_{2})$ is interpreted thus: given a pair of successive waiting times $\tau_{1}$ and $\tau_{2}$, if the corresponding value of $C(\tau_{1},\tau_{2})$ is positive (negative), then it is likely (unlikely) to observe this pair compared to others. If $C(\tau_{1},\tau_{2}) \approx 0$, however, then this pair is no more likely or unlikely to be observed than any other. 

\begin{figure}
\begin{centering}
\includegraphics[width = \columnwidth]{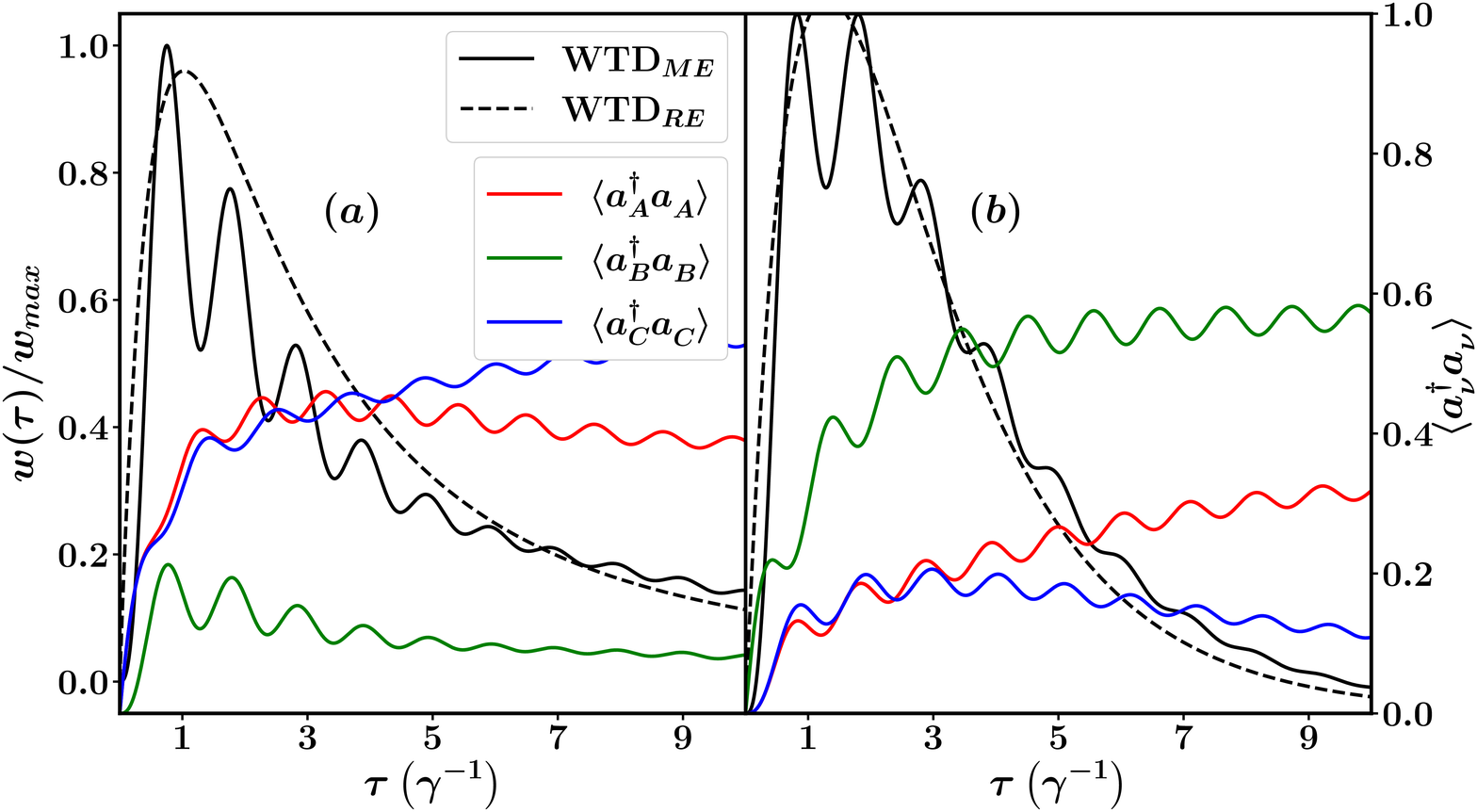}
{\phantomsubcaption\label{fig: 2a}
\phantomsubcaption\label{fig: 2b}}
\caption{WTD (black) calculated from the full BMME, $\text{WTD}_{ME}$, and rate equation, $\text{WTD}_{RE}$, compared to the time-dependent dot occupation probabilities (multi-colored), for (a) the configuration in Fig.(\ref{fig: 1a}) and (b) the configuration in Fig.(\ref{fig: 1b}). Transport is restricted to the single-occupancy regime. The penetrating magnetic flux is $\phi = \pi$, $\varepsilon_{B} = 0$, $\varepsilon_{A} = \Delta$, $\varepsilon_{C} = -\Delta$, $|t_{BC}| = |t_{AB}| = 2\gamma$, $t_{AC} = 1.9\gamma$, $T = 0.075\text{meV}$, $\gamma = 0.01\text{meV}$, and, in order to have unidirectional current, $\mu_{S} = -\mu_{D} = 100\gamma$.}
\label{fig: 2}
\end{centering}
\end{figure}

\section{Results}\label{sec: TQD Results}

\subsection{Spin-independent single-occupancy}

The single-occupancy case, in which the TQD can only be occupied by one excess electron, has received great theoretical attention due to the analytic solutions available in this regime. At $\phi = 0$, for example, Emary \cite{Emary2007b} showed that energy parameters for the configuration in Fig.(\ref{fig: 1a}) can always be adjusted to produce a coherent superposition of dots $A$ and $C$, in which case the dot $B$ occupancy is blocked:
\begin{align}
|\Psi \rangle & = \frac{1}{\sqrt{t_{AC}^{2}+t_{BC}^{2}}}\left(t_{BC}|A\rangle - t_{AB}|C\rangle\right). \label{eq: single-occupancy dark state}
\end{align}
Here, $|\Psi \rangle$, known as a ``dark state", is an eigenvector of $H_{Q}$. To achieve Eq.\eqref{eq: single-occupancy dark state}, the required parameter choices are $\varepsilon_{B} = 0$, $\varepsilon_{A} = \Delta$, and $\varepsilon_{C} = -\Delta$, where 
\begin{align}
\Delta = \frac{|t_{AC}|}{2|t_{AB}||t_{CB}|}\left( |t_{AB}|^{2} - |t_{CB}|^{2} \right).
\end{align}
Unsurprisingly, the formation of a dark state is generally accompanied by super-Poissonian noise, attributed to avalanche tunneling caused by the coherent population blocking. Although super-Poissonian noise is often a symptom of underlying time-correlations, such as when a system displays telegraphic switching \cite{Rudge2019a,Rudge2019c}, in this case renewal theory yields no extra information about the dynamics. 

Because we are limiting the TQD to only one excess electron, the transport in this regime is effectively single-reset; once the TQD empties via a tunneling to the drain, another electron must tunnel in from the source before the next jump to the drain. While each tunneling electron can coherently interfere with itself, therefore, it cannot interfere with other tunneling electrons and the Pearson coefficient is zero for all parameter choices. The single-reset nature is evident in all corresponding WTD plots, such as Fig.(\ref{fig: 2}), where $w(0) = 0$. Even though there are no time-correlated dynamics, however, the WTD itself still provides useful information about the single-occupancy regime.

To see, we first examine Fig.(\ref{fig: 2a}) and Fig.(\ref{fig: 2b}). They show the WTD in black calculated from the BMME, the solid line, and the rate equation for populations only, the dashed line. Plotted alongside these are the quantum-averaged time-dependent dot occupation probabilities:
\begin{align}
\langle a^{\dag}_{\nu}a^{}_{\nu}\rangle & = \text{Tr}\left[a^{\dag}_{\nu}a^{}_{\nu} \rho(t) \right], \:\:\: \text{ with} \\
\rho(t) & = \frac{e^{\left(\mathbf{L} - \mathbf{J}^{F}_{D}\right)\tau}\mathbf{J}^{F}_{D}\bar{\mathbf{P}}}{\left(\mathbf{I},\mathbf{J}^{F}_{D}\bar{\mathbf{P}}\right)}.
\end{align}
In these plots, we have chosen parameters that produce the dark state at $\phi = 0$ for the configuration in Fig.(\ref{fig: 1a}), as outlined at the start of this section. We note, however, that we have to slightly detune $|t_{AC}|$ from the dark state point at $2\gamma$ to $1.9\gamma$, because if $\langle I \rangle (\phi) = 0$, the WTD approaches numerical difficulties. Still, for these plots, the stationary current is largely suppressed: $\frac{\langle I \rangle (\pi)}{\langle I \rangle_{max}} = 0.35$. Accordingly, one can see that the accompanying WTD has a long tail, due to the correspondingly large $\langle \tau \rangle$. Apart from this behavior, the most noticeable features in Fig.(\ref{fig: 2a}) are the coherent oscillations that are present in the WTD calculated from the full BMME and absent in the WTD calculated from just the rate equation.

\begin{figure}
\includegraphics[width = \columnwidth]{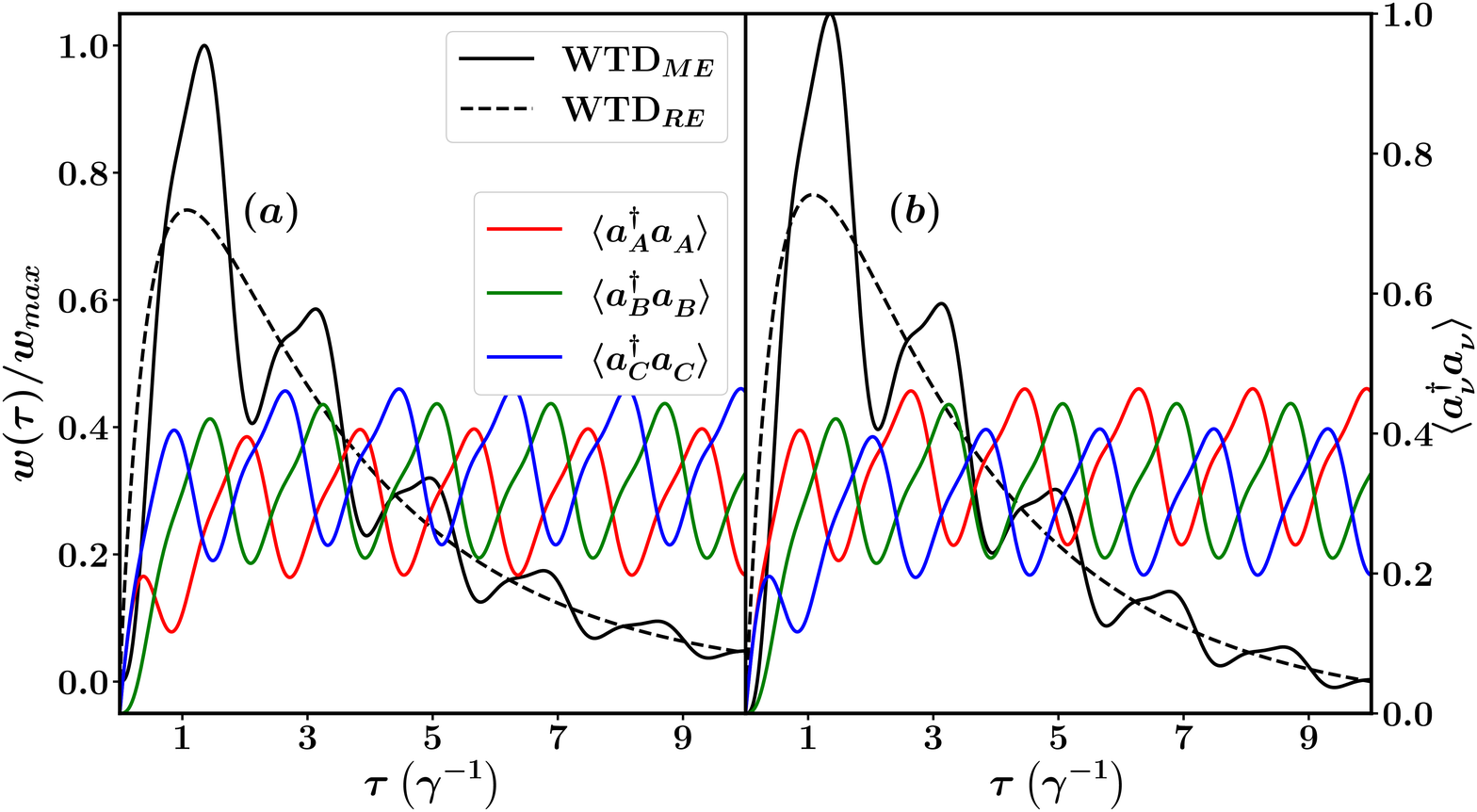}
{\phantomsubcaption\label{fig: 3a}
\phantomsubcaption\label{fig: 3b}}
\caption{WTD and occupation probabilities for the configuration in Fig.(\ref{fig: 1a}). All other parameters are the same as in Fig.(\ref{fig: 2}), except for the magnetic flux, which is now $\phi = \pm\frac{\pi}{2}$ for (a) and (b), respectively.}
\label{fig: 3}
\end{figure}

Such coherent oscillations are crucial to understanding the underlying transport behavior, as one can correlate the WTD behavior directly to the relevant dot occupancies \cite{Engelhardt2019}. For example, in Fig.(\ref{fig: 2a}), one can see that $\text{WTD}_{ME}$ exactly follows the behavior of $\langle a^{\dag}_{B}a^{}_{B} \rangle$. Since this is for the configuration in Fig.(\ref{fig: 1a}), we immediately understand that we can only observe a tunneling to the drain when dot $B$ is occupied. Such analysis demonstrates why the rate equation alone is insufficient to accurately describe the dynamics of triangular TQD; because $\text{WTD}_{RE}$ sees only the `average' occupancy of each dot, it displays no coherent oscillations and thus cannot be matched to individual dot occupancies. 

In Fig.(\ref{fig: 2b}), which displays the WTDs and occupancies for the same $\phi$ and energy parameters but now for the reverse configuration in Fig.(\ref{fig: 1b}), we can apply the same analysis. Here, dots $A$ and $C$ are coupled to the drain, so the WTD peaks correspond to peaks in the occupancies of these two dots. Of note here is that the WTD is bimodal; since dot $B$ is coupled to dots $A$ and $C$ equally and the induced phase difference is $\phi = \pi$, there are two equally probable paths to the drain for the electron to take after entering from the source. At later times, the asymmetric dot coupling causes the occupancy of dot $A$ to correspond with the WTD more closely. One can also see from Fig.(\ref{fig: 2b}) that reversing the dot configuration lifts the dark state; even though we have used exactly the same parameters as in Fig.(\ref{fig: 2a}), the current is now $\frac{\langle I \rangle (\pi)}{\langle I \rangle_{max}} = 0.79$. This is evident from the WTD itself, which displays a shorter tail than Fig.(\ref{fig: 2a}), and the occupancy of dot $B$, which is now greater than dots $A$ and $C$ individually. 

Next, we turn to Fig.(\ref{fig: 3a}) and Fig.(\ref{fig: 3b}), both of which have been plotted for the original configuration in Fig.(\ref{fig: 1a}). The only difference to Fig.(\ref{fig: 2a}) and Fig.(\ref{fig: 2b}) is that in these plots we have chosen $\phi = \pm \frac{\pi}{2}$, respectively. At this phase shift, the dark state is completely lifted and $\langle I \rangle (\pm\frac{\pi}{2})=\langle I \rangle_{max}$. Interestingly, while it may not be immediately obvious, on close inspection it is evident that each WTD peak in these plots has two components: one from the corresponding peak in $\langle a^{\dag}_{B}a^{}_{B}\rangle$ and one from a peak in $\langle a^{\dag}_{C}a^{}_{C} \rangle$ for Fig.(\ref{fig: 2a}) and $\langle a^{\dag}_{A}a^{}_{A} \rangle$ for Fig.(\ref{fig: 2b}). Since a peak in the WTD must correspond to occupation of dot $B$, this second smaller contribution must correspond to a coherent single-electron state formed between $B$ and either $A$ or $C$. The choice is evidently determined by the sign of the magnetic flux; in Fig.(\ref{fig: 2a}), the phase shift is positive, $\phi = +\frac{\pi}{2}$, and the electron moves anti-clockwise from $C \rightarrow B \rightarrow A$, while in Fig.(\ref{fig: 2a}), the phase shift is negative, $\phi = -\frac{\pi}{2}$, and the electron moves clockwise from $A \rightarrow B \rightarrow C$ \cite{Engelhardt2019}. Note that we cannot reproduce this effect for $\phi = \pm \pi$, as in that case $t_{AC}(\pm \pi)  = t_{CA}(\pm \pi)$, so there is no directional difference in dot couplings.

\begin{figure}
\begin{centering}
\begin{subfigure}[b]{0.49\columnwidth}
\includegraphics[width = \textwidth]{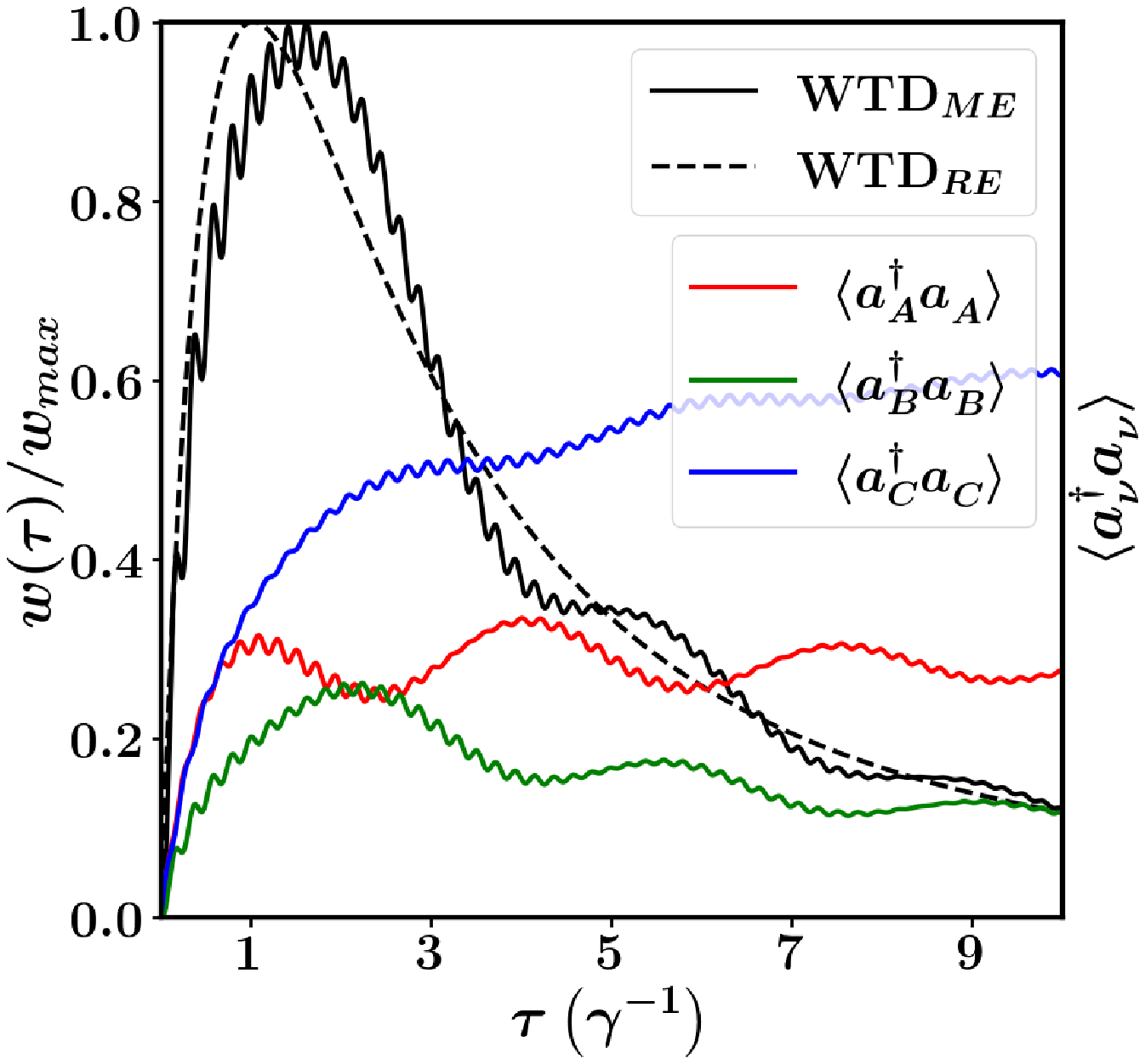}
\caption{}
\label{fig: 4a}
\end{subfigure}
\begin{subfigure}[b]{0.49\columnwidth}
\includegraphics[width = \textwidth]{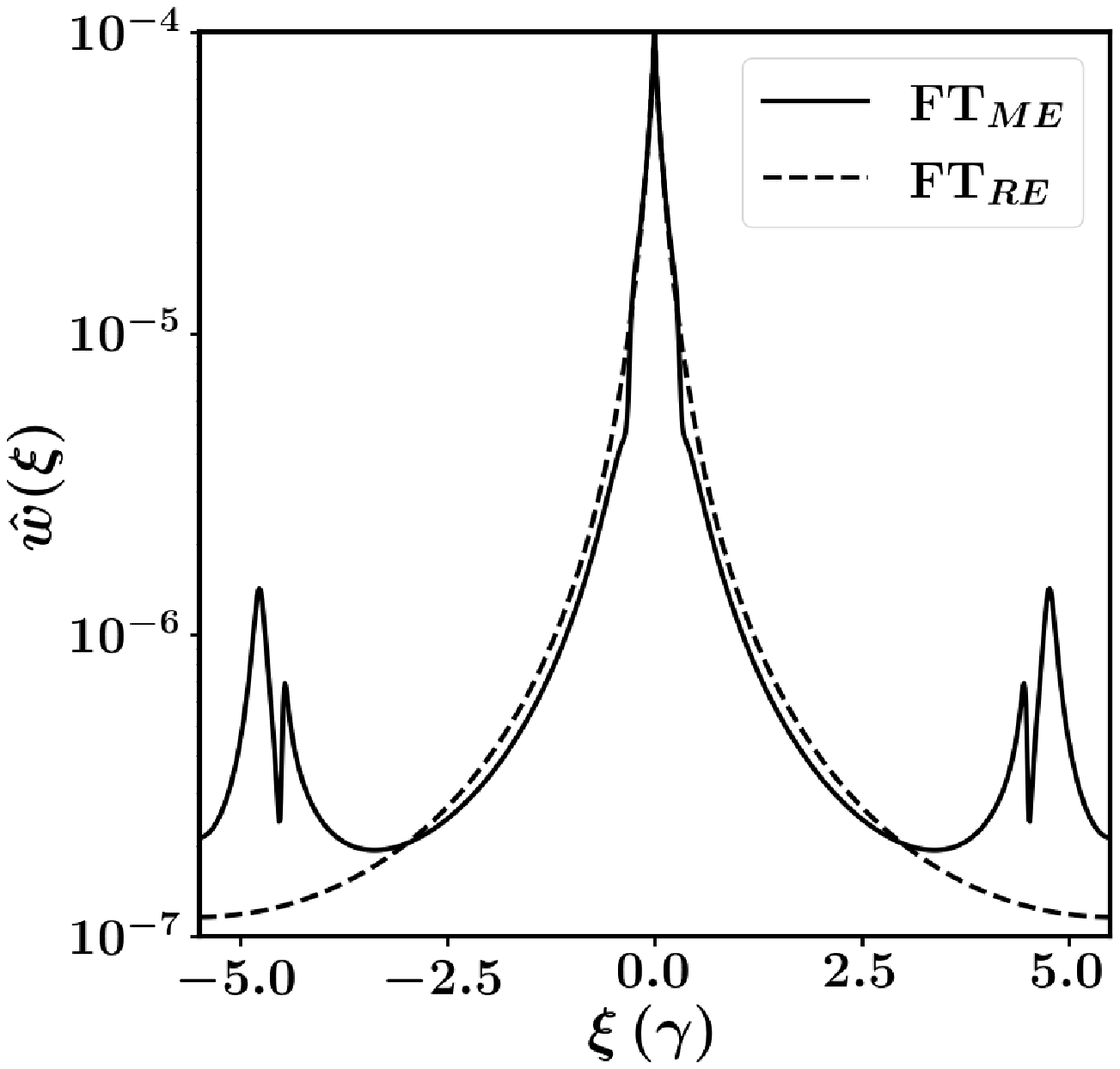}
\caption{}
\label{fig: 4b}
\end{subfigure}
\caption{WTD and occupation probabilities (a) and the corresponding Fourier transform of the WTD, $\hat{w}(\xi)$, (b) for the same parameters as in Fig.(\ref{fig: 2}), except that now the inter-dot couplings are much stronger: $|t_{AB}| = |t_{BC}| = 10\gamma$ and $|t_{AC}| = 9\gamma$.}
\end{centering}
\end{figure}

The final single-occupancy case we consider is that of large inter-dot coupling: $|t| \sim 10\gamma$ in Fig.(\ref{fig: 4a}). For such coupling, the electron is strongly hybridized across the three dots, and so undergoes many inter-dot transitions between tunnelings to the drain. This results in distinct global WTD oscillations, occurring with period $\sim 4\gamma^{-1}$, supplemented by small local oscillations with period $\sim 0.2\gamma^{-1}$. From the corresponding Fourier transform of the WTD, $\hat{w}(\xi)$, which is plotted in Fig.(\ref{fig: 4b}) on a logscale, we see that the small local oscillations are actually resolved into two frequencies at $4.8\gamma = |t_{AC}|/2$ and $5\gamma = |t_{AB}|/2,|t_{BC}|/2$, which evidently correspond to inter-dot transitions. Although not shown, when all $|t_{\nu\nu '}|$ are equal, this double peak collapses into one. The global WTD oscillations, on the other hand, originate from a frequency at $\xi \approx 0.25\gamma$; it is not yet clear what dynamics produces this behavior and this remains an interesting open question for us.

\subsection{Spin-independent triple-occupancy}

\begin{figure}
\begin{centering}
\includegraphics[width = \columnwidth]{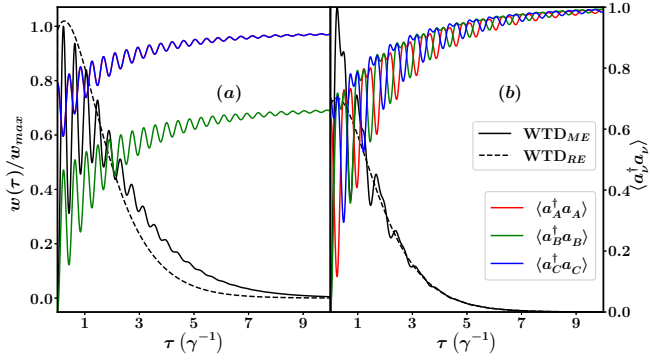}
{\phantomsubcaption\label{fig: 5a}
\phantomsubcaption\label{fig: 5b}}
\caption{WTD and occupation probabilities for the configuration in Fig.(\ref{fig: 1a}) in the triple-occupancy regime. The phase shift in (a) is $\phi = \pi$ and $\phi = \frac{\pi}{2}$ in (b). The inter-dot interaction is $U_{\nu\nu'} = \gamma$, all inter-dot couplings are set to $|t| = 5\gamma$, and all other parameters remain the same as in previous plots. Note that in (a), the $A$ and $C$ dot occupancies are completely in phase, so the red line lies directly beneath the blue one.}
\label{fig: 5}
\end{centering}
\end{figure}

In the spin-independent triple-occupancy regime, the TQD may be occupied by three excess electrons, but there can only be one excess electron on each dot. In this section, we are motivated to explore the same parameter regime that produced the dark state in the single-occupancy regime, again investigating the role of the induced phase shift $\phi$. However, unlike the single-occupancy regime, to observe coherent oscillations now requires an inter-dot coupling greater than the coupling between the TQD and the electrodes: $|t| > \gamma$. If $|t| \sim \gamma$, then the second tunneling to the drain will happen on the same time-scale as transitions between the dots; the resulting WTD is single-peaked. Accordingly, in Fig.(\ref{fig: 5}), we have set $|t| = 5\gamma$.

Although the resulting Fock space is now more complicated, we can still draw relationships between the WTD and corresponding dot occupations as we did for the single-occupancy regime. In Fig.(\ref{fig: 5a}), for example, oscillations in the WTD again correspond to oscillations in the dot $B$ occupancy, which occur at periodic minimums in both the $A$ and $C$ dot occupancies, as this is for the configuration in Fig.(\ref{fig: 1a}). Since $\phi = \pi$ and the dot couplings are symmetric, furthermore, the $A$ and $C$ dot occupancies remain in phase at all times; the red line lies exactly beneath the blue one. By contrast, Fig.(\ref{fig: 5b}) sets $\phi = \frac{\pi}{2}$; the Aharonov-Bohm phase now introduces multiple frequencies into the coherent oscillations and separates the $A$ and $C$ occupancies. 

Apart from the absence of coherent oscillations, in this regime $\text{WTD}_{RE}$ includes another quantitative discrepancy; it predicts that the transport is multiple-reset, $w(\tau) \neq 0$, indicating that dot $B$ can be occupied by more than one electron. Due to the strong intra-dot coupling, $U_{\nu\nu}$, this is evidently impossible, and is thus a feature of the fact that the rate equation only sees `averages' for the occupation of each dot. The WTD calculated from the full BMME, on the other hand, does capture the correct behavior of $w(0) = 0$; after the first jump to the drain dot $B$ must be empty and the system must therefore wait some finite time before the next jump to the drain.

\begin{figure}
\begin{centering}
\includegraphics[width = \columnwidth]{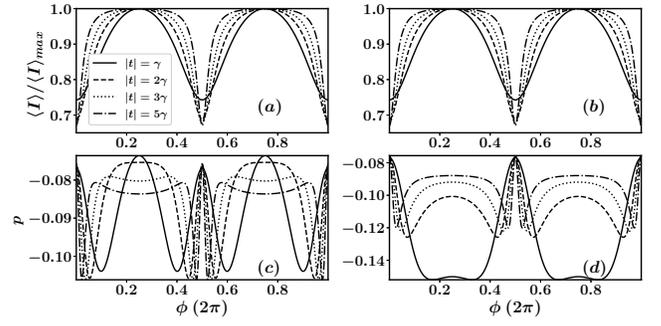}
{\phantomsubcaption\label{fig: 6a}
\phantomsubcaption\label{fig: 6b}
\phantomsubcaption\label{fig: 6c}
\phantomsubcaption\label{fig: 6d}}
\caption{Current as a proportion of its maximum, (a) and (b), and corresponding Pearson correlation coefficient, (c) and (d), all as a function of penetrating magnetic flux $\phi$ and in the triple-occupancy regime for various $|t|$. As in Fig.(\ref{fig: 5}), the inter-dot couplings are all equal to $|t|$ and $U_{\nu\nu '} = \gamma$; otherwise we use the same parameters as all previous plots. The left column is calculated for the configuration in Fig.(\ref{fig: 1a}) and the right column for its mirror in Fig.(\ref{fig: 1b}).}
\label{fig: 6}
\end{centering}
\end{figure}

Despite using the same parameters in Fig.(\ref{fig: 5}) that produced a dark state in the single-occupancy regime, the presence of three electrons largely lifts the coherent population blocking. This can be observed in Figs.(\ref{fig: 6a})-(\ref{fig: 6b}), which calculate the stationary current as a function of the penetrating magnetic flux for both TQD configurations. Similarly to the single-occupancy regime, the  current displays Aharonov-Bohm oscillations, but here the minimum current is still greater than $60\%$ of the maximum. We attribute this to the fact that the Aharonov-Bohm interference is a single-particle effect, at least in the manner we have included it. The two paths available around the TQD allow electrons to accumulate a phase difference with their own wavefunction, but there are still potentially two other electrons with which to form a three-particle wavefunction, diminishing the importance of the Aharonov-Bohm effect. 

\begin{figure}
\begin{centering}
\includegraphics[width = \columnwidth]{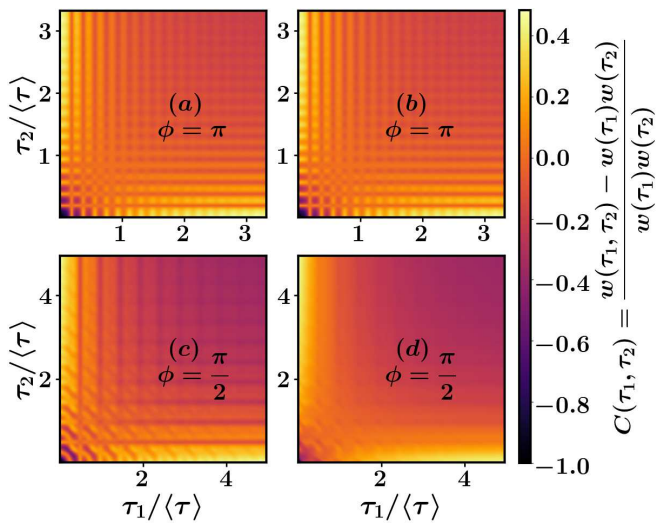}
{\phantomsubcaption\label{fig: 7a}
\phantomsubcaption\label{fig: 7b}
\phantomsubcaption\label{fig: 7c}
\phantomsubcaption\label{fig: 7d}}
\caption{$C(\tau_{1},\tau_{2})$ for multiple configurations and magnetic flux. As in Fig.(\ref{fig: 6}), the columns are organized with Fig.(\ref{fig: 1a}) on the left and Fig.(\ref{fig: 1b}) on the right, while the top and bottom rows set magnetic flux to $\phi = \pi$ and $\phi  = \frac{\pi}{2}$, respectively. Inter-dot dot couplings are $|t| = 5\gamma$ and all other parameters are the same as in Fig.(\ref{fig: 6}).}
\label{fig: 7}
\end{centering}
\end{figure}

Interestingly, we note that the current profiles of both configurations are identical for triple-occupancy. In contrast, the Pearson coefficient, shown in Figs.(\ref{fig: 6c})-(\ref{fig: 6d}), displays quantitatively different behavior between the two configurations; this is a potential method of experimental identification between the two TQD configurations. Aharonov-Bohm interference clearly affects the correlation behavior, as $p$ displays periodic oscillations in $\phi$. Ultimately, this is a very interesting research avenue, as the potential for tuning correlations with magnetic flux could lead to novel information processing methods. At this point, however, the Pearson coefficient is quite small: $|p| \leq 0.1$. In the reverse configuration, we might expect larger correlations; two dots are connected to the drain and so coherent electrons may be detected simultaneously. The Pearson coefficient does actually increase in magnitude for the reverse configuration, but it is still small: $|p| < 0.2$. To analyze further, we turn to $C(\tau_{1},\tau_{2})$, displayed in Fig.(\ref{fig: 7}), which provides correlation information for each pair of successive waiting times.

There are multiple points of interest in Fig.(\ref{fig: 7}); however, we will discuss just a few. We notice immediately that for $\phi = \pi$, both configurations display a highly repeating structure, where the regimes of positive and negative correlations are periodically interspersed with areas of different correlation behavior. Let us examine Fig.(\ref{fig: 7a}), which is for configuration Fig.(\ref{fig: 1a}) and $\phi = \pi$. 

We see that two short waiting times $\tau_{1},\tau_{2} < \langle \tau \rangle$ are in general highly negatively correlated. We can understand this by the following. If $\tau_{1}$ is small, so that we have two electron tunnelings from the TQD to the drain in quick succession, then either the TQD is now fully empty, in which case an electron must tunnel in from the source before another tunneling to the drain, or contains only one remaining excess electron, which must move to dot $B$ before another tunneling to the drain. If there \textit{is} one remaining electron, then we would expect that, as it moves around the TQD, there will be times at which it is occupying dot $B$ and can tunnel to the drain. At these specific times we would expect the correlation relationhip between $\tau_{1}$ and $\tau_{2}$, due only to this mechanism, to be positive. When added to the general negative correlation behavior for $\tau_{1},\tau_{2} < \langle \tau \rangle$, we get an overall correlation value of zero, which is exactly what Fig.(\ref{fig: 7a}) shows. 

If $\tau_{1}$ is long, conversely, then it is likely that the TQD has filled with more electrons from the source, so that now it is likely to observe two tunnelings in quick succession; a long $\tau_{1}$ and short $\tau_{2}$ are positively correlated. In this regime we can still see the periodic structure $C(\tau_{1},\tau_{2})$, again due to electrons moving around the TQD. Now, however, the addition of positive correlations due to this effect increases the value of $C(\tau_{1},\tau_{2})$ overall, which one can see from the periodic brighter spots. 

We see identical behavior in Fig.(\ref{fig: 7b}), which is for the reverse configuration in Fig.(\ref{fig: 1b}). However, if we introduce a phase shift of $\phi = \frac{\pi}{2}$, shown in Fig.(\ref{fig: 7c}) and Fig.(\ref{fig: 7d}), we see that the periodic pattern is both altered and `washed out', in that the distinct repeating structure has been blurred. We can thus move between regimes of highly periodic correlations and comparitively less periodic correlations simply by changing the applied magnetic field, a remarkable degree of control in such systems. A long-term goal in electron transport through nanoscale junctions has been to realize enough control over the system to encode information within the current fluctuations, rather than just the average current. Such a feat would be implicitly related to correlations between successive waiting times, but would require significantly larger values than those found previously. If we were to marry the highly repeating structure we found above with periodically driven transport, say through the current or voltage, then one could conceivably achieve much higher correlation values by matching the driving frequency to the frequency of electron transport around the TQD.

\subsection{Spin-dependent double-occupancy}

\begin{figure}
\begin{centering}
\includegraphics[width = \columnwidth]{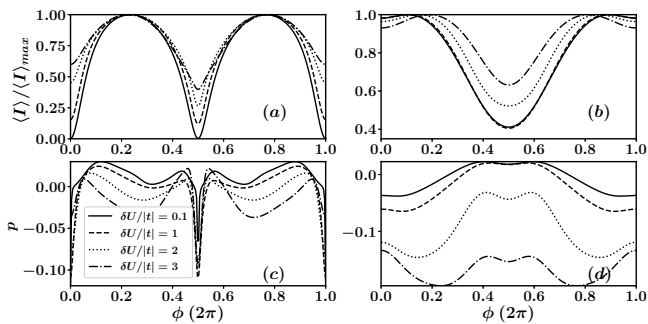}
{\phantomsubcaption\label{fig: 8a}
\phantomsubcaption\label{fig: 8b}
\phantomsubcaption\label{fig: 8c}
\phantomsubcaption\label{fig: 8d}}
\caption{Current as a proportion of its maximum, (a) and (b), and the corresponding Pearson correlation coefficient, (c) and (d), as a function of $\phi$ and for different $\delta U /|t|$. As before, the left column is calculated for the Fig.(\ref{fig: 1a}) configuration and the right column for the Fig.(\ref{fig: 1b}) configuration. All dot energies are set to $\varepsilon_{\nu} = 0$, inter-dot couplings are all $|t| = \gamma$, the inter-dot Coulomb interactions are $U_{AC} = 10\gamma$ and $U_{AB} = U_{BC} = 11\gamma$, and the intra-dot Coulomb interactions are $U_{BB} = 15\gamma$ and $U_{AA} = U_{CC}$. Parameters not mentioned remain the same as in previous plots.}
\label{fig: 8}
\end{centering}
\end{figure}

For spin-dependent double-occupancy, each dot may be occupied by one or two excess electrons, but the TQD as a whole has a maximum occupancy of two. In this regime, P\"{o}ltl et al. \cite{Poltl2009} found that for certain parameters a two-particle dark state forms when the inter-dot repulsion between $A$ and $C$ is equal to the intra-dot repulsion within $A$: $\delta U = U_{AA} - U_{AC} = 0$. Their configuration did not allow tunneling between dot $A$ and $C$, however, and we demonstrate in Fig.(\ref{fig: 8a}) that a dark state exists for the same parameters even when dots $A$ and $C$ are coherently coupled. The dark state can be tuned by $\delta U$ and, now that tunneling between $A$ and $C$ is allowed, the magnetic flux penetrating the ring.

As expected, the correlation between successive waiting times reaches peak magnitude when coherent population blocking is strongest: $\phi = n\pi$ for $\delta U/|t| = 0.1$ in Fig.(\ref{fig: 8a}). The Pearson coefficient, however, is still underwhelming, $|p| < 0.1$, which indicates that a high level of destructive interference does not necessarily produce strong correlations. It is also apparent from Fig.(\ref{fig: 8b}) that significant correlations exist even without the formation of a dark state; the $|p|$ maximum occurs when $\langle I \rangle$ is also a maximum. In this configuration, furthermore, we see the particularly interesting result that correlations can be tuned from positive to negative with $\phi$, as long as $\delta U/|t| \geq 5$.

\begin{figure}
\begin{centering}
\includegraphics[width = \columnwidth]{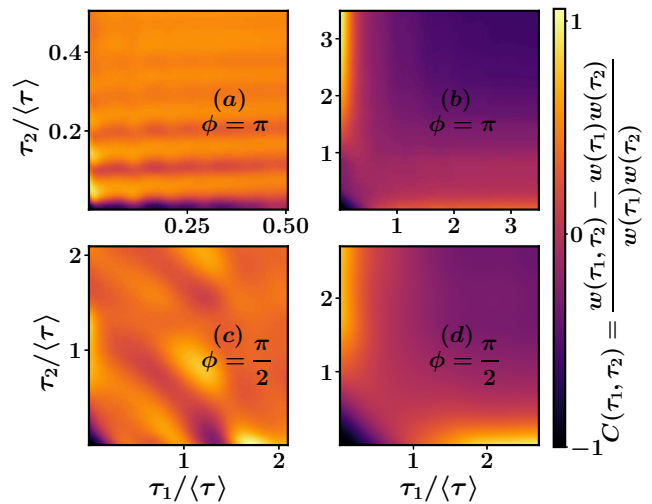}
{\phantomsubcaption\label{fig: 9a}
\phantomsubcaption\label{fig: 9b}
\phantomsubcaption\label{fig: 9c}
\phantomsubcaption\label{fig: 9d}}
\caption{$C(\tau_{1},\tau_{2})$ with the same ordering of configurations and magnetic flux as in Fig.(\ref{fig: 7}). We use the same parameters as in Fig.(\ref{fig: 8}), choosing $\delta U /|t| = 1$ and $\delta U /|t| = 3$ for the left and right columns, respectively.}
\label{fig: 9}
\end{centering}
\end{figure}

To go beyond just the Pearson coefficient, we again turn to $C(\tau_{1},\tau_{2})$ in Fig.(\ref{fig: 9}). Unlike the triple-occupancy regime, in double-occupancy $C(\tau_{1},\tau_{2})$ shows little repeating structure and is not necessarily symmetric under interchange of $\tau_{1}\leftrightarrow\tau_{2}$. In Fig.(\ref{fig: 9a}), for example, there is a band of negatively correlated times at small $\tau_{2}$. That is, if the first waiting time $\tau_{1}$ is less than $0.5\langle \tau \rangle$, it is highly unlikely the waiting time until the next jump $\tau_{2}$ will be extremely short. We can attribute this behavior to there being three quantum jumps involved but only two electrons allowed in the TQD; if $\tau_{1} < 0.5\langle \tau \rangle$, it is likely that the TQD fully empties before the third quantum jump. One might expect, conversely, that observing $\tau_{1} > \langle\tau\rangle$ would then be positively correlated with $\tau_{2} < \langle \tau \rangle$. However, in this configuration and for $\phi = \pi$, $C(\tau_{1},\tau_{2})$ decays for $\tau_{1},\tau_{2} > 0.5\langle \tau \rangle$, hence why we have plotted for $\tau_{1},\tau_{2} < 0.5\langle\tau\rangle$ only.

Fig.(\ref{fig: 9c}) shows that when one imposes a phase difference via $\phi = \frac{\pi}{2}$, then all structure from $C(\tau_{1},\tau_{2})$ disappears but that correlations become stronger and also last longer. For $\tau_{2}<0.25\langle\tau\rangle$ and at $\tau_{1} \approx 1.3\langle\tau\rangle$ and $\tau_{1} \approx 1.75\langle\tau\rangle$, we also see strong positive and negative correlations neighboring each other with only a small uncorrelated area inbetween. This is strange behavior compared to other plots of $C(\tau_{1},\tau_{2})$, in which there are usually large zones of uncorrelatd $\tau_{1},\tau_{2}$ between areas of strong positive and negative correlations.

Indeed, one can observe this in Fig.(\ref{fig: 9b}) for the opposite configuration. An initial waiting time much longer than the average, for example $\tau_{1} = 3\langle\tau\rangle$, is positively correlated with a short second waiting time $\tau_{2} < 0.5\langle \tau \rangle$, as there has been time for the TQD to fill. As expected, the opposite scenario of short $\tau_{1}$ and long $\tau_{2}$ is also positively correlated, although we note that again it is not symmetric. Pairs of waiting times that are both shorter or longer than average are correspondingly negatively correlated. Applying a phase difference of $\phi = \frac{\pi}{2}$ now only serves to extend the regions of correlations, but does not change the overall structure. Now that two dots are coupled to the drain, the system can behave in the multiple-reset nature, which results in the more standard $C(\tau_{1},\tau_{2})$.

\section{Conclusion}\label{sec: Conclusion}

The TQD is a rich experimental and theoretical system displaying quantum phenomena, such as Aharonov-Bohm interference and coherent population blocking. Although it has previously been subjected to fluctuation analysis via FCS in the long-time limit, there is little literature exploring fluctuating-time statistics in this regime. In this paper, therefore, we applied the WTD to a triangular TQD under two geometries in three regimes: spin-independent single- and triple-occupancy and spin-dependent double-occupancy:
\begin{itemize}
\item In the single-occupancy regime, where only one excess electron is allowed in the TQD, we found that coherent oscillations present in the WTD itself, which correlate to occupation probabilities of dots coupled to the drain and depend on the orientation of the TQD. 

\item From the time-dependent occupation probabilities, we are able to see the effect of tuning the Aharonov-Bohm interference on electron tunneling direction in the single-occupancy regime. In the regime of strong inter-dot coupling, furthermore, there exist interesting global and local temporal oscillations for which the underlying dynamics are currently an open question. 

\item Although it is well known that a dark state exists in the single-occupancy regime, we showed that triple-occupancy largely removes coherent population blocking. 

\item In this regime, additionally, the Pearson correlation coefficient is non-zero, although still quite small, and it displays similar periodic, albeit more complicated, behavior to the stationary current. The quantity $C(\tau_{1},\tau_{2})$ shows that, in the triple-occupancy regime, correlations have a highly repeating  structure when $\phi = \pi$, but that this structure is largely destroyed by applying $\phi  = \frac{\pi}{2}$. Such behavior could conceivably be used in conjunction with periodically driven transport to produce temporal correlations of greater magnitude and control.

\item The double-occupancy regime, by contrast, displays relatively unstructured correlation behavior. There are also more significant differences between the two TQD configurations in this regime, most likely due to the fact that each dot in the TQD may now be doubly occupied. Overall, the average Pearson correlation coefficient is stronger in this regime too. 

\item Finally, we also demonstrated that the dark state found by P\"{o}ltl et al. \cite{Poltl2009,Poltl2013} and Busl et al. \cite{Busl2010} extends to a TQD where all dots are coupled, and that, like the single-occupancy regime, this dark state can be tuned via a penetrating magnetic flux. 
\end{itemize}

\appendix 

\section{Master equation derivation}\label{app: Master equation derivation}

First, we condense the bath operators into 
\begin{align}
d^{\dag} &  = \sum_{\alpha} d_{\alpha}^{\dag} & \hspace{-1cm} \text{ and } \hspace{0.4cm} & &  d & = \sum_{\alpha} d_{\alpha} \nonumber \\
& = \sum_{\alpha}\sum_{\mathbf{k}_{\alpha}}\: t^{}_{\mathbf{k}_{\alpha}}a^{\dag}_{\mathbf{k}_{\alpha}} & \hspace{-1cm} \hspace{0.4cm} & & & = \sum_{\alpha}\sum_{\mathbf{k}_{\alpha}}\: t^{}_{\mathbf{k}_{\alpha}}a_{\mathbf{k}_{\alpha}}. \label{new bath operators}
\end{align}
Since we are measuring in one electrode only, we will separate the source and drain parts of the master equation, writing explicitly here only the drain contribution after including the explicit form of $V_{I}(t)$ in Eq.\eqref{First n master equation}:
\begin{widetext}
\begin{align}
\dot{\boldsymbol{\rho}}^{(n)}_{I}(t) & = - \int^{t}_{0} d\tau \: \sum_{qq'} \sum_{n'} \text{Tr}_{B^{(n)}}\left[ \mathcal{M}^{}_{I}(t)a^{\dag}_{q,I}(t)d^{}_{D,I}(t) - \mathcal{M}^{\dag}_{I}(t)a_{q,I}(t) d^{\dag}_{D,I}(t),\right. \nonumber \\
&  \:\:\:\:\: \left.\left[\mathcal{M}^{}_{I}(t-\tau)a^{\dag}_{q',I}(t-\tau)d^{}_{D,I}(t-\tau) - \mathcal{M}^{\dag}_{I}(t-\tau)a_{q',I}(t-\tau) d^{\dag}_{D,I}(t-\tau),\boldsymbol{\rho}_{I}^{(n')}(t-\tau)\boldsymbol{\rho}_{S}\boldsymbol{\rho}_{D}| n' \rangle \langle n' |\right]\right]. \label{eq: first ME appendix}
\end{align}
\end{widetext}
We can consider the source and drain contributions to the master equations additively because under the trace any terms containing operators from different electrodes will disappear. In Eq.\eqref{eq: first ME appendix}, we have also applied the weak coupling ansatz from Eq.\eqref{eq: ansatz Born}.

While the nanostructure and bath operators have their usual form in the interaction picture, it is necessary to discuss the measurement operators,
\begin{align}
\mathcal{M}^{}_{I}(t) &  = \sum_{n} \: e^{i(\varepsilon_{n-1}-\varepsilon_{n})t} | n-1 \rangle\langle n | \\
\mathcal{M}^{\dag}_{I}(t) & = \sum_{n} \: e^{i(\varepsilon_{n}-\varepsilon_{n-1})t} | n \rangle\langle n-1 |,
\end{align}
because at this point we assume that the detector expends a negligible amount of energy in recording each tunneling, so that $|\varepsilon_{n-1}-\varepsilon_{n}|t \approx 0$, resulting in $\mathcal{M}^{}_{I}(t) = \mathcal{M}^{}$ and $\mathcal{M}^{\dag}_{I}(t) = \mathcal{M}^{\dag}$. 

Expanding now the commutators from Eq.\eqref{eq: first ME appendix}, excluding terms with $d^{}_{D,I}d^{}_{D,I}$ or $d^{\dag}_{D,I}d^{\dag}_{D,I}$ as they disappear under the trace, and collecting all measurementpr operators and detector states, the master equation reads
\begin{widetext}
\begin{align}
\dot{\boldsymbol{\rho}}^{(n')}_{I}(t) & = \int^{t}_{0} d\tau \: \sum_{qq'}\sum_{n'} \text{Tr}_{B}\left(| n \rangle \langle n |\mathcal{M}^{}\mathcal{M}^{\dag}| n' \rangle \langle n' |a^{\dag}_{q,I}(t)d^{}_{I}(t) a_{q',I}(t-\tau) d^{\dag}_{I}(t-\tau)\boldsymbol{\rho}^{(n')}_{I}(t)\boldsymbol{\rho}_{S}\boldsymbol{\rho}_{D}\right. \nonumber \\
&\:\:\:\: + | n \rangle \langle n |\mathcal{M}^{\dag}\mathcal{M}^{}| n' \rangle \langle n' |a_{q,I}(t)d^{\dag}_{I}(t) a^{\dag}_{q',I}(t-\tau) d_{I}(t-\tau)\boldsymbol{\rho}^{(n')}_{I}(t-\tau)\boldsymbol{\rho}_{S}\boldsymbol{\rho}_{D} \nonumber \\ 
& \:\:\:\: - | n \rangle \langle n |\mathcal{M}^{}| n' \rangle \langle n' | \mathcal{M}^{\dag}a^{\dag}_{q,I}(t)d_{I}(t)\boldsymbol{\rho}^{(n')}_{I}(t-\tau)\boldsymbol{\rho}_{S}\boldsymbol{\rho}_{D}a_{q',I}(t-\tau) d^{\dag}_{I}(t-\tau) \nonumber \\
&\:\:\:\: - | n \rangle \langle n |\mathcal{M}^{\dag}| n' \rangle \langle n' | \mathcal{M}^{} a_{q,I}(t)d^{\dag}_{I}(t)\boldsymbol{\rho}^{(n')}_{I}(t-\tau)\boldsymbol{\rho}_{S}\boldsymbol{\rho}_{D}a^{\dag}_{q',I}(t-\tau) d_{I}(t-\tau) \nonumber \\
&\:\:\:\: - | n \rangle \langle n |\mathcal{M}^{}| n' \rangle \langle n' | \mathcal{M}^{\dag} a^{\dag}_{q',I}(t-\tau)d_{I}(t-\tau)\boldsymbol{\rho}^{(n')}_{I}(t-\tau)\boldsymbol{\rho}_{S}\boldsymbol{\rho}_{D}a_{q,I}(t) d^{\dag}_{I}(t)  \nonumber \\ 
&\:\:\:\: - | n \rangle \langle n |\mathcal{M}^{\dag}| n' \rangle \langle n' | \mathcal{M}^{} a_{q',I}(t-\tau)d^{\dag}_{I}(t-\tau)\boldsymbol{\rho}^{(n')}_{I}(t-\tau)\boldsymbol{\rho}_{S}\boldsymbol{\rho}_{D}a^{\dag}_{q,I}(t) d_{I}(t) \nonumber \\ 
&\:\:\:\: + | n \rangle \langle n | n' \rangle \langle n' |\mathcal{M}^{}\mathcal{M}^{\dag}\boldsymbol{\rho}^{(n')}_{I}(t-\tau)\boldsymbol{\rho}_{S}\boldsymbol{\rho}_{D}a^{\dag}_{q',I}(t-\tau)d^{}_{I}(t-\tau) a_{q,I}(t) d^{\dag}_{I}(t) \nonumber \\ 
&\:\:\:\: \left. + | n \rangle \langle n | n' \rangle \langle n' |\mathcal{M}^{}\mathcal{M}^{\dag}\boldsymbol{\rho}^{(n')}_{I}(t-\tau)\boldsymbol{\rho}_{S}\boldsymbol{\rho}_{D}a_{q',I}(t-\tau)d^{\dag}_{I}(t-\tau) a^{\dag}_{q,I}(t) d_{I}(t) \right). \label{eq: Second n master equation}
\end{align}
\end{widetext}

This removes the summation over $n'$ and transforms the master equation to
\begin{widetext}
\begin{align}
\dot{\boldsymbol{\rho}}^{(n)}_{I}(t) & = \int^{t}_{0} d\tau \: \sum_{qq'} \text{Tr}_{B}\left\{| n \rangle \langle n | \left(a^{\dag}_{q,I}(t)d^{}_{I}(t) a_{q',I}(t-\tau) d^{\dag}_{I}(t-\tau)\boldsymbol{\rho}^{(n)}_{I}(t-\tau)\boldsymbol{\rho}_{S}\boldsymbol{\rho}_{D}\right.\right. \nonumber \\
&\:\:\:\: + a_{q,I}(t)d^{\dag}_{I}(t) a^{\dag}_{q',I}(t-\tau) d_{I}(t-\tau)\boldsymbol{\rho}^{(n)}_{I}(t-\tau)\boldsymbol{\rho}_{S}\boldsymbol{\rho}_{D}  -  a^{\dag}_{q,I}(t)d_{I}(t)\boldsymbol{\rho}^{(n)}_{I}(t-\tau)\boldsymbol{\rho}_{S}\boldsymbol{\rho}_{D}a_{q',I}(t-\tau) d^{\dag}_{I}(t-\tau) \nonumber \\
&\:\:\:\: -   a_{q,I}(t)d^{\dag}_{I}(t)\boldsymbol{\rho}^{(n)}_{I}(t-\tau)\boldsymbol{\rho}_{S}\boldsymbol{\rho}_{D}a^{\dag}_{q',I}(t-\tau) d_{I}(t-\tau)  -   a^{\dag}_{q',I}(t-\tau)d_{I}(t-\tau)\boldsymbol{\rho}^{(n)}_{I}(t-\tau)\boldsymbol{\rho}_{S}\boldsymbol{\rho}_{D}a_{q,I}(t) d^{\dag}_{I}(t)  \nonumber \\ 
&\:\:\:\: -   a_{q',I}(t-\tau)d^{\dag}_{I}(t-\tau)\boldsymbol{\rho}^{(n)}_{I}(t-\tau)\boldsymbol{\rho}_{S}\boldsymbol{\rho}_{D}a^{\dag}_{q,I}(t) d_{I}(t) +  \boldsymbol{\rho}^{(n)}_{I}(t-\tau)\boldsymbol{\rho}_{S}\boldsymbol{\rho}_{D}a^{\dag}_{q',I}(t-\tau)d^{}_{I}(t-\tau) a_{q,I}(t) d^{\dag}_{I}(t) \nonumber \\ 
&\:\:\:\: \left.\left. + \boldsymbol{\rho}^{(n)}_{I}(t-\tau)\boldsymbol{\rho}_{S}\boldsymbol{\rho}_{D}a_{q',I}(t-\tau)d^{\dag}_{I}(t-\tau) a^{\dag}_{q,I}(t) d_{I}(t) \right)\right\}. \label{eq: third n master equation}
\end{align}
\end{widetext}
The trace over the detector space is removed by the projector $| n \rangle \langle n |$, so that $\text{Tr}_{B}\left[\hdots\right]$ yields the standard bath-correlation functions:
\begin{align}
G_{\alpha}^{>}(\tau) & = \text{Tr}_{B}\left(\rho_{B}d_{\alpha,I}(t)d^{\dag}_{I}(t-\tau)\right) \nonumber \\
G_{\alpha}^{<}(\tau) & = \text{Tr}_{B}\left(\rho_{B}d^{\dag}_{I}(t)d_{\alpha,I}(t-\tau)\right).
\end{align}
Finally, we apply Markovianity after returning from the interaction picture, so that $\boldsymbol{\rho}(t-\tau) \approx \boldsymbol{\rho}(t)$ and 
\begin{widetext}
\begin{align}
\dot{\boldsymbol{\rho}}^{(n)}(t) & = \left[H_{Q},\boldsymbol{\rho}^{(n)}(t)(t)\right] - i\sum_{qq'} \int^{\infty}_{0} d\tau \: \Big(\sum_{\alpha}\left[G^{>}_{\alpha}(\tau)a^{\dag}_{q}e^{-iH_{Q}\tau} a_{q'}e^{iH_{Q}\tau} \boldsymbol{\rho}^{(n)}(t)  + G^{<}_{\alpha}(\tau) a_{q}e^{-iH_{Q}\tau} a^{\dag}_{q'}e^{iH_{Q}\tau}\boldsymbol{\rho}^{(n)}(t)\right]  \nonumber \\ 
& \:\:\:\: -\left[G^{>}_{S}(\tau) e^{-iH_{Q}\tau} a_{q'} e^{iH_{Q}\tau} \boldsymbol{\rho}^{(n)}(t) a^{\dag}_{q}  + G^{<}_{S}(\tau) e^{-iH_{Q}\tau} a^{\dag}_{q'} e^{iH_{Q}\tau} \boldsymbol{\rho}^{(n)}(t) a_{q} + G^{>}_{D}(\tau) e^{-iH_{Q}\tau} a_{q'} e^{iH_{Q}\tau} \boldsymbol{\rho}^{(n-1)}(t) a^{\dag}_{q} \right.  \nonumber \\
& \:\:\:\:  \left.  + G^{<}_{D}(\tau) e^{-iH_{Q}\tau} a^{\dag}_{q'} e^{iH_{Q}\tau} \boldsymbol{\rho}^{(n+1)}(t) a_{q} \right] + \text{h.c.}\Big).  \label{Third n master equation}
\end{align}
\end{widetext}

Performing the integral and defining lesser, $\Sigma_{\alpha}^{<}(\omega_{mn})$, and greater, $\Sigma_{\alpha}^{>}(\omega_{mn})$ self-energies as 
\begin{align}
\Sigma_{\alpha}^{<}(\omega_{mn}) & = i \int^{\infty}_{0} d\tau \: e^{-i\omega_{nm}\tau}G_{\alpha}^{<}(-\tau) \nonumber \\
\Sigma_{\alpha}^{>}(\omega_{mn}) & = -i \int^{\infty}_{0} d\tau \: e^{-i\omega_{nm}\tau}G_{\alpha}^{>}(-\tau), 
\end{align}
we get the master equation in Eq.\eqref{n resolved ME final}.

\section{Analytic Lamb shifts}\label{app: Analytic Lamb shifts}

Under the wide-band limit, we can calculate the Lamb shifts analytically via residue theory and the digamma function $\psi(x)$:
\begin{align}
\Delta^{<}_{\alpha}(\omega) & = \frac{\gamma^{\alpha}}{2\pi}\Re \left\{\psi \left(\frac{1}{2} + \frac{i}{2\pi T}(\omega - \mu_{\alpha}\right) \right\} \label{eq: lamb shift lesser digamma} \\
\Delta^{>}_{\alpha}(\omega) & = -\frac{\gamma^{\alpha}}{2\pi}\Re \left\{\psi \left(\frac{1}{2} + \frac{i}{2\pi T}(\omega - \mu_{\alpha}\right) \right\}. \label{eq: lamb shift greater digamma}
\end{align}
Since the integrals in Eq.\eqref{eq: lesser lamb shift}-Eq.\eqref{eq: greater lamb shift} are formally divergent, Eq.\eqref{eq: lamb shift lesser digamma} and Eq.\eqref{eq: lamb shift greater digamma} also contain divergent logarithmic terms: $-\lim\limits_{W\rightarrow\infty} \ln \left[\frac{W}{2\pi T}\right]$. We can safely ignore these terms, however, because self-energies always appear in the master equation as pairs of differences, $\Sigma^{<,>}_{\alpha}(\omega_{ij}) - \Sigma^{<,>}_{\alpha}(\omega_{kl})$, so these terms will always cancel.

\section{TQD triple-occupancy}\label{app: TQD triple-occupancy}

In defining the individual equations of motion, it is expedient to define coefficients for the overlap between the transformed single- and two-particle energy eigenstates as 
\begin{align}
D_{1i,1j} & = \sum_{\nu = \{A,C\}} \langle 0 | a_{\nu}^{} | 1i \rangle \langle 1j | a^{\dag}_{\nu} | 0 \rangle  \\
G_{2r,2s} & = \sum_{\nu = \{A,C\}} \langle 3 | a_{\nu}^{\dag} | 2r \rangle \langle 2s | a^{}_{\nu} | 3 \rangle  \\
E_{2r,2s,1i,1j} &  = \sum_{\nu = \{A,C\}} \langle 1i | a^{}_{\nu} | 2r \rangle \langle 2s | a^{\dag}_{\nu} | 1j \rangle \\
F_{2r,2s,1i,1j} &  = \langle 1i | a^{}_{B} | 2r \rangle \langle 2s | a^{\dag}_{B} | 1j \rangle.
\end{align}
The master equations for each element are 
\begin{widetext}
\begin{align}
\dot{\rho}_{0,0} & = -i\rho_{0,0} \left(\sum_{1i}\: D_{1i,1i}\left(\Sigma^{<}_{S}(\omega_{1i,0}) - \Sigma^{<}_{S}(\omega_{1i,0})^{*}\right) + |c_{1i,B}|^{2} \left(\Sigma^{<}_{D}(\omega_{1i,0}) - \Sigma^{<}_{D}(\omega_{1i,0})^{*}\right)\right. \nonumber \\
& \:\:\:\:\: - i\sum_{1j}\sum_{1i} \rho_{1i,1j} \left[D_{1i,1j}\left(\Sigma_{S}^{>}(\omega_{1j,0})^{*} - \Sigma_{S}^{>}(\omega_{1i,0})\right) + e^{i\chi}c^{}_{1i,B}c_{1j,B}^{*}\left(\Sigma_{D}^{>}(\omega_{1j,0})^{*} - \Sigma_{D}^{>}(\omega_{1i,0})\right)\right],
\end{align}
\begin{align}
\dot{\rho}_{1i,1j} & =  -i\omega_{1i,1j} - i\rho_{0,0} \left[D_{1j,1i}\left(\Sigma_{S}^{<}(\omega_{1j,0}) - \Sigma_{S}^{<}(\omega_{1i,0})^{*}\right) + e^{-i\chi}c^{}_{1j,B}c_{1i,B}^{*}\left(\Sigma_{D}^{<}(\omega_{1j,0}) - \Sigma_{D}^{<}(\omega_{1i,0})^{*}\right) \right] \nonumber \\
& - i\Bigg[\sum_{1k} \Big(\rho_{1k,1j}\Big[D_{1k,1i}  \Sigma_{S}^{>}(\omega_{1k,0}) + c^{}_{1k,B}c^{*}_{1i,B}\Sigma_{D}^{>}(\omega_{1k,0}) + \sum_{2r}  \left( E_{2r,2r,1i,1k}  \Sigma_{S}^{<}(\omega_{2r,1k})^{*}\right.  \nonumber \\
& \left.+ F_{2r,2r,1i,1k}  \Sigma_{D}^{<}(\omega_{2r,1k})^{*}\right)\Big] - \rho_{1i,1k}\Big[D_{1j,1k}  \Sigma_{S}^{>}(\omega_{1k,0})^{*} + c^{}_{1j,B}c^{*}_{1k,B}\Sigma_{D}^{>}(\omega_{1k,0})^{*} \nonumber \\
& + \sum_{2r}  \left( E_{2r,2r,1k,1j}  \Sigma_{S}^{<}(\omega_{2r,1k}) + F_{2r,2r,1k,1j}  \Sigma_{D}^{<}(\omega_{2r,1k})\right)\Big]\Big) + \sum_{2s}\sum_{2r} \rho_{2r,2s}\big[E_{2r,2s,1i,1j}\left(\Sigma_{S}^{>}(\omega_{2s,1j})^{*} \right. \nonumber \\
& \left. - \Sigma_{S}^{>}(\omega_{2r,1i})\right) + e^{i\chi} F_{2r,2s,1i,1j}\left(\Sigma_{D}^{>}(\omega_{2s,1j})^{*} - \Sigma_{D}^{>}(\omega_{2r,1i})\right)\big]\Bigg],
\end{align}
\begin{align}
\dot{\rho}_{2r,2s} & = -i\omega_{2r,2s} -i \rho_{3,3} \left[G_{2s,2r}\left(\Sigma_{S}^{>}(\omega_{3,2s})^{*} - \Sigma_{S}^{>}(\omega_{3,2r})\right) + e^{i\chi}c^{}_{2r,AC}c_{2s,AC}^{*}\left(\Sigma_{D}^{>}(\omega_{3,2s})^{*} - \Sigma_{D}^{>}(\omega_{3,2r})\right) \right] \nonumber \\
& - i\Bigg[\sum_{2t} \Big(\rho_{2t,2s}\Big[G_{2t,2r}  \Sigma_{S}^{<}(\omega_{3,2t})^{*} + c^{}_{2t,AC}c^{*}_{2r,AC} \Sigma_{D}^{<}(\omega_{3,2t})^{*} + \sum_{1i}  \left( E_{2t,2r,1i,1i}  \Sigma_{S}^{>}(\omega_{2t,1i})\right. \nonumber \\
& \left.+ F_{2t,2r,1i,1i} \Sigma_{D}^{>}(\omega_{2t,1i}) \right)\Big] - \rho_{2r,2t}\Big[G_{2s,2t}  \Sigma_{S}^{<}(\omega_{3,2t}) + c^{}_{2s,AC}c^{*}_{2t,AC}\Sigma_{D}^{<}(\omega_{3,2t}) \nonumber \\ 
& + \sum_{1i}  \left( E_{2s,2t,1i,1i}  \Sigma_{S}^{>}(\omega_{2t,1i})^{*} + F_{2s,2t,1i,1i}  \Sigma_{D}^{>}(\omega_{2t,1i})^{*}\right)\Big]\Big) + \sum_{1j}\sum_{1i} \rho_{1i,1j}\big[E_{2s,2r,1j,1i}\left(\Sigma_{S}^{<}(\omega_{2s,1j})\right. \nonumber \\
& \left. - \Sigma_{S}^{<}(\omega_{2r,1i})^{*}\right) + e^{-i\chi} F_{2s,2r,1j,1i}\left(\Sigma_{D}^{<}(\omega_{2s,1j}) - \Sigma_{D}^{<}(\omega_{2r,1i})^{*}\right)\big]\Bigg], \:\:\: \text{ and }
\end{align}
\begin{align}
\dot{\rho}_{3,3} & = -i\rho_{3,3} \left(\sum_{2r}\: G_{2r,2r}\left(\Sigma^{>}_{S}(\omega_{3,2r}) - \Sigma^{>}_{S}(\omega_{3,2r})^{*}\right) + |c^{}_{2r,AC}|^{2}\left(\Sigma^{>}_{D}(\omega_{3,2r}) - \Sigma^{>}_{D}(\omega_{3,2r})^{*}\right)\right) \nonumber \\
& \:\:\:\:\: -i\sum_{2s}\sum_{2r} \rho_{2r,2s} \left[G_{2r,2s}\left(\Sigma_{S}^{<}(\omega_{3,2s}) - \Sigma_{S}^{<}(\omega_{3,2r})^{*}\right) + e^{-i\chi}c^{}_{1i,B}c_{1j,B}^{*}\left(\Sigma_{D}^{<}(\omega_{3,2s}) - \Sigma_{D}^{<}(\omega_{3,2r})^{*}\right)\right]. 
\end{align}
\end{widetext}
For the configuration in Fig.(\ref{fig: 1b}), $\Sigma_{S}(\omega) \leftrightarrow \Sigma_{D}(\omega)$.


\begin{thebibliography}{82}%
\makeatletter
\providecommand \@ifxundefined [1]{%
 \@ifx{#1\undefined}
}%
\providecommand \@ifnum [1]{%
 \ifnum #1\expandafter \@firstoftwo
 \else \expandafter \@secondoftwo
 \fi
}%
\providecommand \@ifx [1]{%
 \ifx #1\expandafter \@firstoftwo
 \else \expandafter \@secondoftwo
 \fi
}%
\providecommand \natexlab [1]{#1}%
\providecommand \enquote  [1]{``#1''}%
\providecommand \bibnamefont  [1]{#1}%
\providecommand \bibfnamefont [1]{#1}%
\providecommand \citenamefont [1]{#1}%
\providecommand \href@noop [0]{\@secondoftwo}%
\providecommand \href [0]{\begingroup \@sanitize@url \@href}%
\providecommand \@href[1]{\@@startlink{#1}\@@href}%
\providecommand \@@href[1]{\endgroup#1\@@endlink}%
\providecommand \@sanitize@url [0]{\catcode `\\12\catcode `\$12\catcode
  `\&12\catcode `\#12\catcode `\^12\catcode `\_12\catcode `\%12\relax}%
\providecommand \@@startlink[1]{}%
\providecommand \@@endlink[0]{}%
\providecommand \url  [0]{\begingroup\@sanitize@url \@url }%
\providecommand \@url [1]{\endgroup\@href {#1}{\urlprefix }}%
\providecommand \urlprefix  [0]{URL }%
\providecommand \Eprint [0]{\href }%
\providecommand \doibase [0]{http://dx.doi.org/}%
\providecommand \selectlanguage [0]{\@gobble}%
\providecommand \bibinfo  [0]{\@secondoftwo}%
\providecommand \bibfield  [0]{\@secondoftwo}%
\providecommand \translation [1]{[#1]}%
\providecommand \BibitemOpen [0]{}%
\providecommand \bibitemStop [0]{}%
\providecommand \bibitemNoStop [0]{.\EOS\space}%
\providecommand \EOS [0]{\spacefactor3000\relax}%
\providecommand \BibitemShut  [1]{\csname bibitem#1\endcsname}%
\let\auto@bib@innerbib\@empty
\bibitem [{\citenamefont {Engelhardt}\ and\ \citenamefont
  {Cao}(2019)}]{Engelhardt2019}%
  \BibitemOpen
  \bibfield  {author} {\bibinfo {author} {\bibfnamefont {G.}~\bibnamefont
  {Engelhardt}}\ and\ \bibinfo {author} {\bibfnamefont {J.}~\bibnamefont
  {Cao}},\ }\href {\doibase 10.1103/PhysRevB.99.075436} {\bibfield  {journal}
  {\bibinfo  {journal} {Phys. Rev. B}\ }\textbf {\bibinfo {volume} {99}},\
  \bibinfo {pages} {075436} (\bibinfo {year} {2019})}\BibitemShut {NoStop}%
\bibitem [{\citenamefont {Guo}\ \emph {et~al.}(2007)\citenamefont {Guo},
  \citenamefont {Whalley}, \citenamefont {Klare}, \citenamefont {Huang},
  \citenamefont {O'Brien}, \citenamefont {Steigerwald},\ and\ \citenamefont
  {Nuckolls}}]{Guo2007}%
  \BibitemOpen
  \bibfield  {author} {\bibinfo {author} {\bibfnamefont {X.}~\bibnamefont
  {Guo}}, \bibinfo {author} {\bibfnamefont {A.}~\bibnamefont {Whalley}},
  \bibinfo {author} {\bibfnamefont {J.~E.}\ \bibnamefont {Klare}}, \bibinfo
  {author} {\bibfnamefont {L.}~\bibnamefont {Huang}}, \bibinfo {author}
  {\bibfnamefont {S.}~\bibnamefont {O'Brien}}, \bibinfo {author} {\bibfnamefont
  {M.}~\bibnamefont {Steigerwald}}, \ and\ \bibinfo {author} {\bibfnamefont
  {C.}~\bibnamefont {Nuckolls}},\ }\href {\doibase 10.1021/nl070245a}
  {\bibfield  {journal} {\bibinfo  {journal} {Nano Lett.}\ }\textbf {\bibinfo
  {volume} {7}},\ \bibinfo {pages} {1119} (\bibinfo {year} {2007})}\BibitemShut
  {NoStop}%
\bibitem [{\citenamefont {Cuevas}\ and\ \citenamefont
  {Scheer}(2010)}]{Scheer2010}%
  \BibitemOpen
  \bibfield  {author} {\bibinfo {author} {\bibfnamefont {J.}~\bibnamefont
  {Cuevas}}\ and\ \bibinfo {author} {\bibfnamefont {E.}~\bibnamefont
  {Scheer}},\ }\href {https://www.worldscientific.com/doi/abs/10.1142/7434}
  {\emph {\bibinfo {title}
  {\href{https://www.worldscientific.com/doi/abs/10.1142/7434}{Molecular
  Electronics: An Introduction to Theory and Experiment}}}}\ (\bibinfo
  {publisher} {World Scientific, Singapore},\ \bibinfo {year}
  {2010})\BibitemShut {NoStop}%
\bibitem [{\citenamefont {Xiang}\ \emph {et~al.}(2016)\citenamefont {Xiang},
  \citenamefont {Wang}, \citenamefont {Jia}, \citenamefont {Lee},\ and\
  \citenamefont {Guo}}]{Xiang2016}%
  \BibitemOpen
  \bibfield  {author} {\bibinfo {author} {\bibfnamefont {D.}~\bibnamefont
  {Xiang}}, \bibinfo {author} {\bibfnamefont {X.}~\bibnamefont {Wang}},
  \bibinfo {author} {\bibfnamefont {C.}~\bibnamefont {Jia}}, \bibinfo {author}
  {\bibfnamefont {T.}~\bibnamefont {Lee}}, \ and\ \bibinfo {author}
  {\bibfnamefont {X.}~\bibnamefont {Guo}},\ }\href {\doibase
  10.1021/acs.chemrev.5b00680} {\bibfield  {journal} {\bibinfo  {journal}
  {Chem. Rev.}\ }\textbf {\bibinfo {volume} {116}},\ \bibinfo {pages} {4318}
  (\bibinfo {year} {2016})}\BibitemShut {NoStop}%
\bibitem [{\citenamefont {Koch}\ \emph {et~al.}(2006)\citenamefont {Koch},
  \citenamefont {von Oppen},\ and\ \citenamefont {{Andreev}}}]{Koch2006}%
  \BibitemOpen
  \bibfield  {author} {\bibinfo {author} {\bibfnamefont {J.}~\bibnamefont
  {Koch}}, \bibinfo {author} {\bibfnamefont {F.}~\bibnamefont {von Oppen}}, \
  and\ \bibinfo {author} {\bibfnamefont {A.~V.}\ \bibnamefont {{Andreev}}},\
  }\href {\doibase 10.1103/PhysRevB.74.205438} {\bibfield  {journal} {\bibinfo
  {journal} {Phys. Rev. B}\ }\textbf {\bibinfo {volume} {74}},\ \bibinfo
  {pages} {205438} (\bibinfo {year} {2006})}\BibitemShut {NoStop}%
\bibitem [{\citenamefont {Schinabeck}\ \emph {et~al.}(2016)\citenamefont
  {Schinabeck}, \citenamefont {Erpenbeck}, \citenamefont {H\"artle},\ and\
  \citenamefont {Thoss}}]{Schinabeck2016}%
  \BibitemOpen
  \bibfield  {author} {\bibinfo {author} {\bibfnamefont {C.}~\bibnamefont
  {Schinabeck}}, \bibinfo {author} {\bibfnamefont {A.}~\bibnamefont
  {Erpenbeck}}, \bibinfo {author} {\bibfnamefont {R.}~\bibnamefont {H\"artle}},
  \ and\ \bibinfo {author} {\bibfnamefont {M.}~\bibnamefont {Thoss}},\ }\href
  {\doibase 10.1103/PhysRevB.94.201407} {\bibfield  {journal} {\bibinfo
  {journal} {Phys. Rev. B}\ }\textbf {\bibinfo {volume} {94}},\ \bibinfo
  {pages} {201407(R)} (\bibinfo {year} {2016})}\BibitemShut {NoStop}%
\bibitem [{\citenamefont {Schinabeck}\ \emph {et~al.}(2018)\citenamefont
  {Schinabeck}, \citenamefont {H\"{a}rtle},\ and\ \citenamefont
  {Thoss}}]{Schinabeck2018}%
  \BibitemOpen
  \bibfield  {author} {\bibinfo {author} {\bibfnamefont {C.}~\bibnamefont
  {Schinabeck}}, \bibinfo {author} {\bibfnamefont {R.}~\bibnamefont
  {H\"{a}rtle}}, \ and\ \bibinfo {author} {\bibfnamefont {M.}~\bibnamefont
  {Thoss}},\ }\href {\doibase 10.1103/PhysRevB.97.235429} {\bibfield  {journal}
  {\bibinfo  {journal} {Phys. Rev. B}\ }\textbf {\bibinfo {volume} {97}},\
  \bibinfo {pages} {235429} (\bibinfo {year} {2018})}\BibitemShut {NoStop}%
\bibitem [{\citenamefont {Ihn}\ \emph {et~al.}(2007)\citenamefont {Ihn},
  \citenamefont {Sigrist}, \citenamefont {Ensslin}, \citenamefont
  {Wegscheider},\ and\ \citenamefont {Reinwald}}]{Ihn2007}%
  \BibitemOpen
  \bibfield  {author} {\bibinfo {author} {\bibfnamefont {T.}~\bibnamefont
  {Ihn}}, \bibinfo {author} {\bibfnamefont {M.}~\bibnamefont {Sigrist}},
  \bibinfo {author} {\bibfnamefont {K.}~\bibnamefont {Ensslin}}, \bibinfo
  {author} {\bibfnamefont {W.}~\bibnamefont {Wegscheider}}, \ and\ \bibinfo
  {author} {\bibfnamefont {M.}~\bibnamefont {Reinwald}},\ }\href {\doibase
  10.1088/1367-2630/9/5/111} {\bibfield  {journal} {\bibinfo  {journal} {New J.
  Phys.}\ }\textbf {\bibinfo {volume} {9}},\ \bibinfo {pages} {111} (\bibinfo
  {year} {2007})}\BibitemShut {NoStop}%
\bibitem [{\citenamefont {Strambini}\ \emph {et~al.}(2009)\citenamefont
  {Strambini}, \citenamefont {Piazza}, \citenamefont {Biasiol}, \citenamefont
  {Sorba},\ and\ \citenamefont {Beltram}}]{Strambini2009}%
  \BibitemOpen
  \bibfield  {author} {\bibinfo {author} {\bibfnamefont {E.}~\bibnamefont
  {Strambini}}, \bibinfo {author} {\bibfnamefont {V.}~\bibnamefont {Piazza}},
  \bibinfo {author} {\bibfnamefont {G.}~\bibnamefont {Biasiol}}, \bibinfo
  {author} {\bibfnamefont {L.}~\bibnamefont {Sorba}}, \ and\ \bibinfo {author}
  {\bibfnamefont {F.}~\bibnamefont {Beltram}},\ }\href {\doibase
  10.1103/PhysRevB.79.195443} {\bibfield  {journal} {\bibinfo  {journal} {Phys.
  Rev. B}\ }\textbf {\bibinfo {volume} {79}},\ \bibinfo {pages} {195443}
  (\bibinfo {year} {2009})}\BibitemShut {NoStop}%
\bibitem [{\citenamefont {Kobayashi}\ \emph {et~al.}(2002)\citenamefont
  {Kobayashi}, \citenamefont {Aikawa}, \citenamefont {Katsumoto},\ and\
  \citenamefont {Iye}}]{Kobayashi2002}%
  \BibitemOpen
  \bibfield  {author} {\bibinfo {author} {\bibfnamefont {K.}~\bibnamefont
  {Kobayashi}}, \bibinfo {author} {\bibfnamefont {H.}~\bibnamefont {Aikawa}},
  \bibinfo {author} {\bibfnamefont {S.}~\bibnamefont {Katsumoto}}, \ and\
  \bibinfo {author} {\bibfnamefont {Y.}~\bibnamefont {Iye}},\ }\href {\doibase
  10.1143/JPSJ.71.2094} {\bibfield  {journal} {\bibinfo  {journal} {J. Phys.
  Soc. Jpn}\ }\textbf {\bibinfo {volume} {71}},\ \bibinfo {pages} {2094}
  (\bibinfo {year} {2002})}\BibitemShut {NoStop}%
\bibitem [{\citenamefont {Gaudreau}\ \emph {et~al.}(2006)\citenamefont
  {Gaudreau}, \citenamefont {Studenikin}, \citenamefont {Sachrajda},
  \citenamefont {Zawadzki}, \citenamefont {Kam}, \citenamefont {Lapointe},
  \citenamefont {Korkusinski},\ and\ \citenamefont {Hawrylak}}]{Gaudreau2006}%
  \BibitemOpen
  \bibfield  {author} {\bibinfo {author} {\bibfnamefont {L.}~\bibnamefont
  {Gaudreau}}, \bibinfo {author} {\bibfnamefont {S.~A.}\ \bibnamefont
  {Studenikin}}, \bibinfo {author} {\bibfnamefont {A.~S.}\ \bibnamefont
  {Sachrajda}}, \bibinfo {author} {\bibfnamefont {P.}~\bibnamefont {Zawadzki}},
  \bibinfo {author} {\bibfnamefont {A.}~\bibnamefont {Kam}}, \bibinfo {author}
  {\bibfnamefont {J.}~\bibnamefont {Lapointe}}, \bibinfo {author}
  {\bibfnamefont {M.}~\bibnamefont {Korkusinski}}, \ and\ \bibinfo {author}
  {\bibfnamefont {P.}~\bibnamefont {Hawrylak}},\ }\href {\doibase
  10.1103/PhysRevLett.97.036807} {\bibfield  {journal} {\bibinfo  {journal}
  {Phys. Rev. Lett.}\ }\textbf {\bibinfo {volume} {97}},\ \bibinfo {pages}
  {036807} (\bibinfo {year} {2006})}\BibitemShut {NoStop}%
\bibitem [{\citenamefont {Korkusinski}\ \emph {et~al.}(2007)\citenamefont
  {Korkusinski}, \citenamefont {Gimenez}, \citenamefont {Hawrylak},
  \citenamefont {Gaudreau}, \citenamefont {Studenikin},\ and\ \citenamefont
  {Sachrajda}}]{Korkusinski2007}%
  \BibitemOpen
  \bibfield  {author} {\bibinfo {author} {\bibfnamefont {M.}~\bibnamefont
  {Korkusinski}}, \bibinfo {author} {\bibfnamefont {I.~P.}\ \bibnamefont
  {Gimenez}}, \bibinfo {author} {\bibfnamefont {P.}~\bibnamefont {Hawrylak}},
  \bibinfo {author} {\bibfnamefont {L.}~\bibnamefont {Gaudreau}}, \bibinfo
  {author} {\bibfnamefont {S.~A.}\ \bibnamefont {Studenikin}}, \ and\ \bibinfo
  {author} {\bibfnamefont {A.~S.}\ \bibnamefont {Sachrajda}},\ }\href {\doibase
  10.1103/PhysRevB.75.115301} {\bibfield  {journal} {\bibinfo  {journal} {Phys.
  Rev. B}\ }\textbf {\bibinfo {volume} {75}},\ \bibinfo {pages} {115301}
  (\bibinfo {year} {2007})}\BibitemShut {NoStop}%
\bibitem [{\citenamefont {Gaudreau}\ \emph {et~al.}(2007)\citenamefont
  {Gaudreau}, \citenamefont {Sachrajda}, \citenamefont {Studenikin},
  \citenamefont {Zawadzki}, \citenamefont {Kam},\ and\ \citenamefont
  {Lapointe}}]{Gaudreau2007}%
  \BibitemOpen
  \bibfield  {author} {\bibinfo {author} {\bibfnamefont {L.}~\bibnamefont
  {Gaudreau}}, \bibinfo {author} {\bibfnamefont {A.~S.}\ \bibnamefont
  {Sachrajda}}, \bibinfo {author} {\bibfnamefont {S.}~\bibnamefont
  {Studenikin}}, \bibinfo {author} {\bibfnamefont {P.}~\bibnamefont
  {Zawadzki}}, \bibinfo {author} {\bibfnamefont {A.}~\bibnamefont {Kam}}, \
  and\ \bibinfo {author} {\bibfnamefont {J.}~\bibnamefont {Lapointe}},\ }\href
  {\doibase 10.1063/1.2730161} {\bibfield  {journal} {\bibinfo  {journal} {AIP
  Conf. Proc.}\ }\textbf {\bibinfo {volume} {893}},\ \bibinfo {pages} {857}
  (\bibinfo {year} {2007})}\BibitemShut {NoStop}%
\bibitem [{\citenamefont {Gaudreau}\ \emph {et~al.}(2009)\citenamefont
  {Gaudreau}, \citenamefont {Sachrajda}, \citenamefont {Studenikin},
  \citenamefont {Kam}, \citenamefont {Delgado}, \citenamefont {Shim},
  \citenamefont {Korkusinski},\ and\ \citenamefont {Hawrylak}}]{Gaudreau2009}%
  \BibitemOpen
  \bibfield  {author} {\bibinfo {author} {\bibfnamefont {L.}~\bibnamefont
  {Gaudreau}}, \bibinfo {author} {\bibfnamefont {A.~S.}\ \bibnamefont
  {Sachrajda}}, \bibinfo {author} {\bibfnamefont {S.}~\bibnamefont
  {Studenikin}}, \bibinfo {author} {\bibfnamefont {A.}~\bibnamefont {Kam}},
  \bibinfo {author} {\bibfnamefont {F.}~\bibnamefont {Delgado}}, \bibinfo
  {author} {\bibfnamefont {Y.~P.}\ \bibnamefont {Shim}}, \bibinfo {author}
  {\bibfnamefont {M.}~\bibnamefont {Korkusinski}}, \ and\ \bibinfo {author}
  {\bibfnamefont {P.}~\bibnamefont {Hawrylak}},\ }\href {\doibase
  10.1103/PhysRevB.80.075415} {\bibfield  {journal} {\bibinfo  {journal} {Phys.
  Rev. B}\ }\textbf {\bibinfo {volume} {80}},\ \bibinfo {pages} {075415}
  (\bibinfo {year} {2009})}\BibitemShut {NoStop}%
\bibitem [{\citenamefont {van~der Wiel}\ \emph {et~al.}(2002)\citenamefont
  {van~der Wiel}, \citenamefont {De~Franceschi}, \citenamefont {Elzerman},
  \citenamefont {Fujisawa}, \citenamefont {Tarucha},\ and\ \citenamefont
  {Kouwenhoven}}]{vanderWiel2002}%
  \BibitemOpen
  \bibfield  {author} {\bibinfo {author} {\bibfnamefont {W.~G.}\ \bibnamefont
  {van~der Wiel}}, \bibinfo {author} {\bibfnamefont {S.}~\bibnamefont
  {De~Franceschi}}, \bibinfo {author} {\bibfnamefont {J.~M.}\ \bibnamefont
  {Elzerman}}, \bibinfo {author} {\bibfnamefont {T.}~\bibnamefont {Fujisawa}},
  \bibinfo {author} {\bibfnamefont {S.}~\bibnamefont {Tarucha}}, \ and\
  \bibinfo {author} {\bibfnamefont {L.~P.}\ \bibnamefont {Kouwenhoven}},\
  }\href {\doibase 10.1103/RevModPhys.75.1} {\bibfield  {journal} {\bibinfo
  {journal} {Rev. Mod. Phys.}\ }\textbf {\bibinfo {volume} {75}},\ \bibinfo
  {pages} {1} (\bibinfo {year} {2002})}\BibitemShut {NoStop}%
\bibitem [{\citenamefont {Schr\"{o}er}\ \emph {et~al.}(2007)\citenamefont
  {Schr\"{o}er}, \citenamefont {Greentree}, \citenamefont {Gaudreau},
  \citenamefont {Eberl}, \citenamefont {Hollenberg}, \citenamefont {Kotthaus},\
  and\ \citenamefont {Ludwig}}]{Schroer2007}%
  \BibitemOpen
  \bibfield  {author} {\bibinfo {author} {\bibfnamefont {D.}~\bibnamefont
  {Schr\"{o}er}}, \bibinfo {author} {\bibfnamefont {A.~D.}\ \bibnamefont
  {Greentree}}, \bibinfo {author} {\bibfnamefont {L.}~\bibnamefont {Gaudreau}},
  \bibinfo {author} {\bibfnamefont {K.}~\bibnamefont {Eberl}}, \bibinfo
  {author} {\bibfnamefont {L.~C.~L.}\ \bibnamefont {Hollenberg}}, \bibinfo
  {author} {\bibfnamefont {J.~P.}\ \bibnamefont {Kotthaus}}, \ and\ \bibinfo
  {author} {\bibfnamefont {S.}~\bibnamefont {Ludwig}},\ }\href {\doibase
  10.1103/PhysRevB.76.075306} {\bibfield  {journal} {\bibinfo  {journal} {Phys.
  Rev. B}\ }\textbf {\bibinfo {volume} {76}},\ \bibinfo {pages} {075306}
  (\bibinfo {year} {2007})}\BibitemShut {NoStop}%
\bibitem [{\citenamefont {Rogge}\ and\ \citenamefont {Haug}(2008)}]{Rogge2008}%
  \BibitemOpen
  \bibfield  {author} {\bibinfo {author} {\bibfnamefont {M.~C.}\ \bibnamefont
  {Rogge}}\ and\ \bibinfo {author} {\bibfnamefont {R.~J.}\ \bibnamefont
  {Haug}},\ }\href {\doibase 10.1103/PhysRevB.77.193306} {\bibfield  {journal}
  {\bibinfo  {journal} {Phys. Rev. B}\ }\textbf {\bibinfo {volume} {77}},\
  \bibinfo {pages} {193306} (\bibinfo {year} {2008})}\BibitemShut {NoStop}%
\bibitem [{\citenamefont {Amaha}\ \emph {et~al.}(2008)\citenamefont {Amaha},
  \citenamefont {Hatano}, \citenamefont {Kubo}, \citenamefont {Tokura},
  \citenamefont {Austing},\ and\ \citenamefont {Tarucha}}]{Amaha2008}%
  \BibitemOpen
  \bibfield  {author} {\bibinfo {author} {\bibfnamefont {S.}~\bibnamefont
  {Amaha}}, \bibinfo {author} {\bibfnamefont {T.}~\bibnamefont {Hatano}},
  \bibinfo {author} {\bibfnamefont {T.}~\bibnamefont {Kubo}}, \bibinfo {author}
  {\bibfnamefont {Y.}~\bibnamefont {Tokura}}, \bibinfo {author} {\bibfnamefont
  {D.~G.}\ \bibnamefont {Austing}}, \ and\ \bibinfo {author} {\bibfnamefont
  {S.}~\bibnamefont {Tarucha}},\ }\href {\doibase
  https://doi.org/10.1016/j.physe.2007.09.205} {\bibfield  {journal} {\bibinfo
  {journal} {Phys. E}\ }\textbf {\bibinfo {volume} {40}},\ \bibinfo {pages}
  {1322 } (\bibinfo {year} {2008})}\BibitemShut {NoStop}%
\bibitem [{\citenamefont {Amaha}\ \emph {et~al.}(2013)\citenamefont {Amaha},
  \citenamefont {Izumida}, \citenamefont {Hatano}, \citenamefont {Teraoka},
  \citenamefont {Tarucha}, \citenamefont {Gupta},\ and\ \citenamefont
  {Austing}}]{Amaha2013}%
  \BibitemOpen
  \bibfield  {author} {\bibinfo {author} {\bibfnamefont {S.}~\bibnamefont
  {Amaha}}, \bibinfo {author} {\bibfnamefont {W.}~\bibnamefont {Izumida}},
  \bibinfo {author} {\bibfnamefont {T.}~\bibnamefont {Hatano}}, \bibinfo
  {author} {\bibfnamefont {S.}~\bibnamefont {Teraoka}}, \bibinfo {author}
  {\bibfnamefont {S.}~\bibnamefont {Tarucha}}, \bibinfo {author} {\bibfnamefont
  {J.~A.}\ \bibnamefont {Gupta}}, \ and\ \bibinfo {author} {\bibfnamefont
  {D.~G.}\ \bibnamefont {Austing}},\ }\href {\doibase
  10.1103/PhysRevLett.110.016803} {\bibfield  {journal} {\bibinfo  {journal}
  {Phys. Rev. Lett.}\ }\textbf {\bibinfo {volume} {110}},\ \bibinfo {pages}
  {016803} (\bibinfo {year} {2013})}\BibitemShut {NoStop}%
\bibitem [{\citenamefont {Pioro-Ladrière}\ \emph {et~al.}(2008)\citenamefont
  {Pioro-Ladrière}, \citenamefont {Obata}, \citenamefont {Tokura},
  \citenamefont {Shin}, \citenamefont {Kubo}, \citenamefont {Yoshida},
  \citenamefont {Taniyama},\ and\ \citenamefont {Tarucha}}]{PioroLadriere2008}%
  \BibitemOpen
  \bibfield  {author} {\bibinfo {author} {\bibfnamefont {M.}~\bibnamefont
  {Pioro-Ladrière}}, \bibinfo {author} {\bibfnamefont {T.}~\bibnamefont
  {Obata}}, \bibinfo {author} {\bibfnamefont {Y.}~\bibnamefont {Tokura}},
  \bibinfo {author} {\bibfnamefont {Y.-S.}\ \bibnamefont {Shin}}, \bibinfo
  {author} {\bibfnamefont {T.}~\bibnamefont {Kubo}}, \bibinfo {author}
  {\bibfnamefont {K.}~\bibnamefont {Yoshida}}, \bibinfo {author} {\bibfnamefont
  {T.}~\bibnamefont {Taniyama}}, \ and\ \bibinfo {author} {\bibfnamefont
  {S.}~\bibnamefont {Tarucha}},\ }\href {\doibase 10.1038/nphys1053} {\bibfield
   {journal} {\bibinfo  {journal} {Nat. Phys}\ }\textbf {\bibinfo {volume}
  {4}},\ \bibinfo {pages} {776} (\bibinfo {year} {2008})}\BibitemShut {NoStop}%
\bibitem [{\citenamefont {Takakura}\ \emph {et~al.}(2010)\citenamefont
  {Takakura}, \citenamefont {Pioro-Ladrière}, \citenamefont {Obata},
  \citenamefont {Shin}, \citenamefont {Brunner}, \citenamefont {Yoshida},
  \citenamefont {Taniyama},\ and\ \citenamefont {Tarucha}}]{Takakura2010}%
  \BibitemOpen
  \bibfield  {author} {\bibinfo {author} {\bibfnamefont {T.}~\bibnamefont
  {Takakura}}, \bibinfo {author} {\bibfnamefont {M.}~\bibnamefont
  {Pioro-Ladrière}}, \bibinfo {author} {\bibfnamefont {T.}~\bibnamefont
  {Obata}}, \bibinfo {author} {\bibfnamefont {Y.-S.}\ \bibnamefont {Shin}},
  \bibinfo {author} {\bibfnamefont {R.}~\bibnamefont {Brunner}}, \bibinfo
  {author} {\bibfnamefont {K.}~\bibnamefont {Yoshida}}, \bibinfo {author}
  {\bibfnamefont {T.}~\bibnamefont {Taniyama}}, \ and\ \bibinfo {author}
  {\bibfnamefont {S.}~\bibnamefont {Tarucha}},\ }\href {\doibase
  10.1063/1.3518919} {\bibfield  {journal} {\bibinfo  {journal} {Appl. Phys.
  Lett.}\ }\textbf {\bibinfo {volume} {97}},\ \bibinfo {pages} {212104}
  (\bibinfo {year} {2010})}\BibitemShut {NoStop}%
\bibitem [{\citenamefont {Russ}\ and\ \citenamefont
  {Burkard}(2017)}]{Russ2017}%
  \BibitemOpen
  \bibfield  {author} {\bibinfo {author} {\bibfnamefont {M.}~\bibnamefont
  {Russ}}\ and\ \bibinfo {author} {\bibfnamefont {G.}~\bibnamefont {Burkard}},\
  }\href {\doibase 10.1088/1361-648x/aa761f} {\bibfield  {journal} {\bibinfo
  {journal} {J. Phys. Condens. Matter}\ }\textbf {\bibinfo {volume} {29}},\
  \bibinfo {pages} {393001} (\bibinfo {year} {2017})}\BibitemShut {NoStop}%
\bibitem [{\citenamefont {Laird}\ \emph {et~al.}(2010)\citenamefont {Laird},
  \citenamefont {Taylor}, \citenamefont {DiVincenzo}, \citenamefont {Marcus},
  \citenamefont {Hanson},\ and\ \citenamefont {Gossard}}]{Laird2010}%
  \BibitemOpen
  \bibfield  {author} {\bibinfo {author} {\bibfnamefont {E.~A.}\ \bibnamefont
  {Laird}}, \bibinfo {author} {\bibfnamefont {J.~M.}\ \bibnamefont {Taylor}},
  \bibinfo {author} {\bibfnamefont {D.~P.}\ \bibnamefont {DiVincenzo}},
  \bibinfo {author} {\bibfnamefont {C.~M.}\ \bibnamefont {Marcus}}, \bibinfo
  {author} {\bibfnamefont {M.~P.}\ \bibnamefont {Hanson}}, \ and\ \bibinfo
  {author} {\bibfnamefont {A.~C.}\ \bibnamefont {Gossard}},\ }\href {\doibase
  10.1103/PhysRevB.82.075403} {\bibfield  {journal} {\bibinfo  {journal} {Phys.
  Rev. B}\ }\textbf {\bibinfo {volume} {82}},\ \bibinfo {pages} {075403}
  (\bibinfo {year} {2010})}\BibitemShut {NoStop}%
\bibitem [{\citenamefont {S\'{a}nchez}\ \emph {et~al.}(2014)\citenamefont
  {S\'{a}nchez}, \citenamefont {Granger}, \citenamefont {Gaudreau},
  \citenamefont {Kam}, \citenamefont {Pioro-Ladri\`{e}re}, \citenamefont
  {Studenikin}, \citenamefont {Zawadzki}, \citenamefont {Sachrajda},\ and\
  \citenamefont {Platero}}]{Sanchez2014}%
  \BibitemOpen
  \bibfield  {author} {\bibinfo {author} {\bibfnamefont {R.}~\bibnamefont
  {S\'{a}nchez}}, \bibinfo {author} {\bibfnamefont {G.}~\bibnamefont
  {Granger}}, \bibinfo {author} {\bibfnamefont {L.}~\bibnamefont {Gaudreau}},
  \bibinfo {author} {\bibfnamefont {A.}~\bibnamefont {Kam}}, \bibinfo {author}
  {\bibfnamefont {M.}~\bibnamefont {Pioro-Ladri\`{e}re}}, \bibinfo {author}
  {\bibfnamefont {S.~A.}\ \bibnamefont {Studenikin}}, \bibinfo {author}
  {\bibfnamefont {P.}~\bibnamefont {Zawadzki}}, \bibinfo {author}
  {\bibfnamefont {A.~S.}\ \bibnamefont {Sachrajda}}, \ and\ \bibinfo {author}
  {\bibfnamefont {G.}~\bibnamefont {Platero}},\ }\href {\doibase
  10.1103/PhysRevLett.112.176803} {\bibfield  {journal} {\bibinfo  {journal}
  {Phys. Rev. Lett.}\ }\textbf {\bibinfo {volume} {112}},\ \bibinfo {pages}
  {176803} (\bibinfo {year} {2014})}\BibitemShut {NoStop}%
\bibitem [{\citenamefont {\L{}uczak}\ and\ \citenamefont
  {Bu\l{}ka}(2014)}]{Luczak2014}%
  \BibitemOpen
  \bibfield  {author} {\bibinfo {author} {\bibfnamefont {J.}~\bibnamefont
  {\L{}uczak}}\ and\ \bibinfo {author} {\bibfnamefont {B.~R.}\ \bibnamefont
  {Bu\l{}ka}},\ }\href {\doibase 10.1103/PhysRevB.90.165427} {\bibfield
  {journal} {\bibinfo  {journal} {Phys. Rev. B}\ }\textbf {\bibinfo {volume}
  {90}},\ \bibinfo {pages} {165427} (\bibinfo {year} {2014})}\BibitemShut
  {NoStop}%
\bibitem [{\citenamefont {\L{}uczak}\ and\ \citenamefont
  {Bu\l{}ka}(2016)}]{Luczak2016}%
  \BibitemOpen
  \bibfield  {author} {\bibinfo {author} {\bibfnamefont {J.}~\bibnamefont
  {\L{}uczak}}\ and\ \bibinfo {author} {\bibfnamefont {B.~R.}\ \bibnamefont
  {Bu\l{}ka}},\ }\href {\doibase 10.1007/s11128-016-1480-z} {\bibfield
  {journal} {\bibinfo  {journal} {Quantum Inf. Process.}\ }\textbf {\bibinfo
  {volume} {16}},\ \bibinfo {pages} {10} (\bibinfo {year} {2016})}\BibitemShut
  {NoStop}%
\bibitem [{\citenamefont {Mitchell}\ \emph {et~al.}(2009)\citenamefont
  {Mitchell}, \citenamefont {Jarrold},\ and\ \citenamefont
  {Logan}}]{Mitchell2009}%
  \BibitemOpen
  \bibfield  {author} {\bibinfo {author} {\bibfnamefont {A.~K.}\ \bibnamefont
  {Mitchell}}, \bibinfo {author} {\bibfnamefont {T.~F.}\ \bibnamefont
  {Jarrold}}, \ and\ \bibinfo {author} {\bibfnamefont {D.~E.}\ \bibnamefont
  {Logan}},\ }\href {\doibase 10.1103/PhysRevB.79.085124} {\bibfield  {journal}
  {\bibinfo  {journal} {Phys. Rev. B}\ }\textbf {\bibinfo {volume} {79}},\
  \bibinfo {pages} {085124} (\bibinfo {year} {2009})}\BibitemShut {NoStop}%
\bibitem [{\citenamefont {Seo}\ \emph {et~al.}(2013)\citenamefont {Seo},
  \citenamefont {Choi}, \citenamefont {Lee}, \citenamefont {Kim}, \citenamefont
  {Chung}, \citenamefont {Sim}, \citenamefont {Umansky},\ and\ \citenamefont
  {Mahalu}}]{Seo2013}%
  \BibitemOpen
  \bibfield  {author} {\bibinfo {author} {\bibfnamefont {M.}~\bibnamefont
  {Seo}}, \bibinfo {author} {\bibfnamefont {H.~K.}\ \bibnamefont {Choi}},
  \bibinfo {author} {\bibfnamefont {S.-Y.}\ \bibnamefont {Lee}}, \bibinfo
  {author} {\bibfnamefont {N.}~\bibnamefont {Kim}}, \bibinfo {author}
  {\bibfnamefont {Y.}~\bibnamefont {Chung}}, \bibinfo {author} {\bibfnamefont
  {H.-S.}\ \bibnamefont {Sim}}, \bibinfo {author} {\bibfnamefont
  {V.}~\bibnamefont {Umansky}}, \ and\ \bibinfo {author} {\bibfnamefont
  {D.}~\bibnamefont {Mahalu}},\ }\href {\doibase
  10.1103/PhysRevLett.110.046803} {\bibfield  {journal} {\bibinfo  {journal}
  {Phys. Rev. Lett.}\ }\textbf {\bibinfo {volume} {110}},\ \bibinfo {pages}
  {046803} (\bibinfo {year} {2013})}\BibitemShut {NoStop}%
\bibitem [{\citenamefont {Jiang}\ \emph {et~al.}(2005)\citenamefont {Jiang},
  \citenamefont {Sun},\ and\ \citenamefont {Wang}}]{Jiang2005}%
  \BibitemOpen
  \bibfield  {author} {\bibinfo {author} {\bibfnamefont {Z.-t.}\ \bibnamefont
  {Jiang}}, \bibinfo {author} {\bibfnamefont {Q.-f.}\ \bibnamefont {Sun}}, \
  and\ \bibinfo {author} {\bibfnamefont {Y.}~\bibnamefont {Wang}},\ }\href
  {\doibase 10.1103/PhysRevB.72.045332} {\bibfield  {journal} {\bibinfo
  {journal} {Phys. Rev. B}\ }\textbf {\bibinfo {volume} {72}},\ \bibinfo
  {pages} {045332} (\bibinfo {year} {2005})}\BibitemShut {NoStop}%
\bibitem [{\citenamefont {\v{Z}itko}\ \emph {et~al.}(2006)\citenamefont
  {\v{Z}itko}, \citenamefont {Bon\v{c}a}, \citenamefont {Ram\v{s}ak},\ and\
  \citenamefont {Rejec}}]{Zitko2006}%
  \BibitemOpen
  \bibfield  {author} {\bibinfo {author} {\bibfnamefont {R.}~\bibnamefont
  {\v{Z}itko}}, \bibinfo {author} {\bibfnamefont {J.}~\bibnamefont
  {Bon\v{c}a}}, \bibinfo {author} {\bibfnamefont {A.}~\bibnamefont
  {Ram\v{s}ak}}, \ and\ \bibinfo {author} {\bibfnamefont {T.}~\bibnamefont
  {Rejec}},\ }\href {\doibase 10.1103/PhysRevB.73.153307} {\bibfield  {journal}
  {\bibinfo  {journal} {Phys. Rev. B}\ }\textbf {\bibinfo {volume} {73}},\
  \bibinfo {pages} {153307} (\bibinfo {year} {2006})}\BibitemShut {NoStop}%
\bibitem [{\citenamefont {Numata}\ \emph {et~al.}(2009)\citenamefont {Numata},
  \citenamefont {Nisikawa}, \citenamefont {Oguri},\ and\ \citenamefont
  {Hewson}}]{Numata2009}%
  \BibitemOpen
  \bibfield  {author} {\bibinfo {author} {\bibfnamefont {T.}~\bibnamefont
  {Numata}}, \bibinfo {author} {\bibfnamefont {Y.}~\bibnamefont {Nisikawa}},
  \bibinfo {author} {\bibfnamefont {A.}~\bibnamefont {Oguri}}, \ and\ \bibinfo
  {author} {\bibfnamefont {A.~C.}\ \bibnamefont {Hewson}},\ }\href {\doibase
  10.1103/PhysRevB.80.155330} {\bibfield  {journal} {\bibinfo  {journal} {Phys.
  Rev. B}\ }\textbf {\bibinfo {volume} {80}},\ \bibinfo {pages} {155330}
  (\bibinfo {year} {2009})}\BibitemShut {NoStop}%
\bibitem [{\citenamefont {Cheng}\ \emph {et~al.}(2017)\citenamefont {Cheng},
  \citenamefont {Wang}, \citenamefont {Wei}, \citenamefont {Zhu},\ and\
  \citenamefont {Yan}}]{Cheng2017}%
  \BibitemOpen
  \bibfield  {author} {\bibinfo {author} {\bibfnamefont {Y.~X.}\ \bibnamefont
  {Cheng}}, \bibinfo {author} {\bibfnamefont {Y.~D.}\ \bibnamefont {Wang}},
  \bibinfo {author} {\bibfnamefont {J.~H.}\ \bibnamefont {Wei}}, \bibinfo
  {author} {\bibfnamefont {Z.~G.}\ \bibnamefont {Zhu}}, \ and\ \bibinfo
  {author} {\bibfnamefont {Y.~J.}\ \bibnamefont {Yan}},\ }\href {\doibase
  10.1103/PhysRevB.95.155417} {\bibfield  {journal} {\bibinfo  {journal} {Phys.
  Rev. B}\ }\textbf {\bibinfo {volume} {95}},\ \bibinfo {pages} {155417}
  (\bibinfo {year} {2017})}\BibitemShut {NoStop}%
\bibitem [{\citenamefont {L\'{o}pez}\ \emph {et~al.}(2013)\citenamefont
  {L\'{o}pez}, \citenamefont {Rejec}, \citenamefont {Martinek},\ and\
  \citenamefont {\v{Z}itko}}]{Lopez2013}%
  \BibitemOpen
  \bibfield  {author} {\bibinfo {author} {\bibfnamefont {R.}~\bibnamefont
  {L\'{o}pez}}, \bibinfo {author} {\bibfnamefont {T.}~\bibnamefont {Rejec}},
  \bibinfo {author} {\bibfnamefont {J.}~\bibnamefont {Martinek}}, \ and\
  \bibinfo {author} {\bibfnamefont {R.}~\bibnamefont {\v{Z}itko}},\ }\href
  {\doibase 10.1103/PhysRevB.87.035135} {\bibfield  {journal} {\bibinfo
  {journal} {Phys. Rev. B}\ }\textbf {\bibinfo {volume} {87}},\ \bibinfo
  {pages} {035135} (\bibinfo {year} {2013})}\BibitemShut {NoStop}%
\bibitem [{\citenamefont {Vernek}\ \emph {et~al.}(2009)\citenamefont {Vernek},
  \citenamefont {B\"{u}sser}, \citenamefont {Martins}, \citenamefont {Anda},
  \citenamefont {Sandler},\ and\ \citenamefont {Ulloa}}]{Vernek2009}%
  \BibitemOpen
  \bibfield  {author} {\bibinfo {author} {\bibfnamefont {E.}~\bibnamefont
  {Vernek}}, \bibinfo {author} {\bibfnamefont {C.~A.}\ \bibnamefont
  {B\"{u}sser}}, \bibinfo {author} {\bibfnamefont {G.~B.}\ \bibnamefont
  {Martins}}, \bibinfo {author} {\bibfnamefont {E.~V.}\ \bibnamefont {Anda}},
  \bibinfo {author} {\bibfnamefont {N.}~\bibnamefont {Sandler}}, \ and\
  \bibinfo {author} {\bibfnamefont {S.~E.}\ \bibnamefont {Ulloa}},\ }\href
  {\doibase 10.1103/PhysRevB.80.035119} {\bibfield  {journal} {\bibinfo
  {journal} {Phys. Rev. B}\ }\textbf {\bibinfo {volume} {80}},\ \bibinfo
  {pages} {035119} (\bibinfo {year} {2009})}\BibitemShut {NoStop}%
\bibitem [{\citenamefont {Oguri}\ \emph {et~al.}(2011)\citenamefont {Oguri},
  \citenamefont {Amaha}, \citenamefont {Nishikawa}, \citenamefont {Numata},
  \citenamefont {Shimamoto}, \citenamefont {Hewson},\ and\ \citenamefont
  {Tarucha}}]{Oguri2011}%
  \BibitemOpen
  \bibfield  {author} {\bibinfo {author} {\bibfnamefont {A.}~\bibnamefont
  {Oguri}}, \bibinfo {author} {\bibfnamefont {S.}~\bibnamefont {Amaha}},
  \bibinfo {author} {\bibfnamefont {Y.}~\bibnamefont {Nishikawa}}, \bibinfo
  {author} {\bibfnamefont {T.}~\bibnamefont {Numata}}, \bibinfo {author}
  {\bibfnamefont {M.}~\bibnamefont {Shimamoto}}, \bibinfo {author}
  {\bibfnamefont {A.~C.}\ \bibnamefont {Hewson}}, \ and\ \bibinfo {author}
  {\bibfnamefont {S.}~\bibnamefont {Tarucha}},\ }\href {\doibase
  10.1103/PhysRevB.83.205304} {\bibfield  {journal} {\bibinfo  {journal} {Phys.
  Rev. B}\ }\textbf {\bibinfo {volume} {83}},\ \bibinfo {pages} {205304}
  (\bibinfo {year} {2011})}\BibitemShut {NoStop}%
\bibitem [{\citenamefont {Kuzmenko}\ \emph {et~al.}(2006)\citenamefont
  {Kuzmenko}, \citenamefont {Kikoin},\ and\ \citenamefont
  {Avishai}}]{Kuzmenko2006a}%
  \BibitemOpen
  \bibfield  {author} {\bibinfo {author} {\bibfnamefont {T.}~\bibnamefont
  {Kuzmenko}}, \bibinfo {author} {\bibfnamefont {K.}~\bibnamefont {Kikoin}}, \
  and\ \bibinfo {author} {\bibfnamefont {Y.}~\bibnamefont {Avishai}},\ }\href
  {\doibase 10.1103/PhysRevLett.96.046601} {\bibfield  {journal} {\bibinfo
  {journal} {Phys. Rev. Lett.}\ }\textbf {\bibinfo {volume} {96}},\ \bibinfo
  {pages} {046601} (\bibinfo {year} {2006})}\BibitemShut {NoStop}%
\bibitem [{\citenamefont {Saraga}\ and\ \citenamefont
  {Loss}(2003)}]{Saraga2003}%
  \BibitemOpen
  \bibfield  {author} {\bibinfo {author} {\bibfnamefont {D.~S.}\ \bibnamefont
  {Saraga}}\ and\ \bibinfo {author} {\bibfnamefont {D.}~\bibnamefont {Loss}},\
  }\href {\doibase 10.1103/PhysRevLett.90.166803} {\bibfield  {journal}
  {\bibinfo  {journal} {Phys. Rev. Lett.}\ }\textbf {\bibinfo {volume} {90}},\
  \bibinfo {pages} {166803} (\bibinfo {year} {2003})}\BibitemShut {NoStop}%
\bibitem [{\citenamefont {Groth}\ \emph {et~al.}(2006)\citenamefont {Groth},
  \citenamefont {Michaelis},\ and\ \citenamefont {Beenakker}}]{Groth2006}%
  \BibitemOpen
  \bibfield  {author} {\bibinfo {author} {\bibfnamefont {C.~W.}\ \bibnamefont
  {Groth}}, \bibinfo {author} {\bibfnamefont {B.}~\bibnamefont {Michaelis}}, \
  and\ \bibinfo {author} {\bibfnamefont {C.~W.~J.}\ \bibnamefont {Beenakker}},\
  }\href {\doibase 10.1103/PhysRevB.74.125315} {\bibfield  {journal} {\bibinfo
  {journal} {Phys. Rev. B}\ }\textbf {\bibinfo {volume} {74}},\ \bibinfo
  {pages} {125315} (\bibinfo {year} {2006})}\BibitemShut {NoStop}%
\bibitem [{\citenamefont {Michaelis}\ \emph {et~al.}(2006)\citenamefont
  {Michaelis}, \citenamefont {Emary},\ and\ \citenamefont
  {Beenakker}}]{Michaelis2006}%
  \BibitemOpen
  \bibfield  {author} {\bibinfo {author} {\bibfnamefont {B.}~\bibnamefont
  {Michaelis}}, \bibinfo {author} {\bibfnamefont {C.}~\bibnamefont {Emary}}, \
  and\ \bibinfo {author} {\bibfnamefont {C.~W.~J.}\ \bibnamefont {Beenakker}},\
  }\href {\doibase 10.1209/epl/i2005-10458-6} {\bibfield  {journal} {\bibinfo
  {journal} {EPL}\ }\textbf {\bibinfo {volume} {73}},\ \bibinfo {pages} {677}
  (\bibinfo {year} {2006})}\BibitemShut {NoStop}%
\bibitem [{\citenamefont {Emary}(2007)}]{Emary2007b}%
  \BibitemOpen
  \bibfield  {author} {\bibinfo {author} {\bibfnamefont {C.}~\bibnamefont
  {Emary}},\ }\href {\doibase 10.1103/PhysRevB.76.245319} {\bibfield  {journal}
  {\bibinfo  {journal} {Phys. Rev. B}\ }\textbf {\bibinfo {volume} {76}},\
  \bibinfo {pages} {245319} (\bibinfo {year} {2007})}\BibitemShut {NoStop}%
\bibitem [{\citenamefont {P\"{o}ltl}\ \emph {et~al.}(2009)\citenamefont
  {P\"{o}ltl}, \citenamefont {Emary},\ and\ \citenamefont
  {Brandes}}]{Poltl2009}%
  \BibitemOpen
  \bibfield  {author} {\bibinfo {author} {\bibfnamefont {C.}~\bibnamefont
  {P\"{o}ltl}}, \bibinfo {author} {\bibfnamefont {C.}~\bibnamefont {Emary}}, \
  and\ \bibinfo {author} {\bibfnamefont {T.}~\bibnamefont {Brandes}},\ }\href
  {\doibase 10.1103/PhysRevB.80.115313} {\bibfield  {journal} {\bibinfo
  {journal} {Phys. Rev. B}\ }\textbf {\bibinfo {volume} {80}},\ \bibinfo
  {pages} {115313} (\bibinfo {year} {2009})}\BibitemShut {NoStop}%
\bibitem [{\citenamefont {P\"oltl}\ \emph {et~al.}(2013)\citenamefont
  {P\"oltl}, \citenamefont {Emary},\ and\ \citenamefont {Brandes}}]{Poltl2013}%
  \BibitemOpen
  \bibfield  {author} {\bibinfo {author} {\bibfnamefont {C.}~\bibnamefont
  {P\"oltl}}, \bibinfo {author} {\bibfnamefont {C.}~\bibnamefont {Emary}}, \
  and\ \bibinfo {author} {\bibfnamefont {T.}~\bibnamefont {Brandes}},\ }\href
  {\doibase 10.1103/PhysRevB.87.045416} {\bibfield  {journal} {\bibinfo
  {journal} {Phys. Rev. B}\ }\textbf {\bibinfo {volume} {87}},\ \bibinfo
  {pages} {045416} (\bibinfo {year} {2013})}\BibitemShut {NoStop}%
\bibitem [{\citenamefont {Busl}\ \emph {et~al.}(2010)\citenamefont {Busl},
  \citenamefont {S\'{a}nchez},\ and\ \citenamefont {Platero}}]{Busl2010}%
  \BibitemOpen
  \bibfield  {author} {\bibinfo {author} {\bibfnamefont {M.}~\bibnamefont
  {Busl}}, \bibinfo {author} {\bibfnamefont {R.}~\bibnamefont {S\'{a}nchez}}, \
  and\ \bibinfo {author} {\bibfnamefont {G.}~\bibnamefont {Platero}},\ }\href
  {\doibase 10.1103/PhysRevB.81.121306} {\bibfield  {journal} {\bibinfo
  {journal} {Phys. Rev. B}\ }\textbf {\bibinfo {volume} {81}},\ \bibinfo
  {pages} {121306(R)} (\bibinfo {year} {2010})}\BibitemShut {NoStop}%
\bibitem [{\citenamefont {Kostyrko}\ and\ \citenamefont
  {Bu\l{}ka}(2009)}]{Kostyrko2009}%
  \BibitemOpen
  \bibfield  {author} {\bibinfo {author} {\bibfnamefont {T.}~\bibnamefont
  {Kostyrko}}\ and\ \bibinfo {author} {\bibfnamefont {B.~R.}\ \bibnamefont
  {Bu\l{}ka}},\ }\href {\doibase 10.1103/PhysRevB.79.075310} {\bibfield
  {journal} {\bibinfo  {journal} {Phys. Rev. B}\ }\textbf {\bibinfo {volume}
  {79}},\ \bibinfo {pages} {075310} (\bibinfo {year} {2009})}\BibitemShut
  {NoStop}%
\bibitem [{\citenamefont {Weymann}\ \emph {et~al.}(2011)\citenamefont
  {Weymann}, \citenamefont {Bu\l{}ka},\ and\ \citenamefont
  {Barna\'{s}}}]{Weymann2011}%
  \BibitemOpen
  \bibfield  {author} {\bibinfo {author} {\bibfnamefont {I.}~\bibnamefont
  {Weymann}}, \bibinfo {author} {\bibfnamefont {B.~R.}\ \bibnamefont
  {Bu\l{}ka}}, \ and\ \bibinfo {author} {\bibfnamefont {J.}~\bibnamefont
  {Barna\'{s}}},\ }\href {\doibase 10.1103/PhysRevB.83.195302} {\bibfield
  {journal} {\bibinfo  {journal} {Phys. Rev. B}\ }\textbf {\bibinfo {volume}
  {83}},\ \bibinfo {pages} {195302} (\bibinfo {year} {2011})}\BibitemShut
  {NoStop}%
\bibitem [{\citenamefont {Noiri}\ \emph {et~al.}(2017)\citenamefont {Noiri},
  \citenamefont {Takakura}, \citenamefont {Obata}, \citenamefont {Otsuka},
  \citenamefont {Nakajima}, \citenamefont {Yoneda},\ and\ \citenamefont
  {Tarucha}}]{Noiri2017}%
  \BibitemOpen
  \bibfield  {author} {\bibinfo {author} {\bibfnamefont {A.}~\bibnamefont
  {Noiri}}, \bibinfo {author} {\bibfnamefont {T.}~\bibnamefont {Takakura}},
  \bibinfo {author} {\bibfnamefont {T.}~\bibnamefont {Obata}}, \bibinfo
  {author} {\bibfnamefont {T.}~\bibnamefont {Otsuka}}, \bibinfo {author}
  {\bibfnamefont {T.}~\bibnamefont {Nakajima}}, \bibinfo {author}
  {\bibfnamefont {J.}~\bibnamefont {Yoneda}}, \ and\ \bibinfo {author}
  {\bibfnamefont {S.}~\bibnamefont {Tarucha}},\ }\href {\doibase
  10.1103/PhysRevB.96.155414} {\bibfield  {journal} {\bibinfo  {journal} {Phys.
  Rev. B}\ }\textbf {\bibinfo {volume} {96}},\ \bibinfo {pages} {155414}
  (\bibinfo {year} {2017})}\BibitemShut {NoStop}%
\bibitem [{\citenamefont {Niklas}\ \emph {et~al.}(2017)\citenamefont {Niklas},
  \citenamefont {Trottmann}, \citenamefont {Donarini},\ and\ \citenamefont
  {Grifoni}}]{Niklas2017}%
  \BibitemOpen
  \bibfield  {author} {\bibinfo {author} {\bibfnamefont {M.}~\bibnamefont
  {Niklas}}, \bibinfo {author} {\bibfnamefont {A.}~\bibnamefont {Trottmann}},
  \bibinfo {author} {\bibfnamefont {A.}~\bibnamefont {Donarini}}, \ and\
  \bibinfo {author} {\bibfnamefont {M.}~\bibnamefont {Grifoni}},\ }\href
  {\doibase 10.1103/PhysRevB.95.115133} {\bibfield  {journal} {\bibinfo
  {journal} {Phys. Rev. B}\ }\textbf {\bibinfo {volume} {95}},\ \bibinfo
  {pages} {115133} (\bibinfo {year} {2017})}\BibitemShut {NoStop}%
\bibitem [{\citenamefont {Bu\l{}ka}\ \emph {et~al.}(2011)\citenamefont
  {Bu\l{}ka}, \citenamefont {Kostyrko},\ and\ \citenamefont
  {\L{}uczak}}]{Bulka2011}%
  \BibitemOpen
  \bibfield  {author} {\bibinfo {author} {\bibfnamefont {B.~R.}\ \bibnamefont
  {Bu\l{}ka}}, \bibinfo {author} {\bibfnamefont {T.}~\bibnamefont {Kostyrko}},
  \ and\ \bibinfo {author} {\bibfnamefont {J.}~\bibnamefont {\L{}uczak}},\
  }\href {\doibase 10.1103/PhysRevB.83.035301} {\bibfield  {journal} {\bibinfo
  {journal} {Phys. Rev. B}\ }\textbf {\bibinfo {volume} {83}},\ \bibinfo
  {pages} {035301} (\bibinfo {year} {2011})}\BibitemShut {NoStop}%
\bibitem [{\citenamefont {Lai}\ \emph {et~al.}(2018)\citenamefont {Lai},
  \citenamefont {Ventra}, \citenamefont {Scheibner},\ and\ \citenamefont
  {Chien}}]{Lai2018}%
  \BibitemOpen
  \bibfield  {author} {\bibinfo {author} {\bibfnamefont {C.-Y.}\ \bibnamefont
  {Lai}}, \bibinfo {author} {\bibfnamefont {M.~D.}\ \bibnamefont {Ventra}},
  \bibinfo {author} {\bibfnamefont {M.}~\bibnamefont {Scheibner}}, \ and\
  \bibinfo {author} {\bibfnamefont {C.-C.}\ \bibnamefont {Chien}},\ }\href
  {\doibase 10.1209/0295-5075/123/47002} {\bibfield  {journal} {\bibinfo
  {journal} {{EPL}}\ }\textbf {\bibinfo {volume} {123}},\ \bibinfo {pages}
  {47002} (\bibinfo {year} {2018})}\BibitemShut {NoStop}%
\bibitem [{\citenamefont {Brandes}(2008)}]{Brandes2008}%
  \BibitemOpen
  \bibfield  {author} {\bibinfo {author} {\bibfnamefont {T.}~\bibnamefont
  {Brandes}},\ }\href {\doibase 10.1002/andp.200810306} {\bibfield  {journal}
  {\bibinfo  {journal} {Ann. Phys. (Berlin)}\ }\textbf {\bibinfo {volume}
  {17}},\ \bibinfo {pages} {477} (\bibinfo {year} {2008})}\BibitemShut
  {NoStop}%
\bibitem [{\citenamefont {Srinivas}\ and\ \citenamefont
  {Davies}(2010)}]{Srinivas2010}%
  \BibitemOpen
  \bibfield  {author} {\bibinfo {author} {\bibfnamefont {M.}~\bibnamefont
  {Srinivas}}\ and\ \bibinfo {author} {\bibfnamefont {E.}~\bibnamefont
  {Davies}},\ }\href {http://www.tandfonline.com/doi/abs/10.1080/713820643}
  {\bibfield  {journal} {\bibinfo  {journal} {Opt. Acta}\ }\textbf {\bibinfo
  {volume} {28}},\ \bibinfo {pages} {981} (\bibinfo {year} {2010})}\BibitemShut
  {NoStop}%
\bibitem [{\citenamefont {Thomas}\ and\ \citenamefont
  {Flindt}(2013)}]{Thomas2013}%
  \BibitemOpen
  \bibfield  {author} {\bibinfo {author} {\bibfnamefont {K.~H.}\ \bibnamefont
  {Thomas}}\ and\ \bibinfo {author} {\bibfnamefont {C.}~\bibnamefont
  {Flindt}},\ }\href {\doibase 10.1103/PhysRevB.87.121405} {\bibfield
  {journal} {\bibinfo  {journal} {Phys. Rev. B}\ }\textbf {\bibinfo {volume}
  {87}},\ \bibinfo {pages} {121405(R)} (\bibinfo {year} {2013})}\BibitemShut
  {NoStop}%
\bibitem [{\citenamefont {Kosov}(2017)}]{Kosov2017b}%
  \BibitemOpen
  \bibfield  {author} {\bibinfo {author} {\bibfnamefont {D.~S.}\ \bibnamefont
  {Kosov}},\ }\href {\doibase 10.1063/1.4991038} {\bibfield  {journal}
  {\bibinfo  {journal} {J. Chem. Phys.}\ }\textbf {\bibinfo {volume} {147}},\
  \bibinfo {pages} {104109} (\bibinfo {year} {2017})}\BibitemShut {NoStop}%
\bibitem [{\citenamefont {Dasenbrook}\ \emph {et~al.}(2015)\citenamefont
  {Dasenbrook}, \citenamefont {Hofer},\ and\ \citenamefont
  {Flindt}}]{Dasenbrook2015}%
  \BibitemOpen
  \bibfield  {author} {\bibinfo {author} {\bibfnamefont {D.}~\bibnamefont
  {Dasenbrook}}, \bibinfo {author} {\bibfnamefont {P.~P.}\ \bibnamefont
  {Hofer}}, \ and\ \bibinfo {author} {\bibfnamefont {C.}~\bibnamefont
  {Flindt}},\ }\href {\doibase 10.1103/PhysRevB.91.195420} {\bibfield
  {journal} {\bibinfo  {journal} {Phys. Rev. B}\ }\textbf {\bibinfo {volume}
  {91}},\ \bibinfo {pages} {195420} (\bibinfo {year} {2015})}\BibitemShut
  {NoStop}%
\bibitem [{\citenamefont {Albert}\ \emph {et~al.}(2011)\citenamefont {Albert},
  \citenamefont {Flindt},\ and\ \citenamefont {B\"{u}ttiker}}]{Albert2011}%
  \BibitemOpen
  \bibfield  {author} {\bibinfo {author} {\bibfnamefont {M.}~\bibnamefont
  {Albert}}, \bibinfo {author} {\bibfnamefont {C.}~\bibnamefont {Flindt}}, \
  and\ \bibinfo {author} {\bibfnamefont {M.}~\bibnamefont {B\"{u}ttiker}},\
  }\href {\doibase 10.1103/PhysRevLett.107.086805} {\bibfield  {journal}
  {\bibinfo  {journal} {Phys. Rev. Lett.}\ }\textbf {\bibinfo {volume} {107}},\
  \bibinfo {pages} {086805} (\bibinfo {year} {2011})}\BibitemShut {NoStop}%
\bibitem [{\citenamefont {Albert}\ \emph {et~al.}(2012)\citenamefont {Albert},
  \citenamefont {Haack}, \citenamefont {Flindt},\ and\ \citenamefont
  {B\"uttiker}}]{Albert2012}%
  \BibitemOpen
  \bibfield  {author} {\bibinfo {author} {\bibfnamefont {M.}~\bibnamefont
  {Albert}}, \bibinfo {author} {\bibfnamefont {G.}~\bibnamefont {Haack}},
  \bibinfo {author} {\bibfnamefont {C.}~\bibnamefont {Flindt}}, \ and\ \bibinfo
  {author} {\bibfnamefont {M.}~\bibnamefont {B\"uttiker}},\ }\href {\doibase
  10.1103/PhysRevLett.108.186806} {\bibfield  {journal} {\bibinfo  {journal}
  {Phys. Rev. Lett.}\ }\textbf {\bibinfo {volume} {108}},\ \bibinfo {pages}
  {186806} (\bibinfo {year} {2012})}\BibitemShut {NoStop}%
\bibitem [{\citenamefont {Albert}\ \emph {et~al.}(2016)\citenamefont {Albert},
  \citenamefont {Chevallier},\ and\ \citenamefont {Devillard}}]{Albert2016}%
  \BibitemOpen
  \bibfield  {author} {\bibinfo {author} {\bibfnamefont {M.}~\bibnamefont
  {Albert}}, \bibinfo {author} {\bibfnamefont {D.}~\bibnamefont {Chevallier}},
  \ and\ \bibinfo {author} {\bibfnamefont {P.}~\bibnamefont {Devillard}},\
  }\href {\doibase 10.1016/j.physe.2016.02.017} {\bibfield  {journal} {\bibinfo
   {journal} {Phys. E}\ }\textbf {\bibinfo {volume} {76}},\ \bibinfo {pages}
  {209 } (\bibinfo {year} {2016})}\BibitemShut {NoStop}%
\bibitem [{\citenamefont {Albert}\ and\ \citenamefont
  {Devillard}(2014)}]{Albert2014}%
  \BibitemOpen
  \bibfield  {author} {\bibinfo {author} {\bibfnamefont {M.}~\bibnamefont
  {Albert}}\ and\ \bibinfo {author} {\bibfnamefont {P.}~\bibnamefont
  {Devillard}},\ }\href {\doibase 10.1103/PhysRevB.90.035431} {\bibfield
  {journal} {\bibinfo  {journal} {Phys. Rev. B}\ }\textbf {\bibinfo {volume}
  {90}},\ \bibinfo {pages} {035431} (\bibinfo {year} {2014})}\BibitemShut
  {NoStop}%
\bibitem [{\citenamefont {Chevallier}\ \emph {et~al.}(2016)\citenamefont
  {Chevallier}, \citenamefont {Albert},\ and\ \citenamefont
  {Devillard}}]{Chevallier2016}%
  \BibitemOpen
  \bibfield  {author} {\bibinfo {author} {\bibfnamefont {D.}~\bibnamefont
  {Chevallier}}, \bibinfo {author} {\bibfnamefont {M.}~\bibnamefont {Albert}},
  \ and\ \bibinfo {author} {\bibfnamefont {P.}~\bibnamefont {Devillard}},\
  }\href {\doibase 10.1209/0295-5075/116/27005} {\bibfield  {journal} {\bibinfo
   {journal} {EPL}\ }\textbf {\bibinfo {volume} {116}},\ \bibinfo {pages}
  {27005} (\bibinfo {year} {2016})}\BibitemShut {NoStop}%
\bibitem [{\citenamefont {Haack}\ \emph {et~al.}(2014)\citenamefont {Haack},
  \citenamefont {Albert},\ and\ \citenamefont {Flindt}}]{Haack2014}%
  \BibitemOpen
  \bibfield  {author} {\bibinfo {author} {\bibfnamefont {G.}~\bibnamefont
  {Haack}}, \bibinfo {author} {\bibfnamefont {M.}~\bibnamefont {Albert}}, \
  and\ \bibinfo {author} {\bibfnamefont {C.}~\bibnamefont {Flindt}},\ }\href
  {\doibase 10.1103/PhysRevB.90.205429} {\bibfield  {journal} {\bibinfo
  {journal} {Phys. Rev. B}\ }\textbf {\bibinfo {volume} {90}},\ \bibinfo
  {pages} {205429} (\bibinfo {year} {2014})}\BibitemShut {NoStop}%
\bibitem [{\citenamefont
  {Ptaszy\'{n}ski}(2017{\natexlab{a}})}]{Ptaszynski2017a}%
  \BibitemOpen
  \bibfield  {author} {\bibinfo {author} {\bibfnamefont {K.}~\bibnamefont
  {Ptaszy\'{n}ski}},\ }\href {\doibase 10.1103/PhysRevB.96.035409} {\bibfield
  {journal} {\bibinfo  {journal} {Phys. Rev. B}\ }\textbf {\bibinfo {volume}
  {96}},\ \bibinfo {pages} {035409} (\bibinfo {year}
  {2017}{\natexlab{a}})}\BibitemShut {NoStop}%
\bibitem [{\citenamefont
  {Ptaszy\'{n}ski}(2017{\natexlab{b}})}]{Ptaszynski2017}%
  \BibitemOpen
  \bibfield  {author} {\bibinfo {author} {\bibfnamefont {K.}~\bibnamefont
  {Ptaszy\'{n}ski}},\ }\href {\doibase 10.1103/PhysRevB.95.045306} {\bibfield
  {journal} {\bibinfo  {journal} {Phys. Rev. B}\ }\textbf {\bibinfo {volume}
  {95}},\ \bibinfo {pages} {045306} (\bibinfo {year}
  {2017}{\natexlab{b}})}\BibitemShut {NoStop}%
\bibitem [{\citenamefont {Ptaszy\'{n}ski}(2018)}]{Ptaszynski2018}%
  \BibitemOpen
  \bibfield  {author} {\bibinfo {author} {\bibfnamefont {K.}~\bibnamefont
  {Ptaszy\'{n}ski}},\ }\href {\doibase 10.1103/PhysRevE.97.012127} {\bibfield
  {journal} {\bibinfo  {journal} {Phys. Rev. E}\ }\textbf {\bibinfo {volume}
  {97}},\ \bibinfo {pages} {012127} (\bibinfo {year} {2018})}\BibitemShut
  {NoStop}%
\bibitem [{\citenamefont {Stegmann}\ \emph {et~al.}(2020)\citenamefont
  {Stegmann}, \citenamefont {Sothmann}, \citenamefont {K\"{o}nig},\ and\
  \citenamefont {Flindt}}]{Stegmann2020}%
  \BibitemOpen
  \bibfield  {author} {\bibinfo {author} {\bibfnamefont {P.}~\bibnamefont
  {Stegmann}}, \bibinfo {author} {\bibfnamefont {B.}~\bibnamefont {Sothmann}},
  \bibinfo {author} {\bibfnamefont {J.}~\bibnamefont {K\"{o}nig}}, \ and\
  \bibinfo {author} {\bibfnamefont {C.}~\bibnamefont {Flindt}},\ }\href
  {https://arxiv.org/abs/2004.12603} {\bibfield  {journal} {\bibinfo  {journal}
  {arXiv.org}\ } (\bibinfo {year} {2020})}\BibitemShut {NoStop}%
\bibitem [{\citenamefont {Welack}\ \emph {et~al.}(2009)\citenamefont {Welack},
  \citenamefont {Mukamel},\ and\ \citenamefont {Yan}}]{Welack2009}%
  \BibitemOpen
  \bibfield  {author} {\bibinfo {author} {\bibfnamefont {S.}~\bibnamefont
  {Welack}}, \bibinfo {author} {\bibfnamefont {S.}~\bibnamefont {Mukamel}}, \
  and\ \bibinfo {author} {\bibfnamefont {Y.~J.}\ \bibnamefont {Yan}},\ }\href
  {\doibase 10.1209/0295-5075/85/57008} {\bibfield  {journal} {\bibinfo
  {journal} {EPL}\ }\textbf {\bibinfo {volume} {85}},\ \bibinfo {pages} {57008}
  (\bibinfo {year} {2009})}\BibitemShut {NoStop}%
\bibitem [{\citenamefont {Davis}\ \emph {et~al.}(2021)\citenamefont {Davis},
  \citenamefont {Rudge},\ and\ \citenamefont {Kosov}}]{Davis2021}%
  \BibitemOpen
  \bibfield  {author} {\bibinfo {author} {\bibfnamefont {N.~S.}\ \bibnamefont
  {Davis}}, \bibinfo {author} {\bibfnamefont {S.~L.}\ \bibnamefont {Rudge}}, \
  and\ \bibinfo {author} {\bibfnamefont {D.~S.}\ \bibnamefont {Kosov}},\ }\href
  {\doibase 10.1103/PhysRevB.103.205408} {\bibfield  {journal} {\bibinfo
  {journal} {Phys. Rev. B}\ }\textbf {\bibinfo {volume} {103}},\ \bibinfo
  {pages} {205408} (\bibinfo {year} {2021})}\BibitemShut {NoStop}%
\bibitem [{\citenamefont {Stegmann}\ \emph {et~al.}(2018)\citenamefont
  {Stegmann}, \citenamefont {K\"onig},\ and\ \citenamefont
  {Weiss}}]{Stegmann2018}%
  \BibitemOpen
  \bibfield  {author} {\bibinfo {author} {\bibfnamefont {P.}~\bibnamefont
  {Stegmann}}, \bibinfo {author} {\bibfnamefont {J.}~\bibnamefont {K\"onig}}, \
  and\ \bibinfo {author} {\bibfnamefont {S.}~\bibnamefont {Weiss}},\ }\href
  {\doibase 10.1103/PhysRevB.98.035409} {\bibfield  {journal} {\bibinfo
  {journal} {Phys. Rev. B}\ }\textbf {\bibinfo {volume} {98}},\ \bibinfo
  {pages} {035409} (\bibinfo {year} {2018})}\BibitemShut {NoStop}%
\bibitem [{\citenamefont {Li}\ \emph {et~al.}(2005{\natexlab{a}})\citenamefont
  {Li}, \citenamefont {Cui},\ and\ \citenamefont {Yan}}]{Li2005a}%
  \BibitemOpen
  \bibfield  {author} {\bibinfo {author} {\bibfnamefont {X.~Q.}\ \bibnamefont
  {Li}}, \bibinfo {author} {\bibfnamefont {P.}~\bibnamefont {Cui}}, \ and\
  \bibinfo {author} {\bibfnamefont {Y.~J.}\ \bibnamefont {Yan}},\ }\href
  {\doibase 10.1103/PhysRevLett.94.066803} {\bibfield  {journal} {\bibinfo
  {journal} {Phys. Rev. Lett.}\ }\textbf {\bibinfo {volume} {94}},\ \bibinfo
  {pages} {066803} (\bibinfo {year} {2005}{\natexlab{a}})}\BibitemShut
  {NoStop}%
\bibitem [{\citenamefont {Li}\ \emph {et~al.}(2005{\natexlab{b}})\citenamefont
  {Li}, \citenamefont {Luo}, \citenamefont {Yang}, \citenamefont {Cui},\ and\
  \citenamefont {Yan}}]{Li2005b}%
  \BibitemOpen
  \bibfield  {author} {\bibinfo {author} {\bibfnamefont {X.~Q.}\ \bibnamefont
  {Li}}, \bibinfo {author} {\bibfnamefont {J.}~\bibnamefont {Luo}}, \bibinfo
  {author} {\bibfnamefont {Y.~G.}\ \bibnamefont {Yang}}, \bibinfo {author}
  {\bibfnamefont {P.}~\bibnamefont {Cui}}, \ and\ \bibinfo {author}
  {\bibfnamefont {Y.~J.}\ \bibnamefont {Yan}},\ }\href {\doibase
  10.1103/PhysRevB.71.205304} {\bibfield  {journal} {\bibinfo  {journal} {Phys.
  Rev. B}\ }\textbf {\bibinfo {volume} {71}},\ \bibinfo {pages} {205304}
  (\bibinfo {year} {2005}{\natexlab{b}})}\BibitemShut {NoStop}%
\bibitem [{\citenamefont {Breuer}\ \emph {et~al.}(2002)\citenamefont {Breuer},
  \citenamefont {Petruccione},\ and\ \citenamefont {Petruccione}}]{Breuer2002}%
  \BibitemOpen
  \bibfield  {author} {\bibinfo {author} {\bibfnamefont {H.}~\bibnamefont
  {Breuer}}, \bibinfo {author} {\bibfnamefont {F.}~\bibnamefont {Petruccione}},
  \ and\ \bibinfo {author} {\bibfnamefont {S.}~\bibnamefont {Petruccione}},\
  }\href {https://books.google.com.au/books?id=0Yx5VzaMYm8C} {\emph {\bibinfo
  {title} {\href{https://books.google.com.au/books?id=0Yx5VzaMYm8C}{The Theory
  of Open Quantum Systems}}}}\ (\bibinfo  {publisher} {Oxford University Press,
  London},\ \bibinfo {year} {2002})\BibitemShut {NoStop}%
\bibitem [{\citenamefont {Rudge}\ and\ \citenamefont
  {Kosov}(2018)}]{Rudge2018}%
  \BibitemOpen
  \bibfield  {author} {\bibinfo {author} {\bibfnamefont {S.~L.}\ \bibnamefont
  {Rudge}}\ and\ \bibinfo {author} {\bibfnamefont {D.~S.}\ \bibnamefont
  {Kosov}},\ }\href {\doibase 10.1103/PhysRevB.98.245402} {\bibfield  {journal}
  {\bibinfo  {journal} {Phys. Rev. B}\ }\textbf {\bibinfo {volume} {98}},\
  \bibinfo {pages} {245402} (\bibinfo {year} {2018})}\BibitemShut {NoStop}%
\bibitem [{\citenamefont {Tang}\ \emph {et~al.}(2014)\citenamefont {Tang},
  \citenamefont {Xu},\ and\ \citenamefont {Wang}}]{Tang2014}%
  \BibitemOpen
  \bibfield  {author} {\bibinfo {author} {\bibfnamefont {G.~M.}\ \bibnamefont
  {Tang}}, \bibinfo {author} {\bibfnamefont {F.}~\bibnamefont {Xu}}, \ and\
  \bibinfo {author} {\bibfnamefont {J.}~\bibnamefont {Wang}},\ }\href {\doibase
  10.1103/PhysRevB.89.205310} {\bibfield  {journal} {\bibinfo  {journal} {Phys.
  Rev. B}\ }\textbf {\bibinfo {volume} {89}},\ \bibinfo {pages} {205310}
  (\bibinfo {year} {2014})}\BibitemShut {NoStop}%
\bibitem [{\citenamefont {Bagrets}\ and\ \citenamefont
  {Nazarov}(2003)}]{Bagrets2003}%
  \BibitemOpen
  \bibfield  {author} {\bibinfo {author} {\bibfnamefont {D.~A.}\ \bibnamefont
  {Bagrets}}\ and\ \bibinfo {author} {\bibfnamefont {Y.~V.}\ \bibnamefont
  {Nazarov}},\ }\href {\doibase 10.1103/PhysRevB.67.085316} {\bibfield
  {journal} {\bibinfo  {journal} {Phys. Rev. B}\ }\textbf {\bibinfo {volume}
  {67}},\ \bibinfo {pages} {085316} (\bibinfo {year} {2003})}\BibitemShut
  {NoStop}%
\bibitem [{\citenamefont {Nazarov}(1999)}]{Nazarov1999}%
  \BibitemOpen
  \bibfield  {author} {\bibinfo {author} {\bibfnamefont {Y.~V.}\ \bibnamefont
  {Nazarov}},\ }\href {https://arxiv.org/pdf/cond-mat/9908143.pdf} {\bibfield
  {journal} {\bibinfo  {journal} {arXiv.org}\ } (\bibinfo {year}
  {1999})}\BibitemShut {NoStop}%
\bibitem [{\citenamefont {Bruus}\ and\ \citenamefont
  {Flensberg}(2002)}]{Bruus2002}%
  \BibitemOpen
  \bibfield  {author} {\bibinfo {author} {\bibfnamefont {H.}~\bibnamefont
  {Bruus}}\ and\ \bibinfo {author} {\bibfnamefont {K.}~\bibnamefont
  {Flensberg}},\ }\href {\doibase 10.1088/0305-4470/38/8/B01} {\emph {\bibinfo
  {title} {\href{https://books.google.com.au/books?id=v5vhg1tYLC8C}{Many-Body
  Quantum Theory in Condensed Matter Physics: An Introduction}}}}\ (\bibinfo
  {publisher} {Oxford University Press, Oxford},\ \bibinfo {year}
  {2002})\BibitemShut {NoStop}%
\bibitem [{\citenamefont {Timm}(2008)}]{Timm2008}%
  \BibitemOpen
  \bibfield  {author} {\bibinfo {author} {\bibfnamefont {C.}~\bibnamefont
  {Timm}},\ }\href {\doibase 10.1103/PhysRevB.77.195416} {\bibfield  {journal}
  {\bibinfo  {journal} {Phys. Rev. B}\ }\textbf {\bibinfo {volume} {77}},\
  \bibinfo {pages} {195416} (\bibinfo {year} {2008})}\BibitemShut {NoStop}%
\bibitem [{\citenamefont {Aharonov}\ and\ \citenamefont
  {Bohm}(1959)}]{Aharonov1959}%
  \BibitemOpen
  \bibfield  {author} {\bibinfo {author} {\bibfnamefont {Y.}~\bibnamefont
  {Aharonov}}\ and\ \bibinfo {author} {\bibfnamefont {D.}~\bibnamefont
  {Bohm}},\ }\href {\doibase 10.1103/PhysRev.115.485} {\bibfield  {journal}
  {\bibinfo  {journal} {Phys. Rev.}\ }\textbf {\bibinfo {volume} {115}},\
  \bibinfo {pages} {485} (\bibinfo {year} {1959})}\BibitemShut {NoStop}%
\bibitem [{\citenamefont {Budini}(2011)}]{Budini2011}%
  \BibitemOpen
  \bibfield  {author} {\bibinfo {author} {\bibfnamefont {A.~A.}\ \bibnamefont
  {Budini}},\ }\href {\doibase 10.1103/PhysRevE.84.011141} {\bibfield
  {journal} {\bibinfo  {journal} {Phys. Rev. E}\ }\textbf {\bibinfo {volume}
  {84}},\ \bibinfo {pages} {011141} (\bibinfo {year} {2011})}\BibitemShut
  {NoStop}%
\bibitem [{\citenamefont {Rudge}\ and\ \citenamefont
  {Kosov}(2016{\natexlab{a}})}]{Rudge2016a}%
  \BibitemOpen
  \bibfield  {author} {\bibinfo {author} {\bibfnamefont {S.~L.}\ \bibnamefont
  {Rudge}}\ and\ \bibinfo {author} {\bibfnamefont {D.~S.}\ \bibnamefont
  {Kosov}},\ }\href {\doibase 10.1063/1.4944493} {\bibfield  {journal}
  {\bibinfo  {journal} {J. Chem. Phys.}\ }\textbf {\bibinfo {volume} {144}},\
  \bibinfo {pages} {124105} (\bibinfo {year} {2016}{\natexlab{a}})}\BibitemShut
  {NoStop}%
\bibitem [{\citenamefont {Rudge}\ and\ \citenamefont
  {Kosov}(2016{\natexlab{b}})}]{Rudge2016b}%
  \BibitemOpen
  \bibfield  {author} {\bibinfo {author} {\bibfnamefont {S.~L.}\ \bibnamefont
  {Rudge}}\ and\ \bibinfo {author} {\bibfnamefont {D.~S.}\ \bibnamefont
  {Kosov}},\ }\href {\doibase 10.1103/PhysRevE.94.042134} {\bibfield  {journal}
  {\bibinfo  {journal} {Phys. Rev. E}\ }\textbf {\bibinfo {volume} {94}},\
  \bibinfo {pages} {042134} (\bibinfo {year} {2016}{\natexlab{b}})}\BibitemShut
  {NoStop}%
\bibitem [{\citenamefont {Rudge}\ and\ \citenamefont
  {Kosov}(2019{\natexlab{a}})}]{Rudge2019a}%
  \BibitemOpen
  \bibfield  {author} {\bibinfo {author} {\bibfnamefont {S.~L.}\ \bibnamefont
  {Rudge}}\ and\ \bibinfo {author} {\bibfnamefont {D.~S.}\ \bibnamefont
  {Kosov}},\ }\href {\doibase 10.1103/PhysRevB.99.115426} {\bibfield  {journal}
  {\bibinfo  {journal} {Phys. Rev. B}\ }\textbf {\bibinfo {volume} {99}},\
  \bibinfo {pages} {115426} (\bibinfo {year} {2019}{\natexlab{a}})}\BibitemShut
  {NoStop}%
\bibitem [{\citenamefont {Rudge}\ and\ \citenamefont
  {Kosov}(2019{\natexlab{b}})}]{Rudge2019c}%
  \BibitemOpen
  \bibfield  {author} {\bibinfo {author} {\bibfnamefont {S.~L.}\ \bibnamefont
  {Rudge}}\ and\ \bibinfo {author} {\bibfnamefont {D.~S.}\ \bibnamefont
  {Kosov}},\ }\href {\doibase 10.1103/PhysRevB.100.235430} {\bibfield
  {journal} {\bibinfo  {journal} {Phys. Rev. B}\ }\textbf {\bibinfo {volume}
  {100}},\ \bibinfo {pages} {235430} (\bibinfo {year}
  {2019}{\natexlab{b}})}\BibitemShut {NoStop}%
\end{thebibliography}

%

\end{document}